\documentclass[twoside,12pt]{article}
\usepackage{epsfig}
\usepackage{amssymb}
\newcommand{\be}{\begin{equation}}
\newcommand{\ee}{\end{equation}}
\newcommand{\bea}{\begin{eqnarray}}
\newcommand{\eea}{\end{eqnarray}}
\newcommand{\nn}{\nonumber}
\newcommand{\lb}{\label}

\begin{document}
\pagestyle{myheadings}
\markright{\LaTeX\ guidelines for Elsevier Major Reference Works}

\parindent 0mm
\parskip 6pt


\title{On the running coupling constant in QCD}

\author{G. M. Prosperi, M. Raciti and C. Simolo\\
Dipartimento di Fisica dell'Universit\`a di Milano\\
Istituto Nazionale di Fisica Nucleare, Sezione di Milano\\
Via Celoria 16, I20133 Milano (Italy)}

\date{15 April 2006}

\maketitle

\begin{abstract}
We try to review the main current ideas and points of view on the
running coupling constant in QCD. We begin by recalling briefly the
classic analysis based on the Renormalization Group Equations with
some emphasis
on the exact solutions  for a given number of loops,
in comparison with the usual approximate expressions. We
give particular attention to the problem of eliminating the
unphysical Landau singularities, and of defining a coupling that remains
significant at the infrared scales. We consider various proposal of
couplings directly related to the quark-antiquark potential 
or to other physical quantities (effective charges) and discuss optimization
in the choice of the scale parameter and of the renormalization scheme. 
Our main focus is, however,
on dispersive methods, their application, their relation with non-perturbative
effects. We try also to summarize the main results obtained by Lattice
simulations in particular in various MOM schemes. We conclude briefly recalling the
traditional comparison with the experimental data.
\end{abstract}
\tableofcontents
\vspace{-0.3truecm}
\section{Introduction}
\vspace{-0.3truecm}
The renormalization procedure in field theory consists in a redefinition
of the unrenormalized constants that appear in the Lagrangian, in such a
way that the observable quantities can be kept finite when the
ultraviolet cut off $\Lambda_{\rm UV}$ is removed. In the framework of
any given renormalization scheme (RS), dimensional reasons require the 
introduction of a new quantity $\mu$ with the dimension of a mass. 
Note that observables must be independent of the particular RS and 
of $\mu$ by definition. Intermediate quantities, like renormalized 
coupling constants, masses and field normalization factors depend on RS
and on $\mu$ by construction and a change in RS amounts to a
redefinition of such quantities. Obviously, approximate expressions 
of the observables depend in general on $\mu$ and RS.\\
In QED the masses of the charged particles have a direct physical
meaning, so there exists a natural scale referring to which 
$\mu$ can be fixed. In QCD, due to confinement, such a natural scale
does not exist and RS and $\mu$ have to be chosen with other
criteria, those of simplicity and of convergence.\\
In QCD we have a single coupling constant $g_{\rm s}$, or the usually
more convenient $\alpha_{\rm s}= {g_{\rm s}^2\over 4\pi}$, and various 
quark masses $m_f$ with $f=u,\,d,\,\dots, \,t$. We refer to their 
dependence on $\mu$ in the framework of a given RS ($\alpha_{\rm s}(\mu^2),~
m_f(\mu^2),~\dots$)  as to the running coupling constant, to the 
running masses and so on. $\overline{\rm MS}$ is the simplest and most 
commonly used scheme but, being essentially perturbative, alternative
definitions of $\alpha_{\rm s}(\mu^2)$ may be more appropriate in 
many cases.\\
Even in QED perturbative expansion is believed to be only 
asymptotic. As known this means that the approximation
to the considered quantity improves as the number of terms included
increases, until a maximum number $N_*$ is reached. After this the
single terms become progressively larger and the series loses any
meaning. In QED $\alpha \sim 1/137$ and $N_*\sim1 /\alpha \sim 137$,
so in practice no problem arises from this lack of convergence. However 
in QCD $\alpha_{\rm s}$ is nearly two orders of magnitude larger 
and $N_*$ is of order 1. An appropriate choice of $\mu$
and possibly of RS becomes therefore essential.\\
The $\mu$ dependence of the renormalized quantities is controlled by the
renormalization group (RG) equations.  
Let us concentrate on coupling constant $\alpha_{\rm s}(\mu^2)$ which is 
related to the unrenormalized constant, $ \alpha_s^{\rm ur}$, 
by the usual equation 
\be
\alpha_{\rm s}(\mu^2)\,\, Z_\alpha \left(\alpha_{\rm s}(\mu^2),\,{\mu^2 \over
    \Lambda^2_{\rm UV}} \right )\,=\,\alpha_s^{\rm ur}\,
\lb{intr1}
\ee 
where $\Lambda^2_{\rm UV}$ is the cut-off parameter. 
The corresponding differential equation
\be
\mu^2{d\alpha_{\rm s} \over d\mu^2}=\beta (\alpha_{\rm s})\,,
\lb{intr2}
\ee
can be obtained differentiating eq.(\ref{intr1}).
$\,\beta (\alpha_{\rm s})$ remains obviously finite as $\Lambda_{\rm UV} \to \infty$
and in perturbation theory takes the form
\be
\beta(\alpha_{\rm s})=-\alpha_{\rm s}^2(\beta_0 + \beta_1 \alpha_{\rm
  s} + \beta_2  \alpha_{\rm s}^2 + \dots )\,.
\lb{intr3}
\ee
As is known, the various terms in $\beta_0,~ \beta_1,~ \beta_2,~ 
\dots$ correspond to one loop, two loops, three loops \dots
contributions; $\beta_0$ and $ \beta_1$ are universal in the mass 
independent schemes.\\ 
Note that, for a general RS, $Z_\alpha$, $ \beta(\alpha_{\rm s})$,
$\beta_0,~\beta_1,~\dots$ depend also on quark masses through the
variables $m_f^2/\mu^2$, not explicitly
indicated. However, according to the decoupling theorem, all quarks with
masses much larger than the energy scale of interest (in
particular $m_f \gg \mu$) can be ignored. On the contrary, if $m_f \ll
\mu$, we can often neglect $m_f$. Then, the discussion can be greatly
simplified if for every $\mu$ we divide the quarks in active 
quarks with $m_f=0$ and inactive ones,
which we simply ignore. Within this framework $\beta_0,~
\beta_1,~\dots$ depend on $\mu$ only through the number of active 
quarks $n_f$, which changes by $\pm1$ any time $\mu$ crosses 
a quark threshold $m_f\,$. 
Furthermore the first two coefficients, $\beta_0$ and $\beta_1$, 
are RS independent, while all the others depend on the scheme. 
In the one loop approximation (i.e. keeping only the first term in 
(\ref{intr3})) eq.(\ref{intr2}) gives
\be
\alpha_{\rm s}(\mu^2)={\alpha_{\rm s}(\mu_0^2) \over 1+ 
\beta_0 \,\alpha_{\rm s}(\mu_0^2)\ln(\mu^2/\mu_0^2)}=
\alpha_{\rm s}(\mu_0^2) \sum_{n=0}^\infty \left (-  
\beta_0\, \alpha_{\rm s}(\mu_0^2)\ln {\mu^2 \over \mu_0^2}
\right)^n \,,
\lb{intr4}
\ee
which explicitly expresses $\alpha_{\rm s}$ at the $\mu$ scale as a
function of the same quantity at the $\mu_0$ scale. 
Eq. (\ref{intr4}) clearly shows that a change in the value of $\mu$
consists in a reorganization of the perturbative expansion of any
observable or, what is the same thing, in a resummation of various 
contributions. Setting
\be
\Lambda^2 = \mu_0^2 \exp \left [-{1 \over \beta_0}
{1 \over \alpha_{\rm s}(\mu_0^2)} \right ]\,,
\lb{intr5}
\ee
$\alpha_{\rm s}(\mu^2)$ can be written in terms of the overall scale
$\Lambda$, without any reference to a specific $\mu_0$
\be
\alpha_{\rm s}(\mu^2)= {1 \over \beta_0 \ln(\mu^2/\Lambda^2)}\,.
\lb{intr6}
\ee
As concerns the best choice of $\mu^2$ in a specific calculation, let us
consider the perturbative expansion of an amplitude or observable
$G(q,x)$ of canonical dimension 0. We assume $G$ written in terms of
an overall momentum $q$ and
a set of dimensionless variables (angles, Bjorken variables and so on) 
which we collectively denote by $x$. According to
the process we are considering $q$ may be space-like, $q^2<0$ 
(a momentum transfer), or time-like, $q^2>0$ (an energy). We shall often
set $q^2=\mp Q^2$ ($Q^2$ being as a rule positive) or also $q^2=s$ 
in the time-like case.\\
Under the above mentioned assumption that the active quark masses
can be neglected, we can write
\be
G = g_0\left({Q^2 \over \mu^2},x\right) + 
g_1\left({Q^2 \over \mu^2},x\right)
{\alpha_{\rm s}(\mu^2) \over \pi} + 
g_2\left({Q^2 \over \mu^2},x\right)
\left ({\alpha_{\rm s}(\mu^2)\over \pi}\right )^2 
+ \dots \,.
\lb{intr7}
\ee
As we mentioned, $G$ must be independent of $\mu^2$, so
\be   
\mu^2{d \over d \mu^2} G = \left ( \mu^2{\partial \over \partial
  \mu^2} + \beta (\alpha_{\rm s}){\partial \over \partial\alpha_{\rm s}}\right ) G = 0 \,.
\lb{intr8}
\ee
Then by replacing (\ref{intr7}) and  (\ref{intr3}) in  (\ref{intr8}), 
we obtain a
system of differential equation in the coefficients of (\ref{intr7}), 
whose solution gives (see e.g. sec. 3.3 for details)
\bea
&& g_0 =\overline g_0(x) \qquad \qquad g_1 =\overline g_1(x) \nn \\
&& g_2 =\overline g_2(x) - \pi\beta_0\,\overline g_1(x)
     \ln{Q^2 \over \mu^2}\lb{intr9} \\
&& g_3 =\overline g_3(x) - \left(\pi^2\beta_1 \overline g_1(x)
    +2\pi\beta_0\overline g_2(x)\right ) \ln{Q^2 \over \mu^2}+
     \pi^2\beta_0^2 \overline g_1(x) \ln^2{Q^2 \over \mu^2}
   \nn 
\eea
Let us assume that $\overline g_0(x), ~\overline g_1(x),~\dots$
decrease sufficiently fast in order (\ref{intr7}) to become significant when
$Q^2 \sim \mu^2$. This would clearly no longer be true for
very different values of $Q^2$. We must obviously modify
$\mu^2$ choosing it always on the scale of interest. In particular it
may be convenient to chose exactly $\mu^2 = Q^2$ and then
(\ref{intr7}) takes the form
\be
G(q,x) = \overline g_0(x)+\overline g_1(x) {\alpha_ {\rm s}(Q^2) 
 \over \pi} + 
\overline g_2(x)
\left ({\alpha_{\rm s}(Q^2)\over \pi}\right )^2 
+ \dots \,,
\lb{intr10}
\ee
showing that $\alpha_{\rm s}(Q^2)$ is actually the most convenient
expansion parameter for calculating the quantity $G(q,x)$.\\
Now let us go back to eq. (\ref{intr4}). We have $\beta_0=
{1 \over 4\pi}(11-2n_f/3)$ and, since the 
number of presently known  flavors is 6 , this remains positive 
at least in the entire range so far accessible. Then, for $\mu \to
\infty$ we have $\alpha_{\rm s}(\mu^2) \to 0$ (asymptotic freedom) 
and as we can see (see sec. 2) this remains true even if a larger 
number of terms is taken into account in (\ref{intr3}).\\
Conventionally  the mass of the $Z_0$ boson, $M_Z\sim 91.2$ GeV, 
is used as $\mu_0$ reference value. The world average 
of all determination is currently  
$\alpha_{\rm s}(M_Z^2)=0.1189\pm 0.0010$ \cite{bethke,pdg}.\\
With diminishing $Q^2$ we may expect expansion (\ref{intr7}) or 
(\ref{intr10}) to remain meaningful up to a few GeV if $q$ is 
space-like. The situation requires more attention for 
time-like $q$, $q^2=s>0$, as discussed in sec. 2.5. Then, in fact, in (\ref{intr9})
$\beta_0 \ln {Q^2 \over \mu^2} \to \beta_0 \ln {-s-i0 \over \mu^2} 
= \beta_0 (\ln {s \over \mu^2} -i\pi) $ and, even if we take
$\mu^2=s$ to control the logarithms, now the coefficients
 $\overline g_2,~\overline g_3~ \dots$ are modified for
terms  proportional to powers of $\beta_0 \pi$, which become 
rather large as the order increases (see sec. 2.5). Then the ``space-like'' coupling  
$\alpha_{\rm s}(s)$ remains a good expansion parameter for large $s$,
but for some intermediate $s$ may be more appropriate to define a new
``time-like'' coupling  $\tilde \alpha(s)$ in which the terms
in  $\beta_0 \pi$ are reabsorbed, so that the coefficients remain small. 
This can be easily done starting from eq.(\ref{intr10})
rather than (\ref{intr7}) and considering the analytic continuation of the
coupling rather than the coefficients. Naturally using the
time-like coupling rather than the  space-like one can be
interpreted as a change of RS and, as such, the two expressions become
identical for large $s$.\\
Note furthermore that  $\alpha_{\rm s}(\mu^2)\,$, as given by (\ref{intr6}), 
has a pole in $\mu^2=\Lambda^2\,$. In the $\overline {\rm MS}$
scheme, the expression remains singular even if more loops are taken into 
account in (\ref{intr3}), and the singularity structure changes. Such
singularities are obviously related to lack of convergence of the series
occurring in (\ref{intr4}) but they are clearly
non physical.
In fact in a generic scheme and in a generic gauge $\alpha_{\rm s}(\mu^2)\,$
is a kind of intermediate quantity only indirectly related to actual
observables and it is often chosen with criteria of formal simplicity. It
is just the case for the mentioned $\overline {\rm MS}$ scheme and for the 
Landau gauge to which as a rule we shall refer. 
On the contrary we know that on general grounds 
$G(q,x)$ is analytic in the entire $q^2$ complex plane apart from a cut 
on the real positive axis from some threshold to $+\infty$ and in its exact
expression the above non physical singularities must cancel. Correspondingly, 
eq.(\ref{intr10}) suggests that it is possible to define schemes in which 
$\alpha_{\rm s}(Q^2) $ can be extended again to the entire complex $Q^2$ plane 
apart from a cut on the real negative axis. 
To solve the problem of the elimination of the above spurious singularities
various strategies have been proposed. 
We remind here for instance the Brodsky-Lepage-Mackenzie (BLM) criterion 
\cite{Brodsky:1982gc}, which fixes the scale $\mu$ such that the 
next-to-leading order (NLO) coefficient of perturbative series is independent 
of $n_f\,$, and also the so-called ``Fastest Apparent Convergence'' (FAC) 
technique \cite{Grunberg:1980ja}, that amounts to setting the scale $\mu$ so 
that the NLO coefficient and all the higher ones are  zero. 
The latter is related to the 
{\it effective charges} \cite{Grunberg:1980ja}, \cite{dokshitzer2}-\cite{ALEPH}, 
which consist in defining new coupling constants $\alpha_{\rm eff}(Q^2)$ 
in more strict connection with some physical observable (see sec. 3.2). 
Typically let us assume that the 
quantity $G$ in (\ref{intr10}) is independent of $x$ (or $x$ has been 
fixed on some special value) and set exactly 
\be
G(Q^2)= \overline g_0 + \overline g_1 {\alpha_{\rm eff}(Q^2)  \over \pi}
\,. 
\lb{intr11}
\ee
Effective charges are obviously well defined everywhere in principle
and easily extracted from the data. They depend on the particular
observables chosen and, accordingly, can be either of the time-like or
space-like type. However, they can be related to each other by
referring back to the $ {\rm \overline {MS}} $ scheme. In fact, comparing 
(\ref{intr11}) and (\ref{intr10}), we have
\be
{\alpha_{\rm eff}(Q^2)  \over \pi}=
{\alpha_{\rm \overline {MS}}(Q^2) \over \pi} + 
{\overline g_2 \over \overline g_1} \left (
{\alpha_{\rm \overline {MS}}(Q^2)\over \pi}\right )^2 
+ \dots\,.
\lb{int7}
\ee
Another device is the Optimized Perturbation Theory (OPT)
\cite{stevenson}-\cite{higgs}, which consists in improving the 
convergence of the perturbative expansion by choosing the value 
of $\mu^2$ and of some of the RS parameters (like $\beta_2,~\beta_3$ 
themselves) at every order according to a criterion of minimum 
sensitivity. This has been discussed as a matter of example in some 
details in sec. 3.3, with reference to the theoretical predictions it
yields on the IR behavior of the coupling.\\
Finally another method , summarized in sec. 4, consists in requiring ab initio the desired
analyticity properties for $\alpha_{\rm s}(\mu^2)\,,$ by rewriting
this as a dispersion relation and applying the perturbative theory 
to the spectral function ${\rm Im}\,[\alpha_{\rm s}(-m^2-i0)]$ rather 
than to $\alpha_{\rm s}(\mu^2)$ itself 
\cite{Shirkov:1997wi}-\cite{Krasnikov:1995is}. In this way
Landau singularities are suppressed at their very roots. In
the one loop approximation from eq.(\ref{intr4}) we find
\be
{\rm Im}\,[\alpha_{\rm s}(-m^2-i0)]={\pi\beta_0\alpha^2_{\rm s}(\mu_0^2)
\over \left [1 + \beta_0   
\alpha_{\rm s}(\mu_0^2) \ln(m^2/\mu_0^2) \right ]^2
+(\pi\beta_0\alpha_{\rm s}(\mu_0^2))^2}\,,
\lb{int8}
\ee
which is finite everywhere on the positive $m^2$ axis and can be
considered in some way as a reorganization of the series occurring in the
last term in eq.(\ref{intr4}) after evaluation of the imaginary part.\\
It must be stressed that all these techniques lead to a finite
$\alpha_{\rm s}(0)$ and  are essentially consistent. Furthermore 
all of them use as input the coefficients of the power expansion of 
$\alpha_{\rm s}(\mu^2)$ itself or of specific observables in terms of  
$\alpha_{\rm s}(\mu_0^2)$. In this sense we may call them perturbative
\footnote{However, at the one loop level, the
analytic coupling can be written explicitly in terms of the ordinary
coupling in the form (see later)
$\alpha_{\rm an}=\alpha_{\rm s}+{1\over \beta_0}\left( 1-e^{1
  \over \beta_0 \alpha_{\rm s}}\right )^{-1}$
that has an essential singularity in $\alpha_{\rm s}=0$ and in this
sense can be said non perturbative. On the contrary the relation between 
$ \alpha_{\rm an}(\mu^2)$ and $\alpha_{\rm an}(\mu_0^2)$ 
is analytic.}.\\
However, we believe that there are  intrinsically non perturbative 
effects in the theory. These are in part related to the Gribov problem of 
the multiple solutions of the gauge fixing equation \cite{Gribov:1977wm}, 
and more seriously with the occurence of the singularities in $\alpha_{\rm s}=0$ 
which can not be obtained by any manipulation of perturbative expansions. 
Confinement and string tension, condensates, topologically non trivial 
classical solutions are of this type. In this context, unfortunately, 
we can not rely on any rigorous methods in the continuum. We have achieved 
some intuitions and very partial results, but we are far from a comprehensive 
understanding of the all the matter. Within this framework, what we
can do is to parametrize such non perturbative effects in terms of
some universal quantities that occur in different types of
phenomena. In particular, according to perturbation theory, in high 
energy processes observables should vanish logarithmically with 
the energy (apart from mass corrections). Non perturbative effects 
are expected to appear in terms of powers of the ${A_{\nu} \over 
Q^\nu}$ and ${A^\prime_{\nu} \over Q^\nu}\ln Q^2$ form (see sec. 4.5). 
Such terms should in principle be
detected  at somewhat intermediate energies even if the experimental
situation is not sufficiently clear. In this context the coupling is 
written as the sum of a perturbative $\alpha_{\rm s}^{\rm PT}(Q^2)$  
and a non perturbative part $\alpha_{\rm s}^{\rm NP}(Q^2)$. The latter 
is supposed to vanish fast enough out of the infrared region and the 
coefficients  $A_{\nu}$, $A^\prime_{\nu} $ can be expressed in terms of 
moments of $\alpha^{\rm NP}_{\rm s}(Q^2)$ or of its time-like counterpart
$\tilde \alpha_{\rm s}^{\rm NP}(m^2)$ \cite{dokshitzer2}.\\
The only really first principle method that we have to handle
non perturbative effects is to approximate the continuum with
a lattice and then rely on numerical simulations. As applied to the 
running coupling this method delivers results in substantial agreement with
those obtained in the continuum. The significance of the latter
result is however uncertain, due to finite size limitations.\\  
In this review the emphasis is put on the IR behavior of the
coupling constant and in particular on the analytic methods. 
We have tried to give a unified presentation of the matter. To this end we
often took the liberty of modifying notations used in the original
papers. Due to the extent of the literature on the subject, 
and to space limitation, we had to make appropriate choices. 
In this sense we do not claim our paper to be exhaustive or complete. 
The choice we have made is obviously due to our personal 
taste or to the specific point of view we have adopted. 
In particular in some cases we have discussed with a greater detail
some specific papers as an example of a more general methodology. In
such a context we apologize in advance for giving less attention 
to some papers than to others, even very significant.
As a rule, we have kept to a general theoretical level and
referred to application only occasionally. Apart from a brief
summary of significant results in the concluding remarks, no
attempt was made to discuss the actual methods by which the coupling is
extracted from experiments at the various energies. For this we
refer interested people to specific reviews on the subject such as 
\cite{bethke} and \cite{pdg}.\\
The plan of the paper is as follows. 
In sec. 2 we have summarized the main results of the usual perturbation
theory as reference. For details and bibliography on this subject we 
refer to standard books and reviews (see e.g. 
\cite{monograph,Shirkov:1999hj,Collins,Yndurain}). 
We have tried to
emphasize the importance of exact solutions to renormalization group
equations with a given number of loops in comparison with the usual iterative
solutions appropriate for high energies. This was due to our 
interest in the IR region \footnote{In this connection it should be
recalled that the expansion in numbers of loops is an expansion in
the Planck constant $\hbar$ and is conceptually different, even if related,
to a power coupling expansion.}. Some attention has also been  given to the
need to define a specific time-like coupling. 
In sec. 3 we have considered some purely phenomenological modifications
of the coupling constant expression, suggested essentially by the
quark-antiquark potential theory. We have also discussed some significant 
examples of physical couplings, a classic application of the optimized 
perturbation theory.
Sec. 4 is entirely devoted to the dispersive approach, analytic
coupling constant and relation between space-like and time-like 
expressions, 
the parametrization of non-perturbative effects and the so 
called Analytic Perturbation Theory.
In sec. 5 we have tried to summarize the present state of the
studies on the running coupling on the lattice. 
In sec. 6 we outline our conclusive remarks 
and also discuss briefly some information 
on the coupling under 1 GeV that can be obtained
from relativistic studies of the light
quark-antiquark spectrum and from a parametrization of other non-perturbative contributions.
\section{Perturbation Theory}
\vspace{-0.3truecm}
\subsection{QCD $\beta$ function}
\vspace{-0.3truecm}
As we mentioned, eq. (\ref{intr2}) is derived applying the
operator $\mu^2(d/d\mu^2)$ to eq. (\ref{intr1}). Thus one 
needs to compute the renormalization factor for the coupling 
$Z_\alpha$ and this can be accomplished in several ways. One 
can start from the quark-gluon vertex $Z_{\bar q qg}$ 
together with renormalization factors of quark and gluon 
propagators $Z_{q}$ and $Z_{g}$ to obtain 
$Z_{\alpha}=Z_{\bar q qg}^{2}Z_{q}^{-2}Z_{g}^{-1}$, but other 
choices, as ghost-ghost-gluon vertex or trilinear and quartic 
gluon interactions, equally work. If dimensional regularization 
is used the limit $\Lambda_{\rm UV} \to \infty $ can be anticipate 
as long as the space dimension $D=4-\varepsilon$ is kept 
different from 4. Taking into account the physical dimension of the
fields in $D$-dimensional space the coupling constant acquires a not
vanishing physical dimension, thus it is usual the parametrization 
$\alpha_{\rm s}^{\rm ur}=\mu^{\varepsilon}\alpha_{\rm s}Z_{\alpha}\,$,
where in the $\overline{\rm MS}$ renormalization scheme 
\be
Z_{\alpha}=1+\sum_{n=1}^{\infty}
\varepsilon^{-n}Z^{(n)}_{\alpha}(\alpha_{\rm{s}})
\ee
and then
\be
\beta(\alpha_{\rm{s}})=\frac{1}{2}\alpha_{\rm{s}}^2 \frac{d}{d\alpha_{\rm{s}}}
Z_{\alpha}^{(1)}(\alpha_{\rm{s}})\,.
\lb{beta2}
\ee
From this equation and the explicit perturbative form of 
$Z_{\alpha}^{(1)}$ we have \cite{vanRitbergen:1997va}
\bea
&&\beta_{0}={1 \over 4\pi} \left[11-\frac{2}{3}n_{f}\right]\nonumber\\
&&\beta_{1}={1 \over (4\pi)^2} \left[102-\frac{38}{3}n_{f}\right]\nonumber\\
&&\beta_{2}={1 \over (4\pi)^3} \left[\frac{2857}{2}
   -\frac{5033}{18}n_{f}+\frac{325}{54}n_{f}^2\right]\nonumber\\
&&\beta_{3}={1 \over (4\pi)^4} \left[\Big(\frac{149753}{6}+
   3564\zeta_{3}\Big)-\Big(\frac{1078361}{162}+
  \frac{6508}{27}\zeta_{3}\Big)n_{f}\right . \nonumber\\
&&\qquad+\left . \Big(\frac{50065}{162}+\frac{6472}{81}\zeta_{3}\Big)n_{f}^2
  +\frac{1093}{729}n_f^3\right]
\lb{beta-coeffs}
\vspace{-0.5truecm}
\eea
where $\zeta_\nu$ is the Riemann zeta-function, $\zeta_3\simeq1.202057$. 
The coefficients $\beta_{j}$ generally depend on the RS employed,
but the first two are universal among the massless schemes.
Moreover, in the $\rm{MS}$-scheme the $\beta$-function is 
gauge-independent at any order \cite{Caswell:1974cj}, and in an
arbitrary mass-dependent scheme this feature is preserved only 
at the first order.\\ 
As well known, the universal one-loop coefficient \cite{Gross:1973id} 
has a positive sign provided there is a small enough number of quark 
fields $(n_f\le33/2)$, thus the theory is asymptotically free, that 
is the $\beta$-function has a stable UV fixed point as its argument 
approaches zero; indeed this coefficient is the sum of two 
contributions, the relevant one with respect to asymptotic freedom 
property being the first, which arises from pure gauge field  
effects i.e. from the nonlinear Yang-Mills interaction terms. The 
two-loop coefficient has been computed for the first time in 
\cite{Caswell:1974gg} and is positive up to $n_f=8\,$.\\ 
Higher order approximations are scheme-sensitive, and it is common 
practice to perform computation within MS or $\overline{\mathrm{MS}}$ 
procedures which have the same the $\beta$-function. The first 
calculation of three-loop coefficient is due to \cite{Tarasov:1980au}, 
where the ghost-ghost-gluon combination in the Feynman gauge was used. 
In a more recent work \cite{Larin:1993tp} the quark-gluon vertex 
was instead employed, providing an independent check in an arbitrary 
covariant gauge of the previous result and its gauge-independence. 
Finally, the original four-loop calculation \cite{vanRitbergen:1997va}
has been performed using the ghost-ghost-gluon vertex in a arbitrary
covariant gauge, and for a generic semi-simple compact Lie symmetry 
group. The result turns out to be gauge-independent as expected 
within MS procedure, and involves higher order group invariants 
such as quartic Casimir operators; specialized to the standard 
SU(3) symmetry, the four-loop coefficient is a positive number 
for every positive $n_f$ (see also \cite{Czakon:2004bu} ). 
Finally note that all four coefficients 
are positive up to $n_f=6$ except for $\beta_2$ which is negative
for $6\le n_f\le40\,$. 
\vspace{-0.5truecm}
\subsection{Running coupling}
\vspace{-0.3truecm}
Evolution of QCD running coupling can be gained integrating the
differential equation (\ref{intr2}), that can be rewritten as
\be
\ln {\mu^2 \over \mu_0^2 }= \int_{\alpha_{\rm{s}}(\mu_0^2)}^{\alpha_{\rm{s}}(\mu^2)}
{d \alpha \over \beta(\alpha)}
\lb{RGScoupl}
\ee
The exact one-loop solution (\ref{intr4}) or (\ref{intr6}) is 
obtained by straightforward integration retaining only the first 
term in (\ref{intr3}). We report for reference the second form 
\be
\alpha_{\rm s}(\mu^2)= {1 \over \beta_0 \ln\left(\mu^2/\Lambda^2\right)}\,.
\lb{1loopEC2}
\ee
As yet noted the dimensional scale $\Lambda$ keeps track of the 
initial parametrization $(\mu_0,\alpha_{\rm{s}}(\mu_0^2))$, and it 
is scale-invariant; its value is not predicted by the theory 
but must be extracted from a measurement of $\alpha_{\rm{s}}$ at a 
given reference scale. 
Emergence of a scale parameter, sometimes referred to as dimensional 
transmutation, breaks naive scale invariance of the massless theory, 
and it is commonly believed to be associated with the typical hadron 
size i.e. to the energy range where confinement effects set in. 
Roughly speaking, $\Lambda$ is the scale at which the (one-loop) coupling 
diverges (Landau ghost), and perturbation 
theory becomes meaningless. Further, it is scheme-dependent and receives 
further corrections at each loop level, but for simplicity we use the 
same notation throughout.\\
In the next loop level the integration of (\ref{RGScoupl}) leads to a 
transcendental equation, that is, starting from the two-loop 
approximation to the $\beta$-function in (\ref{intr3}), 
straightforward integration in (\ref{RGScoupl}) yields
\be
\ln\frac{\mu^2}{\mu_0^2}=\textrm{C}+\frac{1}{\beta_0\alpha_{\rm{s}}}+B_1\ln\alpha_{\rm{s}} 
-B_1\ln\left(1+\frac{\beta_1}{\beta_0}\alpha_{\rm{s}}\right)
\lb{RG2loop-int}
\ee
where $B_1=\beta_1/\beta_0^2\,$ and the constant term from the 
lower end points can be again reabsorbed into the 
$\Lambda$-parametrization, with the commonly adopted prescription 
(see e.g.\cite{Bardeen:1978yd,Collins})
\be
\ln\frac{\Lambda^2}{\mu_0^2}=\textrm{C}-B_1\ln\beta_0
\lb{lambdafixing}
\ee
which fixes a specific choice for $\Lambda\,$. Thus we get the 
two-loop implicit solution 
\be
\ln\frac{\mu^2}{\Lambda^2}=\frac{1}{\beta_0\alpha_{\rm{s}}}-B_1\ln\left(B_1+
\frac{1}{\beta_0\alpha_{\rm{s}}}\right)\,,
\lb{RG2loop-impl}
\ee
from which the two-loop scaling constant is immediately read with $\mu=\mu_0\,$. 
To achieve an explicit expression for the running coupling 
at this level one should resort to the many-valued Lambert function 
$W(\zeta)$ defined by 
\be
W(\zeta)\exp\left[W(\zeta)\right]=\zeta\,,
\lb{Lamb}
\ee
which has an infinite number of branches $W_k(\zeta)\,$ $k=0,\pm1,\pm2\dots$ 
such that $W_n^*(\zeta)=W_{-n}(\zeta^*)\,$
(for more details we refer to \cite{Corless:1996}). 
The exact solution to eq.(\ref{RG2loop-impl}), being $B_1$ positive, 
for $0\le n_f\le8$ reads\footnote{Note that if $9\le n_f\le16$
the principal branch $W_0$ is involved, but here and throughout our 
discussion is focused on the physical values $n_f\le6\,$.} (see e.g. 
\cite{Grunberg98})
\be
\alpha_{\rm{ex}}^{(2)}(z)=-\frac{1}{\beta_0B_1}\frac{1}{1+W_{-1}(\zeta)}
\qquad\zeta=-\frac{1}{eB_1}\left(\frac{1}{z}\right)^{1/B_1}
\lb{RG2loop-ex}
\ee
where $z=\mu^2/\Lambda^2\,$, and $W_{-1}(\zeta)$ is the ``physical'' 
branch of the Lambert function, i.e. it defines a regular real values
function for $\zeta\in(-e^{-1},0)\,$ which fulfills the asymptotic 
freedom constraint. Indeed, $W_{-1}(\zeta)$ as a function of complex 
variable has a branch cut along the negative real axis (actually a 
superimposition of two cuts starting from $-\infty$ to $-e^{-1}$ and
to $0$ respectively), and defining it on the cut coming from the upper
complex plane it assumes real values in the interval $(-e^{-1},0)\,$, 
with the asymptotics 
\be
W_{-1}(-\varepsilon)=\ln\varepsilon+O(\ln|\ln\varepsilon|)
\lb{lamb1}
\ee
\be
W_{-1}\left(-\frac{1}{e}+\varepsilon\right)=-1-\sqrt{2e\varepsilon}+O(\varepsilon)
\lb{lamb2}
\ee
as $\varepsilon\to0^+\,$. Outside this region of the real axis 
it takes on complex values, 
indeed it is not a real analytic function. 
Though not easy for practical aims, eq.(\ref{RG2loop-ex}) yields 
the most accurate expression for investigating the IR behavior of 
the running coupling, since it has not been derived by means of deep 
perturbative approximations (aside from the truncation of the two-loop 
$\beta$ function). 
Actually, a frequently used two-loop approximate solution, known as the 
\emph{iterative} solution (\cite{Shirkov:1997wi}), is obtained 
starting from eq.(\ref{RG2loop-int}) together with a single iteration 
of the one-loop formula (\ref{1loopEC2}), that is  
\be
\alpha_{\rm{it}}^{(2)}(z)=\frac{\beta_0^{-1}}{\ln z+B_1\ln(1+B_1^{-1}\ln z)}\,,
\lb{2loopECit}
\ee
where $z=\mu^2/\tilde\Lambda^2\,$, and $\tilde\Lambda$ now defined by
\be
\ln\frac{\tilde\Lambda^2}{\mu_0^2}=\textrm{C}-B_1\ln\frac{\beta_1}{\beta_0}
\lb{delta2}
\ee
in (\ref{RG2loop-int}), related to the standard one (\ref{lambdafixing})
by the factor
\be
\ln(\Lambda/\tilde{\Lambda})=\frac{1}{2}B_1\ln{B_1}\,.
\lb{conv1Lambda}
\ee  
However, the commonly used two-loop solution is an asymptotic formula 
which strictly relies on the smallness of $\alpha_{\rm{s}}\,$ for fairly 
large $\mu^2$, since it amounts on solving eq.(\ref{RG2loop-int}) (with the
choice (\ref{lambdafixing})) where 
the last term in the r.h.s. has been neglected. Again after one iteration 
of the one-loop formula, the result is then re-expanded in powers of $1/L$, 
where $L=\ln z$ and $z$ as before
\be
\alpha_{\rm{s}}^{(2)}(z)=\frac{1}{\beta_0\ln z}\,\left[1-\frac{\beta_1}{\beta_0^2}
\frac{\ln\left(\ln z\right)}{\ln z}\right]\,.
\lb{2loopEC}
\ee 
Eq.(\ref{2loopEC}) is known as the standard two-loop running coupling 
and works only in the deep UV regime, i.e. for $L\gg 1\,$ (see fig.1(a)).\\
Under the same assumptions one can easily derive the three and four-loop 
approximate formulas; at any loop level asymptotic RG solutions are 
obtained as a rule through a recursive recipe involving the previous 
order result as an input in the transcendental equation arising from 
integration of (\ref{RGScoupl}). Starting with the approximate 
implicit solution at four-loop level (obtained by expanding the 
integrand on the r.h.s. of eq.(\ref{RGScoupl}))
\be
\ln\frac{\mu^2}{\mu_0^2}=\textrm{C}+\frac{1}{\beta_0\alpha_{\rm{s}}}+\frac{\beta_1}{\beta_0^2}
\ln\alpha_{\rm{s}}+\frac{\beta_2\beta_0-\beta_1^2}{\beta_0^3}\alpha_{\rm{s}}
+\frac{\beta_1^3-2\beta_0\beta_1\beta_2+\beta_0^2\beta_3}{2\beta_0^4}\alpha_{\rm{s}}^2
+O(\alpha_{\rm{s}}^3)\,,
\lb{RG4loop-int}
\ee
the analogous trick which led to eq.(\ref{2loopEC}) yields the UV 
asymptotic four-loop running coupling in the standard form of an
expansion in inverse powers of the logarithm $L$ for $L\gg1$ (see 
e.g. \cite{Chetyrkin:1997sg})
\bea
&&\alpha_{\rm{s}}^{(4)}(\mu^2)=\frac{1}{\beta_0\,L}\,
\left\{1-\frac{\beta_1}{\beta_0^2}\frac{\ln L}{L}+
\frac{1}{\beta_0^2 L^2}\left[\frac{\beta_1^2}{\beta_0^2}
\left(\ln^2L-\ln L-1\right)+\frac{\beta_2}{\beta_0}\right]+\right.\nn\\
&&\left.\frac{1}{\beta_0^3L^3}\left[\frac{\beta_1^3}{\beta_0^3}\left(
-\ln^3L+\frac{5}{2}\ln^2L+2\ln L -\frac{1}{2}\right)-3\frac{\beta_1\beta_2}
{\beta_0^2}\ln L+\frac{\beta_3}{2\beta_0}\right]\right\}\,
\lb{4loopEC}
\eea
which turns out to be nearly indistinguishable from the three-loop
curve (see fig.1(b)). Here the one-loop solution eq.(\ref{1loopEC2}) has been 
emphasized, and being the leading UV term in (\ref{4loopEC}), it 
defines the asymptotic behavior of the perturbative running coupling,  
i.e. it clearly exhibits the asymptotic freedom property. 
On the other hand, the two and three-loop asymptotic expressions are 
easily read from eq.(\ref{4loopEC}) by keeping only the first two or
three terms respectively inside the curly bracket.
\begin{figure}[t]
\begin{picture}(150,230)
 \put(-10,235){\mbox{\epsfig{file=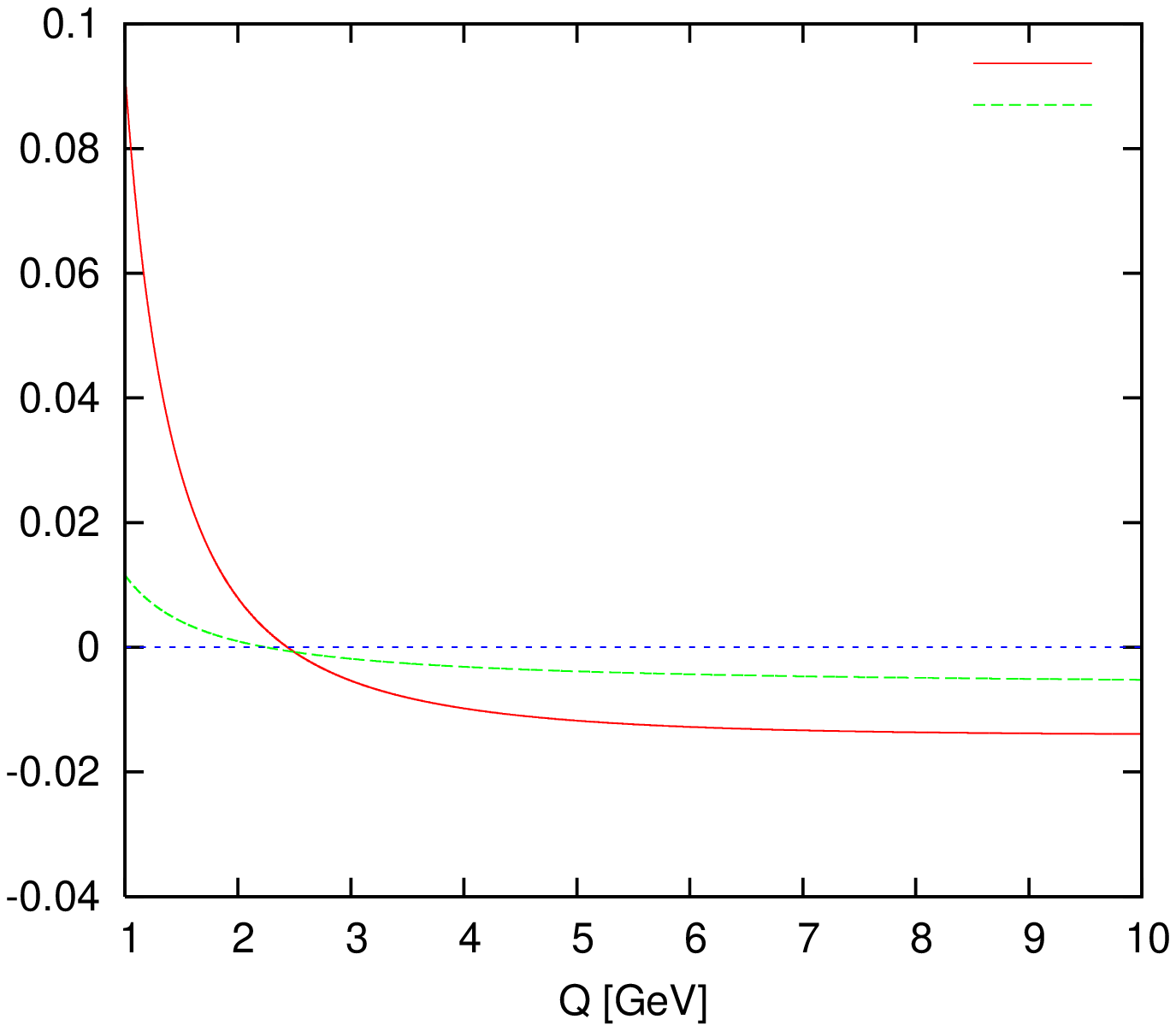,height=8.5cm,width=6.5cm,angle=-90}}}
\put(140,219){ {\tiny $1-\alpha_{\rm ex}/\alpha_{\rm s}$} }  
 \put(140,211){ {\tiny $1-\alpha_{\rm ex}/\alpha_{\rm it}$} } 
\put(240,50){\mbox{\epsfig{file=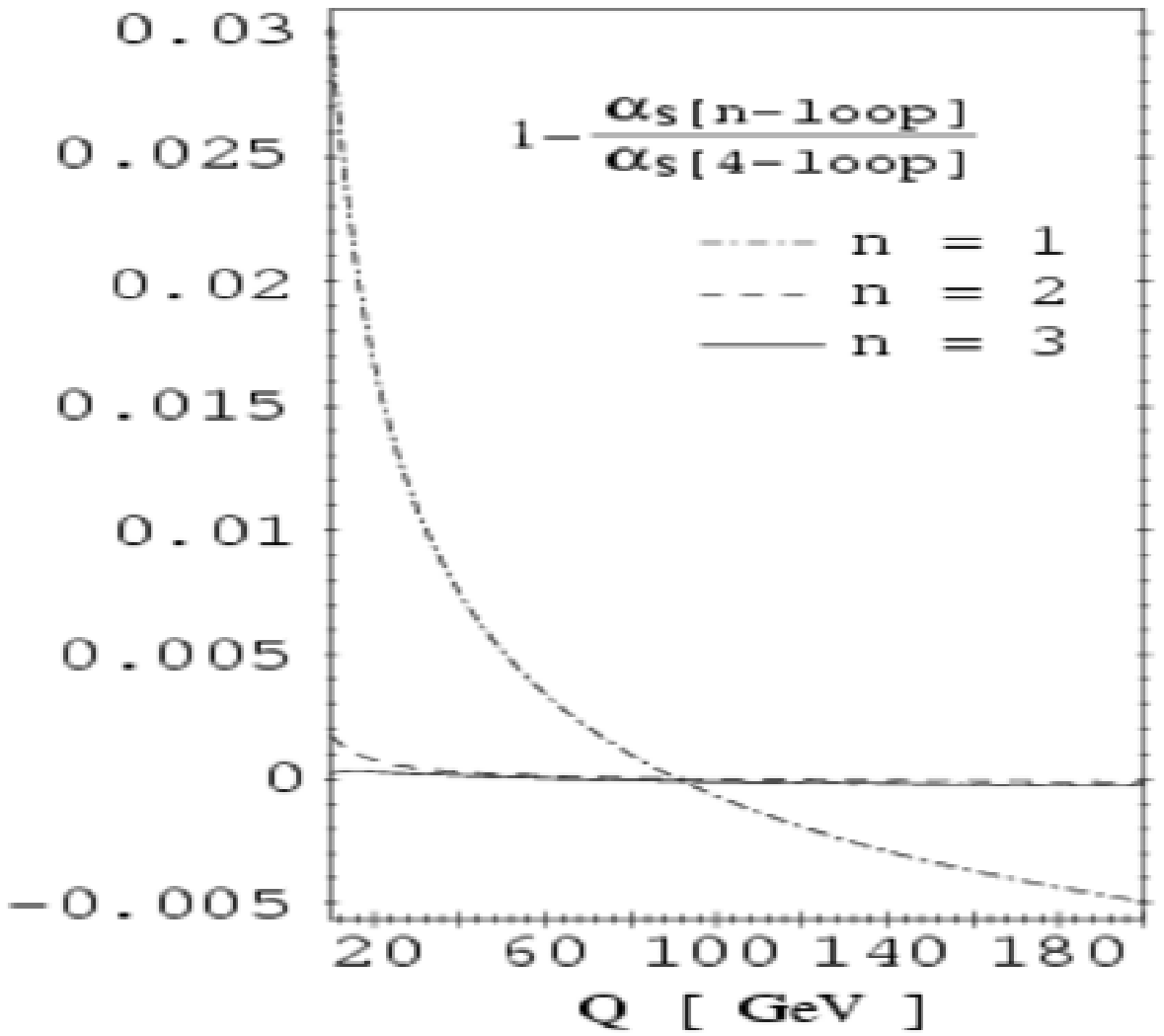,height=6.5cm,width=4.8cm}}}
\put(20,50){ {\footnotesize (a)} } 
\put(273,50){ {\footnotesize (b)}} 
\end{picture}
\vspace{-1.9truecm}
\caption{\footnotesize (a) We show the fractional difference 
between eq.(\ref{RG2loop-ex}) and eqs.(\ref{2loopEC}), (\ref{2loopECit}) (solid and dashed 
line respectively), with $\Lambda=350\,$MeV and $n_f=4\,$. 
(b) From\cite{bethke}: it is displayed the fractional difference between 
the 4-loop and the 1-, 2-, 3-loop approximations to eq.(\ref{4loopEC}), with $n_f=5$
and normalizing conditions for all curves at $\alpha_{\rm s}(M_Z^2)=0.119\,$.
\vspace{-0.3truecm}} 
\label{fig:als-asord}
\end{figure}
We recall that exact integration of the truncated four (or
three)-loop $\beta$-function leads to a more involved structure 
than eq.(\ref{RG4loop-int}), which poses serious difficulties in 
finding its inverse; nevertheless, in \cite{Grunberg98} a useful 
solution has been still worked out at three-loop level via the 
real branch $W_{-1}(\zeta)$ of the Lambert function together with 
the Pade' Approximant of the related $\beta$-function. Moreover, 
in \cite{Kourashev:1999ye} a reliable approximation to higher order 
running coupling has been suggested, via a power expansion in the 
two-loop exact coupling eq.(\ref{RG2loop-ex}), of the form
\be
\alpha_{\rm{s}}^{(k)}(\mu^2)=\sum_{n\ge1} p_n^{(k)}
\left[\alpha_{\rm{ex}}^{(2)}(\mu^2)\right]^n
\lb{alpha^k}
\ee
with $k\ge3$ the loop order, and $p_n^{(k)}$ proper functions of the
coefficients $\beta_j$. Comparison \cite{Kurashev:2003pt} of these 
multi-loop approximants to the coupling, with the relative asymptotic 
formulas, eq.(\ref{4loopEC}) and the three-loop analogue, reveals 
the better agreement of the former ones with the higher-loop exact 
coupling numerically estimated (i.e. starting from the exact 
implicit solution), even at low scales (see also 
\cite{Magradze:2005ab}).\\
Finally, being the definition of the scaling constant $\Lambda$ 
not completely unambiguous, few comments are in order. As yet 
pointed out, starting from the two-loop level an arbitrary 
constant has to be fixed for $\Lambda$ being uniquely defined; 
beside the commonly accepted convention (\ref{lambdafixing}) 
(or (\ref{delta2})), we just mention that other prescriptions 
have been proposed \cite{Radyushkin:1982kg} in order to optimize the 
$1/L$-expansion for the running coupling, while (\ref{lambdafixing}) 
does remain the preferred one as no further terms of order 
$1/L^2$ appear in the two-loop asymptotic formula (\ref{2loopEC}). 
Thus, in the higher-loop levels the scaling constant is analogously  
related to the initial parameterization, and the four (and three) loop 
value reproducing the world average, roughly $\alpha(M_Z^2)=0.119$, is 
$\Lambda^{(n_f=5)}_{\overline{\mathrm{MS}}}=220$ MeV with 
consistently five active flavors \cite{bethke}.\\
A last remark concerns the scheme-dependence of the coupling and
the scale parameter. Restricting ourselves to mass-independent 
RS (as MS-like schemes or trivially any prescription in the 
massless theory), renormalized coupling constants (see 
eq.(\ref{intr1})) in two such different schemes can be related 
perturbatively at any fixed scale $\mu\,$
\be
\alpha_{\rm{s}}'=\alpha_{\rm{s}}\left[1+v_1\frac{\alpha_{\rm{s}}}{\pi}+
v_2\left(\frac{\alpha_{\rm{s}}}{\pi}\right)^2
+\dots\,\right]\,.
\label{alphaconv}
\ee
Then, it is easy to verify that the first two coefficients of 
the relative $\beta$-functions do not change as the 
renormalization prescription is changed, while, for instance,
the third ones are related by 
$\beta'_2=\beta_2-v_1\beta_1+v_2\beta_0-\beta_0v_1^2\,$. 
As a result the running coupling at each loop-level (e.g. 
eq.(\ref{4loopEC})) depends on the RS, through the 
coefficients $\beta_j$ with $j\ge2$ and the initial 
parameterization as well. The latter obviously amounts 
in suitably adjusting the scaling constant $\Lambda\,$,  
and the relation is given exactly by one-loop calculation 
\cite{Gonsalves79}
\begin{equation} 
\ln\left(\Lambda'/\Lambda\right)=\frac{v_1}{2\pi\beta_0}\,,  
\label{conv2lambda}
\end{equation}
which works through all orders; e.g. 
$\Lambda_{\mathrm{\overline{MS}}}/\Lambda_{\mathrm{MS}}=
\exp{[(\ln4\pi-\gamma_e)/2]}\simeq2.66$, and with $n_f=4$  
we roughly have \cite{Gonsalves79}
$\Lambda_{\mathrm{MOM}}/\Lambda_{\mathrm{MS}}\simeq 4.76$ where 
$\Lambda_{\mathrm{MOM}}$ refers to a scheme in which the 
subtraction of the relevant green functions is performed
in the symmetric point and the renormalized coupling is
defined through the three gluon vertex.
\vspace{-0.2truecm}
\subsection{Threshold matching}
\vspace{-0.3truecm}
Quark mass effects, till now ignored, reveal themselves in 
explicit corrections within higher order perturbation theory 
(see e.g. \cite{Bernreuther:1997jn}), and in the energy 
dependence of the effective (running) quark masses as a 
result of the RG improvement \cite{Vermaseren:1997fq}, 
but we will not go into further details being these topics 
beyond the scope of this report.
A direct effect of the quark masses 
on the evolution of the coupling is, as yet noted, through 
the dependence of the $\beta$ coefficients on the number 
$n_f\,$ of active quarks. A quark is active if $m_f\ll\mu\,$, 
where $\mu$ is the renormalization scale, and $m_f$ the 
$\rm{\overline{MS}}$ quark mass (see e.g. 
\cite{pdg}); the definition can be also formulated in terms of 
the pole mass $M_f\,$ that can be related to the former 
\cite{Melnikov:2000qh}. 
Indeed, within 
MS-like RS, decoupling of heavy quarks 
\cite{Appelquist:1974tg} is made explicit by constructing beside 
the full $n_f$ flavors QCD an $(n_f-1)$ effective theory 
below a heavy quark threshold \cite{Bernreuther:1983zp}. 
Then, to have a unique theory on the whole range, the 
two couplings $\alpha_{\rm{s}}^{(n_f)}$ and 
$\alpha_{\rm{s}}^{(n_f-1)}$ must be matched at each heavy quark 
mass scale $\mu^{(n_f)}=O(m_f)\,$. As a result the scaling constant 
$\Lambda$ depends also on $n_f\,$ (see e.g.
\cite{Larin:1994va,Chetyrkin:1997sg,Chetyrkin:1997un}).\\
The most straightforward way is to impose continuity of 
$\alpha_{\rm{s}}$ by means of the matching condition 
$\alpha_{\rm{s}}^{(n_f-1)}(m^2_f)=\alpha_{\rm{s}}^{(n_f)}(m^2_f)$, 
which works up to next-to-leading order. At the one-loop 
level e.g., it translates into
\be
\Lambda^{(n_f)}=\Lambda^{(n_f-1)}\left[\frac{\Lambda^{(n_f-1)}}{m_f}\right]
^{2/(33-2n_f)}\,,
\lb{Lmatch}
\ee
which can be extended up to two-loop, and exhibits explicit 
dependence on the $m_f$ values. Since trivial matching does 
not generally hold in higher orders within $\overline{\rm{MS}}$ 
scheme, a more accurate formula is required in this case (see 
\cite{Chetyrkin:1997sg} and refs. therein); specifically 
to obtain the global evolution of the four-loop coupling the 
proper matching condition reads \cite{Chetyrkin:1997sg}
\be
\alpha_{\rm{s}}^{(n_f-1)}=\alpha_{\rm{s}}^{(n_f)}\left[1+
k_2\left(\frac{\alpha_{\rm{s}}^{(n_f)}}{\pi}\right)^2 +
k_3\left(\frac{\alpha_{\rm{s}}^{(n_f)}}{\pi}\right)^3\right]
\lb{L3match}
\ee
with
\be
k_2=\frac{11}{72}\,,\qquad k_3=\frac{564731}{124416}-
\frac{82043}{27648}\zeta_3-\frac{2633}{31104}(n_f-1)
\lb{k_j} 
\ee
if $\mu^{(n_f)}=m_f\,$ is exactly assumed. With this convention, 
eq.(\ref{L3match}) yields the relationship between the scaling 
constants $\Lambda^{(n_f-1)}$ and $\Lambda^{(n_f)}$ in the 
$\overline{\rm{MS}}$ scheme (see \cite{Chetyrkin:1997sg}). Note
that one can equally fix $\mu^{(n_f)}=M_f\,$, that amounts to a 
proper adjustment of the coefficients in (\ref{L3match}); for 
instance \cite{bethke} starting with (\ref{4loopEC}) and 
$\Lambda^{(n_f=5)}=220\,$MeV, the values  
$\Lambda^{(n_f=4)}=305\,$MeV and $\Lambda^{(n_f=3)}=346\,$MeV are
obtained, with threshold fixed at the pole masses $M_b=4.7\,$GeV
and $M_c=1.5\,$GeV.   
Finally, we observe that eq.(\ref{L3match}) clearly spoils 
continuity of $\alpha_{\rm{s}}\,$; thereby, whenever continuity of the 
global coupling and of its first derivative is mandatory, 
one can resort to a more sophisticated technique 
\cite{Shirkov:1990vw}, which relying upon mass-dependent 
RS yields a smooth transition across thresholds.

\vspace{-0.2truecm}
\subsection{Landau singularities}
\vspace{-0.3truecm}
The running coupling is not an observable by itself but plays the
role of the expansion parameter for physical quantities. At one 
loop-level (cf. eq.(\ref{intr4})), it resums an infinite series of 
leading-logs UV contributions 
and similarly the two-loop solution yields the next-to-leading-logs
approximation; 
thus RG formalism provides an iterative recipe 
for improving perturbative results.\\
However, as a result, the RG resummation significantly modifies the 
analytical structure of the series in the complex plane. As yet noted
the one-loop coupling (\ref{1loopEC2}) is clearly affected by a 
spacelike pole at $\Lambda$ with residue $1/\beta_0\,$ (Landau ghost). 
Adding multi-loop corrections does not overcome the
hurdle. Rather the singularity structure of the higher 
order solutions is more involved due to the log-of-log dependence, 
so that a branch cut adds on to the one-loop single pole in the IR 
domain of the spacelike axis. 
Moreover, at a given loop level Landau singularities sensibly 
depend upon the approximation used. For instance, the two-loop 
iterative formula (\ref{2loopECit}) has a pole at $z=1$ 
($\mu=\tilde\Lambda\,$, see eqs.(\ref{delta2}) and (\ref{conv1Lambda}))  
with residue $1/(2\beta_0)\,$, and a cut for $0<z<\exp{(-B_1)}$ due 
to the double logarithm. On the other hand, when considering the same 
loop approximation (\ref{2loopEC}), we note how the singularity 
in $z=1$ already acquires a stronger character
\be
\alpha_{\rm{s}}^{(2)}(z)\simeq -\frac{B_1}{\beta_0}\frac{\ln(z-1)}{(z-1)^2}
\qquad z\to1
\lb{IR2loop}
\ee
with $z=\mu^2/\Lambda^2\,$, and the cut now runs from 0 to 1.
Analogously the three and four-loop asymptotic solutions, as given by
eq.(\ref{4loopEC}), nearby the Landau ghost respectively becomes  
\be
\alpha_{\rm{s}}^{(3)}(z)\simeq \frac{B^2_1}{\beta_0}\frac{\ln^2(z-1)}{(z-1)^3}\,,
\qquad\alpha_{\rm{s}}^{(4)}(z)\simeq -\frac{B^3_1}{\beta_0}\frac{\ln^3(z-1)}{(z-1)^4}
\qquad z\to1
\lb{IR3loop}
\ee
and are equally affected by an unphysical cut. 
However, the cumbersome singularity structure of the leading 
Landau ghost, and of the unphysical cut as well, are an 
artifact of the UV approximations introduced. Therefore to deal 
with the IR behavior of the running coupling, e.g. at two-loop 
level, it is necessary to face with the exact solution 
(\ref{RG2loop-ex}); clearly it is singular when 
$W_{-1}(\zeta)=-1$ that is at $z=B_1^{-B_1}\,$ 
($\mu^2=B_1^{-B_1}\Lambda^2\,$), corresponding to the branch 
point $\zeta=-1/e$ of the Lambert function. Nearby this point, 
due to the asymptotic (\ref{lamb2}) of $W_{-1}(\zeta)\,$, we have
\be
\alpha_{\rm{ex}}^{(2)}(z)\simeq\frac{1}{\beta_0}\sqrt{\frac{B_1^{-B_1-1}}{2}}
\left[z-B_1^{-B_1}\right]^{-1/2}\,.
\lb{IR2loop-ex}
\ee
i.e., an integrable singularity (note that the awkward factor 
in front of $\Lambda$ in the singular point can be reabsorbed 
into a proper redefinition of the integration constant, through
(\ref{conv1Lambda})).\\
A more detailed investigation about IR singularity structure 
of higher-order perturbative running coupling is performed on
the ground of eq.(\ref{alpha^k}) in the recent work 
\cite{Magradze:2005ab}, where in particular the location of 
Landau singularities is determined as a function of $n_f\,$. 
\vspace{-0.2truecm}
\subsection{Time-like coupling}
\vspace{-0.3truecm}
Until now we have worked exclusively in the space-like region 
to derive the evolution of coupling, namely
we have implicitly admitted the theory renormalized at 
momentum scale with negative squared invariant mass. 
However, in perturbation theory one needs to parameterize 
observables depending on time-like arguments by means of 
an effective parameter, to improve perturbative expansions. 
While this poses no special problem in large energy processes, 
at any finite energy the issue of which should be the most 
suitable parameter in the s-channel must be carefully considered. 
For this purpose we briefly sketch the key points of this 
subject, and start by noting that the standard practice is to 
merely take over to the time-like domain the space-like form 
at any loop level, regardless of the crossing between two 
disconnected regions, thus importing the same IR singular 
structure in a specular way.
Nevertheless, from many early works based upon analysis of 
$e^+e^-$-annihilation data (see e.g. \cite{Moorhouse:1976qq} 
and refs. therein), it is known that this should not be the 
case but far in the asymptotic regime. This is because of the 
appearance of not negligible corrections ($\pi^2$-terms) to 
higher order coefficients of the $\alpha_{\rm{s}}(s)$-expansions, 
due to analytic continuation from space-like to time-like axis. 
The problem has not yet been univocally solved, as it is 
strongly related to the IR non analyticity of perturbative 
running coupling, though it receives a well satisfactory answer 
in the framework of Analytic Perturbation Theory (see secs.4.3 and 4.6).\\
Referring for definiteness to $e^+e^-$-annihilation into 
hadrons, the issue can be stated as follows. Firstly recall 
that the main way (see e.g. \cite{Yndurain}) to deal 
with the ratio
\be
R(s)=\frac{\sigma(e^+e^-\to\mathrm{hadrons})}{\sigma(e^+e^-\to\mu^+\mu^-)}\,,
\lb{Rdef}
\ee
where $s=q^2>0\,$, is to start with the hadron contribution to photon 
polarization tensor in (space-like) momentum space
\be
\Pi_h^{\mu\nu}(q)=(g^{\mu\nu}q^2-q^\mu q^\nu)\Pi_h(-q^2)
\propto\int d^4x e^{iqx}<0|T(j^\mu(x)j^\nu(0))|0>
\lb{Pdef}
\ee
$j^\mu$ being the quark electromagnetic current operator. 
Then, as known, optical theorem ensures the ratio $R(s)$ to be 
straightforwardly related to the absorptive part of the forward 
scattering amplitude $e^+e^-\to e^+e^-\,$. Indeed, being the 
analytical properties of the two-point correlation function 
(\ref{Pdef}) in the cut complex plane $\mathbb{C}-\{q^2>0\}$ 
well established, 
this amounts 
to taking the discontinuity of $\Pi_h(-q^2)$ across the cut
\be
R(s)=\frac{1}{2\pi i}\lim_{\varepsilon\to0}\left[\Pi_h(-s+i\varepsilon)-
\Pi_h(-s-i\varepsilon)\right]\,,
\lb{Rcut}
\ee
having computed the RG improved expansion for 
$\Pi_h(-q^2)$ on the space-like axis ($q^2<0$). To this end 
one formally works with its first logarithmic 
derivative (thus avoiding subtraction constants), the Adler 
$D$-function \cite{Adler:1974gd}, with the same space-like 
argument 
\be
D(-q^2)=-q^2\frac{d\Pi_h(-q^2)}{dq^2}\,,
\lb{Adl}
\ee
which in the $\overline{\mathrm{MS}}$ reads 
\be
D_{\rm{PT}}(Q^2)=3\sum_f Q_f^2\left[1+\frac{\alpha_{\rm{s}}(Q^2)}{\pi}+
d_2\left(\frac{\alpha_{\rm{s}}(Q^2)}{\pi}\right)^2+
d_3\left(\frac{\alpha_{\rm{s}}(Q^2)}{\pi}\right)^3 +\dots\,\right]
\lb{D_ptb}
\ee
being $Q^2=-q^2\,$, $Q_f$ the quark charges, and 
\cite{Chetyrkin:1979bj}, \cite{Gorishnii:1990vf} 
\bea
&&d_2\,\simeq\,1.986-0.115n_f\qquad\nn\\ 
&&d_3\,\simeq\,18.244-4.216n_f+0.086n_f^2-1.24\left(3\sum_fQ_f^2\right)^{-1}
\left(\sum_fQ_f\right)^2\,.\qquad
\lb{D_coefs}
\eea
Then by integration $\Pi_h(Q^2)$ is readily obtained. One should 
be aware that in the massless theory the cut spreads over the 
whole positive axis, 
and when taking into account quark masses the cut starts at the two-pion 
threshold $4m_\pi^2\,$.\\  
Whatever the loop order, the result for $R(s)$ is usually recasted as a series 
in the effective parameter $\alpha_{\rm{s}}(s)$, naively obtained 
by specular reflection, that is by replacing the space-like argument 
$Q^2=-q^2>0$ straightway with the time-like one $s=q^2>0$ 
in the coupling at a given loop order 
(e.g. eq.(\ref{4loopEC})). This final step displays nontrivial 
correction terms starting from $O(\alpha_{\rm{s}}^3)$, which are 
proportional to powers of $\pi$ and are the drawback of the analytic 
continuation of hadronic tensor nearby the time-like axis.
We then have the ordinary perturbative expansion for $R(s)$ (see e.g. 
\cite{Kataev:1995vh})
\be
R_{\rm{PT}}(s)= 3\sum_f Q_f^2 \left[1+\frac{\alpha_{\rm{s}}(s)}{\pi}+
r_2\left(\frac{\alpha_{\rm{s}}(s)}{\pi}\right)^2 +
r_3\left(\frac{\alpha_{\rm{s}}(s)}{\pi}\right)^3 +\dots\,\right]\,
\lb{R_ptb}
\ee 
\be
r_2=d_2\,;\qquad r_3=d_3-\delta_3\,,\quad \delta_3=\frac{\pi^2b_0^2}{48}
\lb{R_coefs}
\ee
with $d_2$ and $d_3$ the same as in (\ref{D_ptb}), and we have used
the shortcut $b_j=(4\pi)^{j+1}\beta_j$ to emphasize the $\pi$-powers. 
The number $\delta_3$ gives to 
the $O(\alpha_{\rm{s}}^3)$ coefficient a strongly negative contribution 
for each $n_f$, e.g. roughly $\simeq14.3$ for $n_f=4$. Higher order 
$\pi^2$-terms were analyzed in detail in \cite{Bjorken:1989xw}, for 
instance the fourth order correction is
\be
\delta_4\equiv d_4-r_4=\frac{\pi^2b_0^2}{16}\left(r_2+\frac{5b_1}{24b_0}\right)
\lb{delta4}
\ee
roughly $\delta_4\simeq120$ for $n_f=4$; analogous (but more 
cumbersome) behavior is found for still higher orders, from which 
it becomes patent the remarkable growth in these correction terms.\\ 
A similar treatment also holds for other 
s-channel observables \cite{Kataev:1995vh}, as the normalized rate 
for $\tau$-decay into hadrons, showing that the effects of analytical 
continuation make the perturbative expansions in the time-like region 
deeply different from Euclidean ones.\\
Since the $\pi^2$-terms play a dominant role in higher order 
coefficients, expansion (\ref{R_ptb}) works only asymptotically at 
large $s$ (that is when the smallness of $\alpha_{\rm{s}}(s)$ scales 
down these large coefficients); thus the 
space-like coupling is not a reliable expansion parameter 
in the s-channel at finite energy, and it is not yet clear which one 
is instead to be used. Actually, as yet noted in pioneer works 
\cite{Pennington:1981cw,Pennington:1983rz}, 
the expression of $R(s)$, resulting from eq.(\ref{Rcut}) together with 
the improved perturbative series for $\Pi_h$, exhibits no natural 
expansion parameter (since both real and imaginary parts of 
$\alpha_{\rm{s}}(Q^2)$ enter into the form of $R$), and the choice of 
such a suitable parameter is essentially a matter of expediency, that 
is it should be selected the one which yields better convergence 
properties.\\
Alongside less meaningful attempts, we mention here the 
analysis \cite{Pennington:1981cw} of the use of $|\alpha_{\rm{s}}(-s)|$ as  
expansion parameter for $R(s)$\,. By sensibly reducing 
higher order terms it yields faster convergence than 
$\alpha_{\rm{s}}(s)$; it further possesses the relevant feature of IR 
freezing, in agreement with contemporary phenomenological models 
(e.g. \cite{Parisi:1980jy}), thus avoiding the hurdle of Landau 
singularities on the time-like domain. Indeed for low $s\,$, $|\alpha_{\rm{s}}(-s)|$   is 
anyway less than 0.33 for $n_f=3\,$. According to this prescription 
the one-loop running coupling in this region should read 
\cite{Pennington:1981cw}
\be
|\alpha_{\rm{s}}(-s)|=\frac{1}{\beta_0}\left[\frac{1}{\ln^2(s/\Lambda^2)+
\pi^2}\right]^{1/2}
\lb{penn}
\ee
and asymptotically, i.e. for $s\gg\Lambda^2 e^{\pi}$
\be
|\alpha_{\rm{s}}(-s)|=\frac{1}{\beta_0}\frac{1}{\ln(s/\Lambda^2)}\left[1
-\frac{\pi^2}{2}\frac{1}{\ln^2(s/\Lambda^2)}+\dots\,\right]
\ee
resembling the UV behavior of the relative space-like coupling 
(\ref{1loopEC2}). Aside from these nice features, this function 
cannot entirely sum up the $\pi^2$-terms.\\
In order to deal with these corrections a somewhat different 
approach, known as RKP (Radyushkin-Krasnikov-Pivovarov) procedure, has been suggested 
\cite{Radyushkin:1982kg,Krasnikov:1982fx} (see also
\cite{Jones:1995rd}), which is firmly based upon the analytical 
properties of the polarization tensor $\Pi_h(-q^2)$ and of the 
related $D$-function (\ref{Adl}), summarized by the dispersion 
relations, respectively 
\be
\Pi_h(-q^2)=\int_0^\infty ds\frac{R(s)}{s-q^2}\,\,,
\lb{dispPi}
\ee
\be
D(-q^2)=-q^2\int_0^\infty ds\frac{R(s)}{(s-q^2)^2}
\lb{dispAdl}
\ee
where $R(s)$ is given by (\ref{Rcut}), and 
$q^2$ lying in $\mathbb{C}-\{q^2=s>0\}\,$. The key point here is the 
inverse of (\ref{dispAdl}) given by the contour integral
\be
R(s)= \frac{i}{2\pi}\int_{s-i\varepsilon}^{s+i\varepsilon}\frac{dq^2}{q^2}D(-q^2)
\lb{invsAdl}
\ee
to be computed along a path in the analyticity region for 
the $D$-function. Eq.(\ref{invsAdl}) can be then generalized to an 
integral transform mapping a space-like argument function into 
a time-like one
\be
R(s)=\Phi\left[D(-q^2)\right]\,,
\lb{phi}
\ee
that can be straightforwardly applied to the perturbative expansion 
(\ref{D_ptb}), provided that the integration contour is always kept 
far enough from IR space-like singularities. This yields $R(s)$ as 
an expansion into the images of $\alpha_{\rm{s}}(Q^2)$ and of its 
powers, through the map $\Phi$ \cite{Radyushkin:1982kg}
\be
R(s)=3\sum_f Q_f^2\left\{1+\sum_{n\ge1}d_n\Phi\left[\left(\frac{\alpha_{\rm{s}}(Q^2)}
{\pi}\right)^n\right]\right\}\,,
\lb{R_rkp}
\ee
where $d_n$ are the same as in (\ref{D_ptb}). Eq.(\ref{R_rkp}) is to 
be compared with the standard 
perturbative expansion (\ref{R_ptb}); here the $\pi^2$-terms are 
entirely summed up, with the drawback that within this framework 
there is no uniquely defined expansion parameter. However, it is 
useful to work out its behavior at $O(\alpha_{\rm{s}})$-approximation 
for $R(s)$, i.e.
\be
\tilde{\alpha}^{(1)}(s)\equiv\Phi\left[\alpha_{\rm{s}}^{(1)}(Q^2)\right]
=\frac{1}{\beta_0}\left\{\frac{1}{2}-\frac{1}{\pi}\arctan
\left[\frac{\ln\left(s/\Lambda^2\right)}{\pi}\right]\right\}
\lb{alpha_RKP}
\ee
easily obtained by applying the integral transformation (\ref{invsAdl}) 
to the one-loop space-like coupling (\ref{1loopEC2}). The related 
time-like form (\ref{alpha_RKP}) is once more free of unphysical 
singularities at low $s\,$, and for $s\gg\Lambda^2e^{\pi}$ as above 
can be expanded into powers of $\pi/\ln(s/\Lambda^2)$ 
\be
\tilde{\alpha}^{(1)}(s)= 
\frac{1}{\pi\beta_0}\left[\frac{\pi}{\ln(s/\Lambda^2)}-\frac{1}{3}\frac{\pi^3}
{\ln^3(s/\Lambda^2)}+\,\dots\,\right]\,,
\lb{UValpha_RKP}
\ee
Then eq.(\ref {UValpha_RKP}) can be recasted as a power series in the
one-loop space-like coupling 
\be
\tilde{\alpha}^{(1)}(s)=\alpha_{\rm{s}}^{(1)}(s)\left[1-
\frac{\pi^2b_0^2}{48}\left(\frac{\alpha_{\rm{s}}^{(1)}(s)}{\pi}\right)^2+\,
\dots\,\right]\,,
\lb{UValpha_RKP2}
\ee
emphasizing that the two couplings differ in three-loop level.
By comparing (\ref{UValpha_RKP2}) with eqs.(\ref{R_ptb}) and 
(\ref{R_coefs}), it becomes clear how RKP resummation of the 
$\pi^2$-terms into the time-like coupling (\ref{alpha_RKP}) 
does work. 
The main shortcoming of this recipe is that by applying to 
(\ref{alpha_RKP}) the inverse transformation to (\ref{phi}), i.e. 
relation (\ref{dispAdl}), we are not back to the original input 
(\ref{1loopEC2}). Obviously, this is because integral transformations
(\ref{invsAdl}) and its inverse are well behaved as long as the 
integrand possesses the correct analytical properties in the cut 
complex plane; actually this is not the case for the space-like 
coupling (\ref{1loopEC2}) and its higher-loop approximations, and 
we are then forced to compute the integral along a path large 
enough to avoid the IR space-like singularities.

\vspace{-0.2truecm}
\section{Infrared behavior}
\vspace{-0.3truecm}
As we told, Landau singularities are an artifact of the formalism
and are related to a breakdown of the perturbative expansion even 
as an asymptotic expansion.
On the basis of general principles of field theory 
$\alpha_{\rm s}(\mu^2)$ should have
singularities in the $\mu^2$ complex plane only on the negative real
axis, from a threshold to $-\infty$. 
This circumstance is irrelevant as long as we consider high energy
processes in which the energy scale
is much larger than $\Lambda\,$, but leads to serious difficulties 
in other problems in which much smaller values
of $\mu$ are involved. Of this type are the bound state calculations, 
the decay of particles, the annihilation processes and even the deep 
inelastic collisions in particular geometries. 
Let us consider e.g. the ratio 
\begin{equation}
  R_\tau = \frac{\Gamma(\tau \rightarrow \nu_\tau + {\rm hadrons})}
{\Gamma(\tau \rightarrow \nu_\tau + e + \bar \nu_e)}
\label{taudecay}
\end{equation}
between the hadronic and leptonic decay width of the
$\tau$ lepton. The non strange contribution to this quantity can be
written in the form
\bea
&&R_\tau^{\rm nst} = {12 \pi S_{\rm EW} |V_{ud}|^2 \over m_\tau^2}
 \int_{m_\pi^2}^{m_\tau^2} dt \left(1-\frac{t}{m_\tau^2}\right)^2\nn\\
&&\qquad\quad\left\{\left(1+ \frac{2t}{m_\tau^2}\right){\rm Im}\, 
\Pi_{ud}^{(1)}(t) +{\rm Im}\, \Pi_{ud}^{(0)}(t) \right\}\, ,
\label{taudecay2}
\eea
where $S_{\rm EW}$ is the electro-weak factor, $V_{ud}$ the relevant 
CKM matrix element, $\Pi_{ud}^{(1)}(t)$ and $\Pi_{ud}^{(0)}(t)$ are the transversal
and the longitudinal part of the appropriate current-current
correlator respectively. In this expression $\alpha_{\rm s}(\mu^2)$ should be known
in principle from the threshold $m_\pi \sim 0.14$ GeV to $m_\tau =
1.78$ GeV. 
As we told, to extrapolate $\alpha_{\rm s}(\mu^2)$ to the infrared region, 
various proposals have been advanced. In this section we shall consider some
purely phenomenological attempts, few examples of the so-called physical couplings
and an optimization procedure.

\vspace{-0.2truecm}
\subsection{Potential inspired approaches}
\vspace{-0.3truecm}
One of the first attempts to modify the expression for  $\alpha_{\rm s}(\mu^2)$
in the infrared region was made in the framework of the
quark-antiquark potential. 
      The potential $V(r)$ between two infinitely heavy quark
and antiquark (static potential) can be defined by the equation
\begin{equation}
 V(r) =  \lim_{T\to\infty}{i \over T}\ln W[\Gamma]
\label{static}
\end{equation}
$\Gamma$ being a rectangular Wilson loop of size $r\times T$ and
\begin{equation}
W [\Gamma] = \bigg\langle{1 \over 3} {\rm Tr}\,{\cal P} \exp \left \{i g 
\oint_\Gamma dx^\mu A_\mu(x)\right \}\bigg\rangle  \,,
\label{wilson}
\end{equation}
${\cal P}$ being the ordering prescription on the path $\Gamma$ and Tr
the trace over the color indices.
The simplest assumption is to write $\ln W$ as the sum of its
perturbative expression and a non perturbative term proportional to
the area $S=rT$ delimited by $\Gamma$
\begin{equation}
i\ln W  = (i\ln W)_{\rm PT} + \sigma S  \,.
\label{ansatz}
\end{equation}
If $(i\ln W)_{\rm PT}$ is evaluated at the first order in the coupling one obtains 
\begin{equation}
 V(r) =  - \frac{4}{3}\frac{\alpha_{\rm s}(\mu^2)}{r} + \sigma r \,,
\label{cornell}
\end{equation}
where $\mu$ should be taken of the order of the typical $\langle1/r\rangle\,$.  
This is the so called Cornell potential, expressed as the
sum of a Coulombian and linear term. As well known, by solving 
the corresponding Schroedinger equation a reasonable reproduction 
of the spin averaged bottonium and charmonium spectrum can already be 
obtained. 
In the momentum representation (\ref{cornell}) can be
written
\begin{eqnarray}
 \tilde V(Q) &\equiv& \langle {\bf k^\prime}|V(r)|{\bf k}\rangle 
={1 \over (2\pi)^3} \int d^3 {\bf r} e^{-i{\bf
   Q\cdot r}} V(r) =  \nonumber \\ 
&=& - {1 \over 2 \pi^2}{4 \over 3}{\alpha_{\rm s}(\mu^2) \over 
    Q^2} - {1 \over \pi^2} {\sigma \over Q^4} \,, 
\label{fourier}
\end{eqnarray}
where ${\bf k}$, ${\bf k^\prime}$ denote the initial and final center
of mass momentum of the quark and ${\bf Q} =
{\bf k^\prime}-{\bf k}$ the momentum transfer
\footnote{More properly ${1 \over Q^4}$ stays for the limit 
  of ${1 \over (Q^2 + \epsilon^2)^2} - {4 \epsilon^2 
\over (Q^2 + \epsilon^2)^3}$ (the Fourier transform of $ -\pi^2 r
  e^{-\epsilon r}$)
for infinitesimal $\epsilon$. The same recipe should be applied to
regularize eqs.(\ref{richardson1}-\ref{richardson3}).}. 
According to the general rule we should identify in (\ref{fourier})
the scale $\mu$ with $Q$. Since however in heavy quarkonia ($b \bar b$ 
and $ c \bar c $) $ Q $ typically ranges between 200 MeV and 2 GeV, we 
come close to Landau singularities and need to use some kind of 
regularization. In this order
of ideas many years ago it was proposed  \cite{richardson} to write
the entire potential as
\begin{equation}
 \tilde V(Q) = - {1 \over 2 \pi^2}{4 \over 3}{\alpha_{\rm V}(Q^2) \over 
    Q^2}
\label{richardson1}
\end{equation}
and to take
\begin{equation}
\alpha_{\rm V}(Q^2) = {1 \over \beta_0 \ln (1 + {Q^2 \over
    \Lambda^2})}\,.
\label{richardson2}
\end{equation}
Note that, defined in this way, $\alpha_{\rm V}(Q^2)$ reproduces the
ordinary one-loop expression (\ref{1loopEC2}) for $Q\to \infty$, while
for $Q \to 0$ it gives
\begin{equation}
\alpha_{\rm V}(Q^2) \to {\Lambda^2 \over \beta_0 Q^2}\, .
\label{richardson3}
\end{equation}
Replacing (\ref{richardson3}) in (\ref{richardson1}) and comparing it
with (\ref{fourier}) we see that (\ref{richardson2}) incorporate the
confining part of the potential (\ref{cornell}) with 
\begin{equation}
\sigma =  {2\Lambda^2 \over 3\beta_0 }\, .
\label{richardson4}
\end{equation}
In (\ref{cornell}) the perturbative part of the potential was
evaluated at the first order in $\alpha_{\rm s}(\mu^2)$. However, 
higher orders corrections have also been considered starting from (\ref{static})
or by other methods \cite{fischler}. As proposed in
\cite{buchmuller}, such corrections can be taken
into account assuming (\ref{richardson1}) as the definition
of a new coupling constant $\alpha_{\rm V}(\mu^2)$ and re-expressing this
in term the ordinary ${\rm\overline{MS}}$ constant. We can write
\footnote{Note that at three loops even terms of the form 
$\left ({\alpha_{\rm \overline {MS}}\over \pi} \right )^4
\ln  \alpha_{\rm \overline {MS}}$ occur \cite{appelquist}.}  
\begin{equation}
\alpha_{\rm V}(\mu^2) = \alpha_{\rm \overline {MS}}(\mu^2) \left[1 + u_1
  {\alpha_{\rm \overline {MS}}(\mu^2)\over \pi} + 
u_2 \left({\alpha_{\rm \overline
  {MS}}(\mu^2) \over \pi}\right)^2 + ...\,\right]\,.
\label{alpha-pt1}
\end{equation}
with
\be
u_1 = {31\over 12} - {5\over 18}n_f\,;\quad
u_2 \cong 28.538 - 4.145 n_f + 0.077 n_f^2
\label{alpha-pt2}
\ee
i. e.  $u_1\cong 1.472$, $u_2\cong 13.190$ for $n_f = 4$. 
Alternatively one can use ordinary one or two-loop equations for the
running $\alpha_{\rm V}(\mu^2)$ but with $\Lambda_{\rm V}$ redefined
as (cf. (\ref{conv2lambda}))
\begin{equation}
\Lambda_{\rm V} = \Lambda_{\rm \overline {MS}}\, e^{u_1 \over 2 \pi \beta_0}
\cong 1.42 \Lambda_{\rm \overline {MS}}
\label{Lambda-pt}
\end{equation}
and this should be the value to be used in (\ref{richardson2}). 
Eq.(\ref{richardson2}) can be re-obtained from
(\ref{RGScoupl}) by the modified $\beta$-function \cite{buchmuller}
\begin{equation}
\beta_{\rm V}(\alpha) = - \beta_0\, \alpha^2 \left (1- e^{-{1
    \over \beta_0 \alpha}}\right ).
\label{beta-pt1}
\end{equation}
More generally one can take
\begin{equation}
{ 1 \over \beta_{\rm V}(\alpha)} = - {1 \over \beta_0\, \alpha^2 
\left (1- e^{-{1 \over \beta_0 \alpha}}\right )} + {\beta_1
    \over \beta_0^2}{1 \over \alpha} \, e^{-l\alpha}\, ,
\label{beta-pt2}
\end{equation}
$l$ being an adjustable parameter. Eq. (\ref{beta-pt2}) reduces to
(\ref{beta-pt1}) for $l\to \infty $, while for finite $l$ produce the
following asymptotic behavior  
\begin{eqnarray}
&\beta_{\rm V}(\alpha) \sim -\beta_0 \alpha^2 - \beta_1 \alpha^3 -
  \dots \qquad & {\rm for}\qquad \alpha \to 0 \\ \label{asympta}
&\beta_{\rm V}(\alpha) \sim -\, \alpha + \dots 
\qquad & {\rm for} \qquad \alpha \to \infty 
\label{alpha-pt22}
\end{eqnarray}
which corresponds for $\alpha_{\rm V}(Q^2)$ to a two-loop expression of
the type (\ref{2loopECit}) (or (\ref{2loopEC})) 
for $Q\to \infty$ or of the form (\ref{richardson3}) for $Q \to 0\, $.\\
Eqs.(\ref{richardson1}-\ref{richardson4}), or the more general
corresponding to (\ref{beta-pt2}), are very appealing. They do 
not introduce any additional parameters in the theory
and for $\Lambda_{\rm V}\sim 0.5$ GeV (and $l=24$) reproduce well
the spin averaged first excited states in the $b \bar b$ and $c \bar
c$ systems. However, they are not satisfactory from other points of view.  
They correspond to assume that all forces including confinement are
due to the exchange of some effective vectorial object. On the contrary,
if eq.(\ref{ansatz}) is generalized to Wilson loops with distorted
contour (more sophisticate ansatzs also exist \cite{dosh, baker}) and 
some more elaborate method is applied, spin dependent and
velocity dependent (relativistic) corrections can be obtained, which
differ from those that would be derived from the exchange of such a 
vectorial particle alone and seem to be preferred by the data
\cite{eichten,bcp,simonov}. The majority of the analysis seems 
rather in favor of assuming a finite limit for $\alpha_{\rm s}(\mu^2)$ 
for $\mu \to 0$ and adding a separate term for confinement as in 
(\ref{cornell}). 
Besides hadron spectroscopy there is other abundant phenomenology that 
seems to be consistent with the existence of a finite IR limit for 
$\alpha_{\rm s}(\mu^2)$. This concerns the hadron-hadron 
scattering and the hadron form
factors, the properties of jets, the transversal momentum
spectrum in $W$ and $Z$ production etc.\\
Coming to the explicit form of $\alpha_{\rm s}(Q^2)$ the simplest
modification of the one-loop expression that has been considered is
the ``hard freeze'' assumption
\be \alpha_{\rm s}(Q^2)  = \left\{
\begin{array}{ll}
{1 \over \beta_0}{1\over \ln (Q^2/\Lambda^2)}  & \mbox{for $Q^2
  \geq Q_0^2$} \\
H \equiv {1 \over \beta_0}{1\over \ln (Q_0^2/\Lambda^2)} & \mbox{for $Q^2
  \leq Q_0^2$} 
\end{array} 
\qquad \label{hard} \right. 
\ee
This equation has been used in hadron spectrum calculations, in a model for
hadron-hadron scattering and in studies on nucleon structure
function \cite{nikolaev92} (see also \cite{Higashijima:1983gx}); the values 
$Q_0\,=$ 0.44 GeV and $\Lambda \,$ corresponding to $\frac{H}{\pi}=0.28\,$GeV have
been found appropriate for the last two applications, a little smaller 
value ${H \over \pi}\, = $ 0.26 GeV comes from other phenomenology. 
Other convenient interpolation formulas between the large $Q$
perturbative expression and a finite $\alpha_{\rm s}(0)$ have been
used again in hadron spectrum studies \cite{isgur} with 
${\alpha_{\rm s}(0) \over \pi}\,\sim\, 0.19-0.25$. In a fully
relativistic treatment in \cite{Zhang} it was found that in order to
obtain a $\pi$ mass so much lighter than the $\rho$ mass a value
${\alpha_{\rm s}(0) \over \pi}\,=\, 0.265$ was necessary. A
similar results was obtained in \cite{Baldicchi:2004wj} with a
one-loop analytic coupling (see later) with $\Lambda=0.18$ GeV corresponding to
${\alpha_{\rm s}(0) \over \pi}= 0.44$ but, what is really
relevant, to an average value of $\alpha_{\rm s}(Q^2) \over \pi$ in 
the interval between 0 and 0.5 GeV about 0.22. 
Finally in many analysis the successful phenomenological hypothesis
was adopted that the gluon acquires an effective dynamical mass $m_g$
\cite{Ball:1995ni,Cornwall:1981zr, Field:2001iu}.  
We shall come back in the following to this point, for the
moment let us observe that to the leading order the following
equation generalizes naively (\ref{richardson2})
\begin{equation}
\alpha_{\rm s}(Q^2) = {1 \over \beta_0 \ln ({4m_g^2+Q^2 \over
    \Lambda^2})}\,.
\label{gluonmass}
\end{equation}
In particular the mentioned hadron-hadron scattering model gives
in this case  $\Lambda\,=\, 0.3$ GeV and $m_g\,=\, 0.37$ GeV 
(${\alpha_{\rm s}(0) \over \pi}\,=\, 0.26$) \cite{Halzen:1992vd}.\\
All the above mentioned attempts are essentially in agreement for
what concerns the qualitative behavior of $\alpha_{\rm s}(Q^2)$ in
the infrared region, even if they differ in the details. The point is that
for many purposes we can simply introduce an
infrared-ultraviolet matching point in the range of variability of
$Q$; let us say  $Q_{\rm I} = 1$ or 2 GeV. Then one can use ordinary 
perturbative expressions of the type (\ref{1loopEC2}) for $Q>Q_{\rm I}$
and treat the quantities 
\begin{equation}
\langle {\alpha_{\rm s} \over \pi}\rangle = {1\over Q_{\rm I}}
 \int _0^{Q_{\rm I}} dQ {\alpha_{\rm s}(Q^2) \over \pi}
\label{dks1}
\end{equation}
or corresponding higher moments as adjustable parameters. Various
event shape in $e^+e^-$ annihilation can be reproduced simply assuming
$Q_{\bf I}=1$ GeV and $\langle {\alpha_{\rm s} \over \pi}\rangle \, 
\approx 0.2 $  \cite{dokshitzer1} . Such procedure is believed to be a 
way to parametrize and derive from experience truly non-perturbative effects
(sec 4.5).

\vspace{-0.2truecm}
\subsection{Physical couplings}
\vspace{-0.3truecm}
By physical couplings we mean any type of {\it effective charges} 
\cite{Grunberg:1980ja, Beneke:1994qe}, defined by (\ref{intr11}), and in 
general any other observable related coupling. 
The advantage of such quantities is that they are in principle well 
defined at every scale and that can be easily extracted from experience. The
disadvantage is that they are observable dependent. However, as we told they
can be expanded in terms of the $\overline {\rm MS}$ coupling (cf.(\ref{int7})) 
and related to each other \cite{Brodsky:1994eh}.\\
First we may consider the typical case of the quantity
$R_{e^+e^-}(s)\,$, (eq.(\ref{Rdef})) and the related Adler function 
eq.(\ref{Adl}). Their perturbative expansions in the $\overline {\rm MS}$
scheme are given by (\ref{R_ptb}) and (\ref{D_ptb}), respectively. 
Correspondingly two different effective charges $\alpha_{\rm R}(s)$
and $\alpha_{\rm D}(Q^2)$ can be defined 
\be
R_{e^+e^-}(s)=3\sum_f Q_f^2 \left[1+{\alpha_{\rm R}(s) \over
    \pi}\right], \quad 
D(Q^2)=3\sum_f Q_f^2 \left[1+{\alpha_{\rm D}(Q^2) \over
    \pi}\right]
\lb{brd1}
\ee
The first of these is of time-like, the second of space-like
type. As apparent from eq.(\ref{dispAdl}), they are related by the
transformation (\ref{phi}), i.e. 
$\alpha_{\rm R}(s)=\Phi\left[\alpha_{\rm D}(Q^2)\right]\,$, 
and are actually the model over which (\ref{alpha_RKP}) is written.\\
A third effective charge, that has been used in 
analysis of the $\tau$
decay, can be defined with reference to the quantity $R_\tau$
(cf.(\ref{taudecay}) and (\ref{taudecay2})), which we discuss in
some more detail as an example. We can consider
separately the contributions due to the vector and the axial currents
and set \cite{brodsky03,ALEPH}
\bea
 R_\tau^{\rm V/A}(s)= 12\pi S_{\rm EW} |V_{ud}|^2 \int_0^s {dt \over s}
 \left(1-\frac{t}{s}\right)^2\qquad\qquad\nn\\
\qquad\left\{\left(1+ \frac{2t}{s}\right) {\rm Im}\, 
 \Pi_{\rm V/A}^{(1)}(t) +{\rm Im}\, \Pi_{\rm V/A}^{(0)}(t) \right\}\,.
\lb{brd2}
\eea
For $s=m_\tau^2$ eq.(\ref{brd2}) gives the ordinary $R_\tau^{V/A}$ for
the $\tau$ lepton, for $s<m_\tau^2$ gives the corresponding expression
for a fictitious $\tau^\prime$ with mass $m_{\tau^\prime} = \sqrt s$. In
analogy with (\ref{brd1}) we can define a vector and axial coupling
$\alpha_\tau^{\rm V/A}(s)$ by 
\be
 R_\tau^{\rm V/A}(s)={ R_\tau^{0}\over 2}\left[1+
{\alpha_\tau^{\rm V/A}(s) \over
 \pi}\right]\,,
\lb{brd3}
\ee
where $R_\tau^{0}$ denotes the same quantity at zero order in the 
strong coupling. It is also defined a global $\alpha_\tau (s)$ by
\be
R_\tau (s) = R_\tau^{\rm V}(s) + R_\tau^{\rm A}(s)=
 R_\tau^{0}\left[1+{\alpha_\tau(s) \over
    \pi}\right]
\lb{brd4}
\ee
and $\alpha_\tau(s)={1 \over 2}\left[\alpha_\tau^{\rm V}(s)
  +\alpha_\tau^{\rm A}(s)\right]\,$. 
The quantity ${\rm Im}\, \Pi_{\rm V}^{(0)}(s)$ may be assumed to
vanish for small quark masses and ${\rm Im}\, \Pi_{\rm A}^{(0)}(s)$ to
be given only by the pion pole, ${\rm Im}\, \Pi_{\rm A}^{(0)}(s)= 
{\pi \over m_\pi^2} \delta (s-m_\pi^2)$. Under the same hypothesis the quantities 
${\rm Im}\, \Pi_{\rm V}^{(1)}(s)$ and ${\rm Im}\, \Pi_{\rm
  A}^{(1)}(s)$ should be identical at the
perturbative level and they are expected to differ asymptotically
only for powers of $1/s$ which have a non-perturbative origin. The same
must be true for $\alpha_\tau^{\rm V}(s)$ and $\alpha_\tau^{\rm
A}(s)$. Note that, due to isospin invariance $\rm Im\Pi_{V}^{(1)}$ is proportional
to the isovectorial component $\rm Im\Pi_{em}^{(\rm I=1)}\,$, and 
$\alpha_\tau^{\rm V} (s)$ and $\alpha_{\rm R}(s)$ are
related by
\be
\alpha_\tau^{\rm V} (s) = 2\,\int_0^s {dt \over s}\left(1-{t \over s} \right)^2
 \left(1+ \frac{2t}{s}\right) \alpha_{\rm R}(t)\,.
\lb{rd5}
\ee
Since the first coefficients of the perturbative expansion of
$R_{e^+e^-}$ and $R_\tau$ in the $ \overline {\rm MS}$ scheme are known, even 
the coefficients $v_1, v_2, v_3$ of (\ref{alphaconv}) and 
$\beta_2^{\tau},~\beta_3^{\tau}$ can be calculated.
Therefore the couplings $\alpha_\tau (s)$ and $\alpha_{\rm R}(s)$ that can
be extracted directly from experience can be immediately translated in
terms of a 
$\alpha_{\overline{\rm MS}}(s)$ value. Note, however, that in the
expressions of $v_3$ and $\beta_3$, which we do not report, there appears 
a quantity $K_4$ that has been only estimated, $K_4=25\pm50$ \cite{pich}. 
We report in fig.\ref{alphas}(a) taken from \cite{brodsky03}
the experimental value for $\alpha_\tau^{\rm V}(s)$, $\alpha_\tau^{\rm A}(s)$ 
and $\alpha_\tau(s)$ as extracted from the data of OPAL collaboration 
\cite{ALEPH}. The results are confronted with the resolution of the RG
equation for the approppriate $\beta_{\tau}(\alpha_{\tau})\,$. 
As it can be seen the 3-loop $\alpha_\tau(s)$ and 4-loop for
$K_4=\pm25$ have a finite limit for $s\to 0\,$; the latter fits data very well 
for $K_4=25$ down to $s\,\sim\, 1\, {\rm GeV}^2$. The strong enhancement of the
experimental  $\alpha_\tau(s)$ below such value can be related to the
pion pole that has not been included in the RG treatment.\\
A last definition of effective charge has been given recently 
\cite{Grunberg:2006jx} in the context of the Sudakov resummation
formalism. 
Let us consider, e.g., the Mellin transform $\hat F_2(Q^2,N)$ of the
structure function  $F_2(Q^2,x)$ in DIS ($x$ being the Bjorken
variable). Even after the mutual cancellation of the infrared 
singularities due to the soft real and virtual gluons, $F_2(Q^2,x)$ has
a logarithmic singularity at any order for $x \to 1$ which makes 
ordinary perturbation theory inapplicable. Such a singularity which is 
translated in the Mellin N-space in $\ln N$ power behavior for $N\to \infty$, can
however be resummed \cite{sudakov}. One can then obtain the following
asymptotic equation
\bea  
&&Q^2{\partial \ln \hat F_2(Q^2,N) \over \partial Q^2}=\qquad\lb{sud1}\\
&&{{\rm C}_F \over \pi}
\left[\int _0^1 dx {x^{N-1}-1 \over 1-x} A_{\cal S}[(1-x)Q^2]+\right. 
\left.H(Q^2) +O\left({1 \over N}\right)\right]\,,\nn
\eea
where $ A_{\cal S}$ and $H$ have the usual 
expansions  
\be
A_{\cal S}[Q^2] = \alpha_{\rm s}(Q^2) [1 + a_1 \alpha_{\rm s}(Q^2)
+\dots ]\,.
\lb{sud2}
\ee
Then $A_{\cal S}[Q^2]$ is assumed as an effective charge (Sudakov charge). 
For large $n_f$ the Borel transform of
(\ref{sud2}) $B[A_{\cal S}](t)$ can be given as an analytic expression
without singularities. The corresponding $A_{\cal S}[Q^2]$ turns out
to be free of Landau singularities but for $Q\to 0$ behaves as $-{1
\over 2 \beta_0}[{\Lambda ^4 \over Q ^4}- {\Lambda ^2 \over Q^2}]$.
This embarrassing singularity can be compensated with a non-perturbative
term and leads to corrections in $1/Q$ (see sec 4.5).\\  
Another very interesting physical coupling is introduced in
\cite{dokshitzer2}.  This is a generalization to QCD
of the Gell-Mann Low effective coupling for QED. It can be 
considered a generalization of the coupling $\alpha_V(Q^2)$ discussed
in sec. 3.1 and it is also related to the pinch scheme  
discussed in \cite{Watson:1996fg}, which is somewhat more involved but
explicitly gauge-independent.\\ 
Let us isolate in a Feynman integrand the factor corresponding to
the exchange of one dressed gluon between two quark lines
\be
{-i \alpha_{\rm s}(\mu^2) \over q^2 + i0} {1 \over 1- 
\Pi[q^2;\alpha_{\rm s}(\mu^2),\mu^2]}\,,
\lb{dks2}
\ee
and set, e.g. in the $\overline{\rm MS}$ scheme
\be
\vspace{-0.2truecm}
\alpha_{\rm SGD}(Q^2)= \left[{Z_{\bar q q g}^{\rm NA}(\mu^2) 
\over Z_{\bar q q g}^{\rm NA}(Q^2)}\right]^2\alpha_{\rm s}(\mu^2)
{1 \over 1- \Pi[-Q^2;\alpha_{\rm s}(\mu^2),\mu^2]}\,,
\lb{dks3}
\ee
where $Z_{\bar q q g}^{\rm NA}(\mu^2)$ is the appropriate
renormalization factor that makes the definition independent of the initial
scale $\mu^2$. Obviously
\be
Z_{\bar q q g}^{\rm NA}(\mu^2) = [Z_\alpha(\mu^2) Z_g(\mu^2)]^{1 \over 2}=
Z_{\bar q q g}(\mu^2)Z_q^{-1}(\mu^2)\,,
\lb{dks4}
\ee
where we have used the same notation as in sec. 2.1.  Note that in
an abelian theory, like QED, $Z_{\bar q q g}^{\rm NA}=1$ due to 
Ward identity. In QCD the quark factor $Z_q$ cancels only 
the ``abelian part'' of the vertex factor $Z_{\bar q q
  g}$. This is the meaning of the superscript NA (non abelian part). 
The index SGD in eq.(\ref{dks3}) means ``single gluon dressing''.\\
From (\ref{dks3}), if $\Pi$ is evaluated in the $\rm \overline{MS}$ scheme, 
we can write, setting $\mu=Q$,
\be
\alpha_{\rm SGD}(Q^2) = \alpha_{\rm \overline{MS}}(Q^2) \left [
1 + k_1 { \alpha_{\rm \overline{MS}}(Q^2)\over \pi} + \dots \right]\,.
\lb{dks40}
\ee
Here $k_1$ depends on the constant terms occurring in the renormalized $\Pi$ 
and it is so gauge dependent; e. g. in the Feynman gauge  
$k_1=(31 - 10n_f/3) /12$. Note, however, that eq. (\ref{dks3}) can be 
related to the pinch scheme and $k_1$ made gauge independent simply
including in the definition of $\Pi$ additional constants coming
from the {\it pinch parts} (selfenergy like parts originating from
contraction of internal lines as a concequence of Ward identities) of 
the vertex and the box diagrams \cite{Watson:1996fg}. In this way we
obtain  $k_1=(67 - 10n_f/3) /12\,$. On the contrary, if we absorb
the pertinent soft gluon corrections in a redefinition of 
$\alpha_{\rm SGD}(Q^2)$, we find $k_1=(67-3\pi^2-10n_f/3) /12\,$ 
\cite{cmw,dokshitzer2}. It is the latter value which is used in connection with the 
considerations of sec. 4.5.  

\vspace{-0.2truecm}
\subsection{Optimized perturbation theory}
\vspace{-0.3truecm}
By optimized perturbation theory it is
generally  meant some kind of perturbative expansion in which the
expansion variable, or the splitting of the Lagrangian in an
unperturbed and a perturbation part, is chosen in dependence on a
number of arbitrary parameters. Such parameters should not appear in the 
exact result, but obviously they occur in any expansion which is 
stopped to a certain maximum term $n$. However, it is immediately 
shown even by very simple examples, that the convergence of the series 
is greatly improved if for every $n$ the parameters are chosen at 
some stationary value (optimized choice) that depends on $n$.\\ 
In QCD, due to the arbitrariness in the choice of the RS, the 
optimization can be required in the variables that control such a 
scheme, e.g. the subtraction point $\mu$  and the 
scheme dependent coefficients $\beta_2,~\beta_3,~\dots\,$. 
Obviously every choice corresponds to a different definition of the 
coupling constant $\alpha_{\rm s}\,$. 
Let us e.g. consider the quantity  $R_{e^+e^-}(s)$ 
defined by eq.(\ref{Rdef}), and rewrite expansion (\ref{R_ptb}) with an arbitrary 
choice of $\mu^2$ (for the moment different from the total energy $s$) and 
in a arbitrary RS
\begin{eqnarray}
&&R_{e^+e^-}(s)= \nonumber\\
&&3\sum_f Q_f^2 \left[1+\frac{\alpha_{\rm s}(\mu^2)}{\pi}+r_2(s)
\left(\frac{\alpha_{\rm s}(\mu^2)}{\pi}\right)^2 +
r_3(s)\left(\frac{\alpha_{\rm s}(\mu^2)}{\pi}\right)^3 + \dots
\right]\,.
\label{OPTR} 
\end{eqnarray}
The quantity $R_{e^+e^-}(s)$ must be RS independent and, if 
we neglect the masses of active quarks, we can write
\begin{eqnarray}
&& \left({\partial \over \partial \tau} + \beta(\alpha_{\rm s}){\partial 
   \over \partial \alpha_{\rm s}} \right) R_{e^+e^-} = 0 \qquad \nonumber \\
&&\left ({\partial \over \partial \beta_j} - \beta(\alpha_{\rm s}) \int _0^{\alpha_{\rm s}}
    d\alpha^\prime {\alpha^{\prime {j+2}}\over [\beta(\alpha^\prime)]^2}
{\partial \over \partial \alpha_{\rm s}} \right ) R_{e^+e^-} = 0\,,
\label{OPTRG}
\end{eqnarray}
where $j=2,3,\dots$, $\tau =\ln(\mu^2/\tilde \Lambda^2)$ (recall eq.(\ref{conv1Lambda}))  
and we have used
\begin{equation}
{\partial \alpha_{\rm s} \over \partial \beta_j}=
-\beta(\alpha_{\rm s}) \int_0^{\alpha_{\rm s}} d\alpha^\prime
{\alpha^{\prime j+2} \over [\beta(\alpha^\prime)]^2} = 
\frac{\alpha_{\rm s}^{j+1}}{\beta_0}\left(\frac{1}{j-1}-\frac{\beta_1}{\beta_0}
\frac{j-2}{j(j-1)}\alpha_{\rm s}+\dots\right)\,,
\label{OPTAUS}
\end{equation}
as it can be seen deriving eq.(\ref{RGScoupl}). 
Eq.(\ref{OPTRG}) can be used, first to obtain $r_2,~r_3,\dots$ in an 
arbitrary RS when we know this quantities in a specific RS (cf. eq.(\ref{intr9})), 
and then to make the 
optimal choice for $\tau,~\beta_2,~\beta_3,\dots$, when we have stopped
expansions (\ref{OPTR}) and (\ref{intr3}) to a certain maximum
order; let us say to the term in $r_3$ in (\ref{OPTR}) and
to three-loop (i.e. to the term in $\beta_2$ ) in (\ref{intr3}).\\
From now on we will use the notations $b_j=(4\pi)^{j+1}\beta_j$ and 
$a=\alpha_{\rm s}/\pi\,$.
Replacing (\ref{OPTR}) in (\ref{OPTRG}), asking that this
equations for 
a given $\mu^2$ are satisfied for an arbitrary value of $a$ we obtain 
differential equations for $r_2,~ r_3,\dots\,$. Restricting to $r_2$  and $r_3$ 
and $j = 2$ we have
\begin{eqnarray}
&&{\partial r_2 \over \partial \tau}= {1 \over 4}b_0  \qquad \qquad \qquad
{\partial r_2 \over \partial b_2} = 0 \qquad\nonumber \\
&&{\partial r_3 \over \partial\tau}={1 \over 2}b_0 r_2 
+ {1\over 16}b_1\qquad
{\partial r_3 \over \partial b_2} = - {1 \over 16}{1 \over b_0} \qquad
\label{OPTRG2}
\end{eqnarray}
Integrating the above equations, we obtain
\begin{eqnarray}
&& r_2={1\over 4}b_0 \tau + \rho_2  \qquad \qquad\nonumber \\
&& r_3={ 1 \over 16} b_0^2 \tau^2 + {1 \over 2}b_0 \rho_2
\tau + { 1 \over 16} b_1 \tau - { 1 \over 16}{ b_2 \over
b_0} + \rho_3^\prime \qquad \qquad\nonumber \\
&&\quad  = \left ( r_2 + {1 \over 8}{b_1 \over b_0} \right)^2 - 
{1 \over 16 }{b_2 \over b_0} + \rho_3 
\label{OPTcoeff} \,,
\end{eqnarray}
where $\rho_2$ and $\rho_3$ are integration constants and so quantities 
independent of $\tau,~b_2,\dots$ and RS independent. They can be 
calculated e.g. equating $b_2,~ r_2,~ r_3$ to their expressions 
$b_2^{\rm \overline {MS}},~r_2^{\rm \overline {MS}},~r_3^{\rm \overline
{MS}}$ in the $\rm \overline {MS}$ scheme as
given by equations (\ref{R_coefs}) and (\ref{D_coefs}) after setting 
$\mu^2 = s$ . We have
\be
\rho_2 =  r_2^{\rm \overline {MS}} - {1\over 4} b_0 
\ln {s \over \tilde \Lambda^2}\,\,,\quad
\rho_3 = r_3^{\rm \overline {MS}} - 
\left ( r_2^{\rm \overline {MS}}+{1 \over 8}{b_1 \over
  b_0} \right )^2 + {1 \over 16}{ b_2^{\rm \overline {MS}}\over b_0}\,.
\label{OPTRGinv}
\ee
Note that $\rho_3$ turns out to be independent of $s\,$, and $r_2$
has the form
\begin{equation}
r_2 = -{1\over 4}b_0 \ln {s \over \tilde \Lambda^2}+ {1\over 4}
b_0 \tau + r_2^{\rm \overline {MS}}\,,
\label{OPTr2}
\end{equation}
while $r_3$ depends on $s$ and $\tau$ only through $r_2$.
Let us now replace  $\beta(a)$ in (\ref{RGScoupl}) with its 3-loop
expression $\beta^{(3)}(a) = -\pi\,a^2(\frac{b_0}{4}+\frac{b_1}{16}a 
+ \frac{b_2}{64}a^2 )$.          
We have
\be
\tau = {4\over b_0 a} + {b_1 \over b_0^2}\ln \left({b_1 a \over b_0}\right) -
{b_1 \over 2b_0^2} \ln \left({16b_0+4b_1 a + b_2 a^2\over b_0}\right) + 
{2b_2 b_0 - b_1^2 \over 2b_0^2}\,f(a,b_2)
\label{running3loop}
\ee
with
\begin{equation}
f(a,b_2)={1 \over \sqrt{D}} \ln {4b_0 + {1 \over 2} a
  (b_1+\sqrt D) \over 4b_0 + {1 \over 2}a (b_1-\sqrt D)}
\label{function}
\end{equation}
and $D=b_1^2-4b_2 b_0\,$. 
Let us make the same replacement in (\ref{OPTRG}) and stop 
(\ref{OPTR}) at the $a^3$ term. By requiring that eq.(\ref{OPTRG})
is now exactly satisfied, we obtain the following equations
\begin{eqnarray}
&& 3 b_0 r_3 + {1 \over 2} b_1 r_2 + {1 \over 16} b_2 
+(3b_1 r_3 + {1 \over 2}b_2 r_2){a\over4} + 3 b_2 r_3 {a^2\over16} = 0 \qquad
 \nonumber \\
&&\left[1 +\left({b_1\over4 b_0} +2r_2 \right)\, a \right] I(a,b_2) - a = 0  \,,
\label{optimization}
\end{eqnarray}
with
\begin{equation}
I(a,b_2)={4b_0 \over D}\left[
{(4b_1^2 - 8b_2b_0)a+b_2b_1 a^2 \over 16b_0 + 4b_1 a
  + b_2 a^2} -
 2b_2b_0 \, f(a,b_2) \right]\,.
\label{I}
\end{equation}
Eq. (\ref{running3loop}) gives $a$ as a function of $\tau$ (or $\mu$) 
and then (\ref{I}) and (\ref{optimization}) become equations in 
$\tau$ and $b_2$ that determine the optimal choice $\overline \tau(s)$ 
and $\overline b_2(s)$ of such quantities for every $s$.
\begin{figure}[t]
\begin{center}
\begin{tabular}{c c c}
\includegraphics[width=17pc]{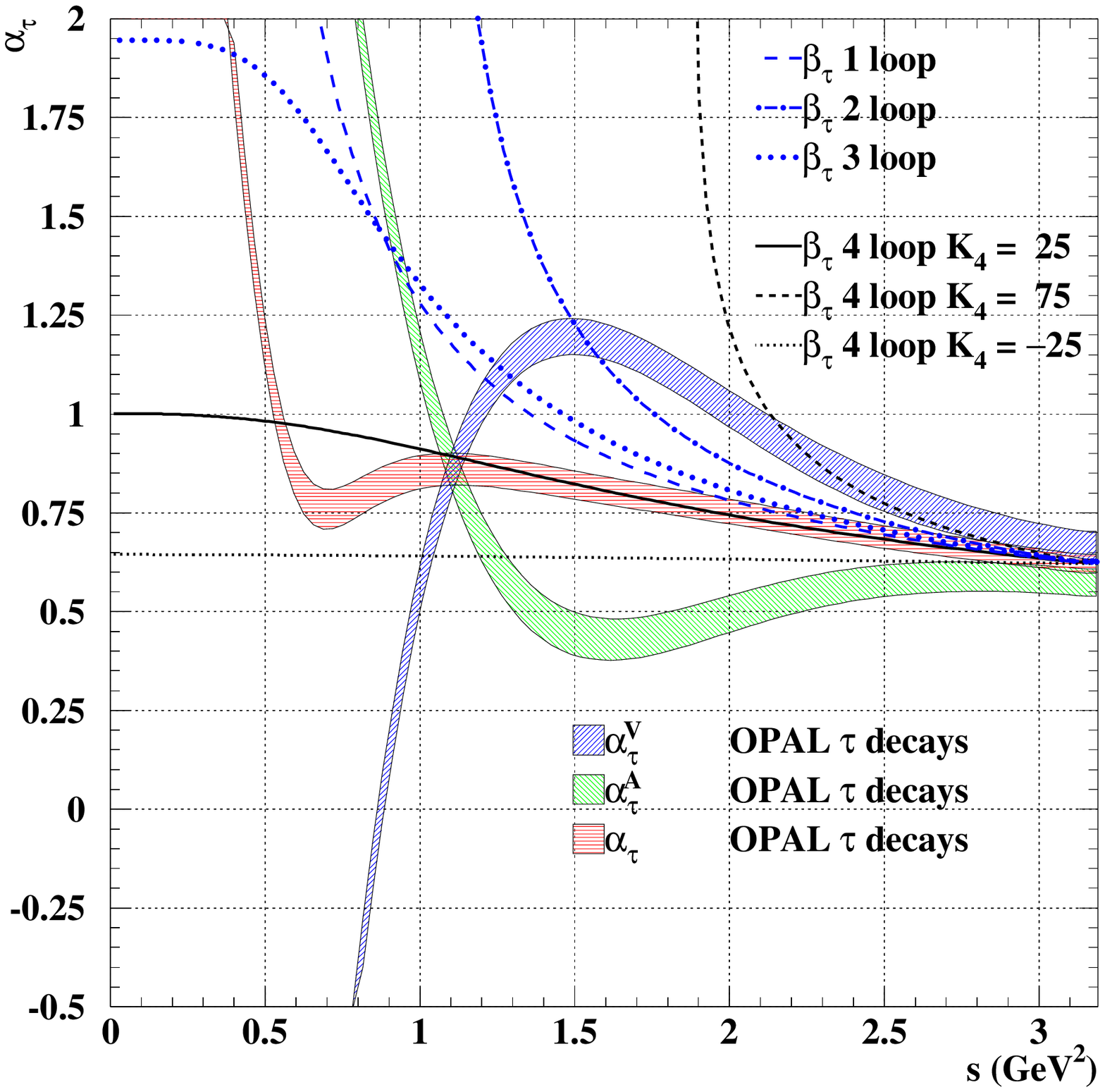}  & 
\hskip -1.2 cm\includegraphics[width=19pc,height=6.8cm]{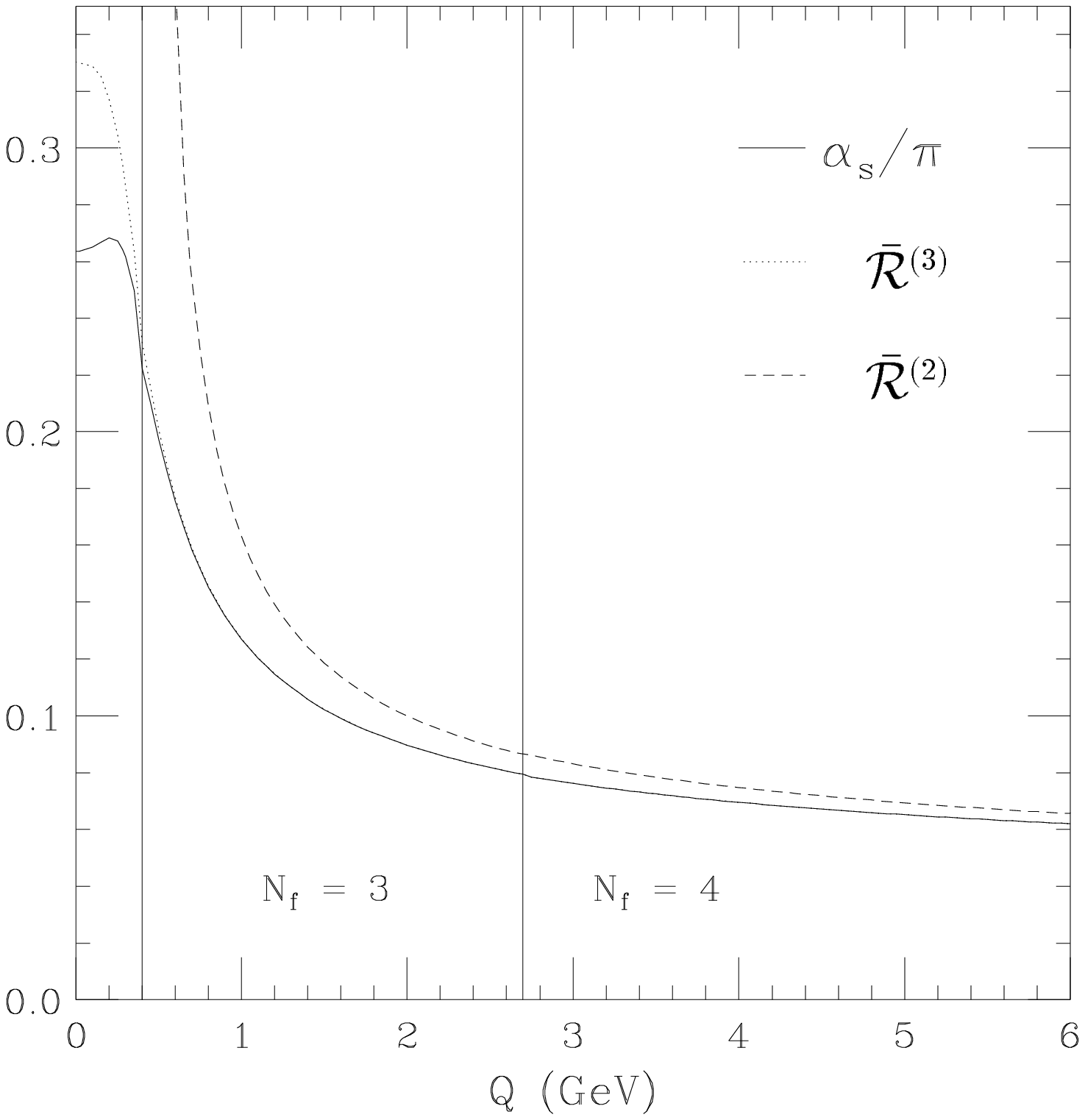} \\
  \footnotesize{(a)} & \footnotesize{(b)} \\
\end{tabular}
\caption{\footnotesize (a) Effective charge $\alpha_{\tau}$ compared with 
solution to RG equation up to four-loop.
(b) The optimized third-order results for the coupling 
$\bar{a}=\alpha_s/\pi$ and the QCD correction $\bar{{\cal R}}^{(3)}$ 
to the ratio $R_{e^+e^-}$. Also shown is the second-order result 
$\bar{{\cal R}}^{(2)}$. Quark thresholds are indicated by vertical lines.
\vspace{-0.5truecm}} 
\label{alphas}
\end{center}
\end{figure}
In this way we obtain an optimized running coupling 
$\overline a(s)= {\alpha_{\rm OPT}(s) \over \pi}$ which together with the optimized 
values $\overline r_2(s),~ \overline r_3(s)$, can be used in
(\ref{OPTRG}) to evaluate the quantity $R_{e^+e^-}(s)$. The
resulting $\alpha_{\rm OPT}(s)$ can be evaluated numerically and is
reported in fig.\ref{alphas}(b) taken from \cite{stevenson} as a function of 
$q = \sqrt s$ for 
$\Lambda_{\rm  \overline {MS}}\,$= 230 MeV  ($\tilde \Lambda_{\rm
 \overline {MS}}\,$= 264 MeV). 
Note that $\alpha_{\rm  OPT}(q^2)$ stays finite as $q$ decreases 
and attains a maximum for $q\sim$200 MeV, after which it remains 
practically constant. Such a maximum value can be shown to be given 
by the equation \cite{stevenson}
\begin{equation}
{\alpha_{\rm  OPT}^* \over \pi}= 
{-b_1 + \sqrt{b_1^2-336 b_0^2\rho_3}\over 24 b_0 \rho_3}
\,. \label{alphamax}
\end{equation}
Taking $n_f = 2$, as appropriate to the range of energy, we have 
from (\ref{OPTRGinv}) $\rho_3 = - 10.911$ and 
 so ${\alpha_{\rm OPT}^* \over \pi}=0.263$.  
Note that the behavior of $\alpha_{\rm OPT}(q^2)$ for $0<q<1$ GeV 
is perfectly consistent with what follows eq.(\ref{dks1}).
In fig.\ref{alphas}(b) are also reported the second and third order results 
for the quantity ${\cal R}$ defined by $R_{e^+e^-} = 3 \sum_f
Q_f^2(1+{\cal R})$ and expressing the QCD correction to
$R_{e^+e^-}\,$. 
The results compare favorably with the 
experimental data appropriately smeared to wash the irregularity due 
to the resonances.
\vspace{-0.5truecm}
\section{The dispersive approach}
\vspace{-0.3truecm}
As mentioned in the introduction we can dispose of the Landau
singularities simply exploiting the general analyticity properties
expected for $\alpha_{\rm s}(Q^2)$ and applying perturbation theory
directly to the spectral function \cite{Shirkov:1997wi,Ginzburg:1966}. 
This idea generalizes to QCD a
method originally introduced in QED \cite{Redmond:1958pe}.\\
Defining the spectral density
\be
\rho(\sigma)=\mathrm{Im}\,\alpha_{\rm{s}}(-\sigma-i0)=\frac{1}{2i}
\left[\alpha_{\rm{s}}(-\sigma-i0)-
\alpha_{\rm{s}}(-\sigma+i0)\right]\,, \lb{dens}
\ee
where $\sigma>0$ and $\alpha_{\rm{s}}(-\sigma)$ is the perturbative 
RG solution at a given loop level, the analytically improved running 
coupling is thus given by \cite{Shirkov:1997wi}
\be
\alpha_{\rm{an}}(Q^2)=\frac{1}{\pi}\int_0^{\infty}d\sigma\,
\frac{\rho(\sigma)}{\sigma+Q^2} \lb{alphaSH}
\ee
whose argument $Q^2=-q^2>0$ now runs over the whole space-like axis 
(we observe here and in the foregoing the identification $\mu^2=Q^2$), 
that is $\alpha_{\rm{an}}(Q^2)$ is free of any space-like unphysical
singularities by construction; moreover, due to the asymptotically 
free nature of the perturbative input the spectral integral
(\ref{alphaSH}) needs no subtractions.\\
Note that different strategies to incorporate analyticity into the
RG formalism, or even to implement the above device, exist as well, 
and they will be briefly reminded in sec. 4.4.

\vspace{-0.2truecm}
\subsection{One-loop analytic coupling}
\vspace{-0.3truecm}
At one-loop level this trick works quite straightforwardly;
starting from the leading-logs expression (\ref{1loopEC2})
the related spectral density (see also eq.(\ref{int8}))
\be
\rho^{(1)}(\sigma)=\frac{\pi\,\beta_0^{-1}}{\ln^2(\sigma/\Lambda^2)+\pi^2}
\lb{dens1} 
\ee
turns out to be explicitly integrable, and eq.(\ref{alphaSH}) 
yields \cite{Shirkov:1997wi} 
\be
\alpha_{\rm{an}}^{(1)}(Q^2)=\frac{1}{\beta_0}\left[\frac{1}{\ln(Q^2/\Lambda^2)}
+\frac{\Lambda^2}{\Lambda^2-Q^2}\right]\,. \lb{alphaSH1} 
\ee 

The analytically generated 
non-per\-tur\-ba\-ti\-ve contribution in (\ref{alphaSH1}) subtracts 
the pole in a minimal way, yielding a ghost-free behavior 
which avoids any adjustable parameter. 
Obviously eq.(\ref{intr5}) for the scaling constant does not work 
anymore and, at one-loop, it has now to be changed to 
\be
\Lambda^2=\mu_0^2\exp\left[-\phi\left(\beta_0\,\alpha_{\rm{s}}(\mu_0^2)
\right)\right]\,,
\lb{lambdaSH1} 
\ee 
where the function $\phi$ is related to the formal inverse of 
(\ref{alphaSH1}) that is, with  
$x=\beta_0\alpha_{\rm{an}}^{(1)}$ and $Q^2/\Lambda^2=\exp\phi(x)$, 
it satisfies 
\be
\frac{1}{\phi(x)}+\frac{1}{1-\exp\phi(x)}=x\,. 
\lb{phiSH} 
\ee
Among the main features of the analytically improved space-like 
coupling, it should be firstly stressed its agreement 
with asymptotic freedom constraint, being the pure perturbative 
contribution ruling in the deep UV region over the ''non-perturbative''  
one. Indeed the latter for $Q^2>\Lambda^2$ can be rewritten  
as the sum of the series 
\be
\Delta_{\rm p}^{(1)}(Q^2)=-\frac{1}{\beta_0}\sum_1^{\infty}
\left(\frac{\Lambda^2}{Q^2}\right)^n\,, 
\lb{npt1} 
\ee 
yielding power correction terms to the one-loop
perturbative coupling (\ref{1loopEC2}) (see sec 4.5). 
On the other hand, in the extreme opposite domain eq.(\ref{alphaSH1}) 
exhibits the infrared freezing value 
$\alpha_{\rm{an}}^{(1)}(0)=1/\beta_0\simeq1.396$ 
choosing consistently $n_f=3$ at low scales; 
this value turns out to be independent of $\Lambda$ and  
universal with respect to higher-loop corrections, i.e. the 
analytic coupling (\ref{alphaSH}) has a remarkably stable 
IR behavior.\\
The beta-function for the one-loop coupling (\ref{alphaSH1})  
reads \cite{Shirkov:1997wi}
\be
\beta^{(1)}_{\rm{an}}(x)=-\frac{1}{\phi^2(x)}+\frac{\exp\phi(x)}
{\left[\exp\phi(x)-1\right]^2}
\lb{anbeta}
\ee
with $\phi$ satisfying eq.(\ref{phiSH}). Despite the implicit form 
of (\ref{anbeta}), its symmetry property 
$\beta_{\rm{an}}^{(1)}(x)=\beta_{\rm{an}}^{(1)}(1-x)$ reveals
the existence of a IR fixed point at $x=1$, corresponding to 
$\alpha_{\rm{an}}(0)=1/\beta_0\,$ (see also \cite{Magradze:1999um}).\\
Finally, we just mention here that the relation between the
$\beta$-function structure and the analytical properties of the
running coupling has been investigated in ref.\cite{Krasnikov:1995is}
within a pure perturbative approach; some resummation tricks for
the asymptotic $\beta$-function have been there outlined, which can
cure the Landau ghost problem, and rely upon freedom in choosing
the RS in higher orders. Another viewpoint is given in 
\cite{Grunberg98}, where the hypothesis of pure perturbative freezing 
has been explored by analyzing the $n_f$-dependence of the 
$\beta$-coefficients (see also eq.(\ref{sud2}) and the following).

\vspace{-0.2truecm}
\subsection{Two-loop and higher orders}
\vspace{-0.3truecm} 
Actually, as discussed in sec. 2.2, the two and 
higher-loop RG equation for invariant coupling has no simple 
exact solution, so that 
rough UV approximations are commonly used 
(eq.(\ref{4loopEC})), leading to a 
cumbersome IR nonanalytical structure. 
Nevertheless, to go further inside analytization and its features  
we follow ref.\cite{Solovtsov:1999in} and choose the two-loop 
iterative solution as given by eq.(\ref{2loopECit}). 
By applying now the analytization recipe the main difficulty
arises from the integration of the related spectral density
\cite{Shirkov:1997wi}
\bea
\rho_{\rm{it}}^{(2)}(\sigma)=\mathrm{Im}\,\alpha_{\rm{it}}^{(2)}(-\sigma)=
\frac{1}{\beta_0}\frac{I(t)}{I^2(t)+R^2(t)}
\qquad t=\ln(\sigma/\Lambda^2)\qquad\qquad\qquad\lb{spect2}\\
I(t)=\pi+B_1\arccos\frac{B_1+t}{\sqrt{(B_1+t)^2+\pi^2}}\,,\,\,\,
R(t)=t+B_1\ln\frac{\sqrt{(B_1+t)^2+\pi^2}}{B_1}\nn
\eea
which does not lead to an explicit final formula, though the relative 
dispersion integral (\ref{alphaSH}) can be handled by numerical tools. 
However, recalling the singularity structure of (\ref{2loopECit}) 
(sec. 2.4), entirely subtracted by analytization, 
the two-loop analytic coupling can be recovered 
by merely adding to (\ref{2loopECit}) two compensating terms,  
cancelling respectively the pole and the cut \cite{Solovtsov:1999in}
\bea
&&\alpha_{\rm{an.it}}^{(2)}(z)=\alpha_{\rm{it}}^{(2)}(z)+\Delta_{\rm p}^{(2)}
+\Delta_{\rm c}^{(2)}
\lb{SH2}\\
&&\Delta^{(2)}_{\rm p}(z)= \frac{1}{2\beta_0}\frac{1}{1-z}\nn\\
&&\Delta^{(2)}_{\rm c}(z)=
\frac{1}{\beta_0}\int_0^{\exp{(-B_1)}}\frac{d\xi}{\xi-z}\,
\frac{B_1}{\left[\ln\xi+B_1\ln\left(-1-B_1^{-1}\ln\xi\right)\right]^2+
\pi^2 B_1^2}\nn\,
\eea
with the dimensionless variable $\xi=\sigma/\Lambda^2$. Despite 
the little handiness of (\ref{SH2}), one can still readily verify 
its limit $\alpha_{\rm{an.it}}^{(2)}(0)=1/\beta_0$; for more 
general arguments concerning universality of the IR freezing value 
through all orders see for instance 
\cite{Solovtsov:1999in, Milton:1997mi} and \cite{Alekseev:2002zn}.
Moreover, the non-perturbative UV tail of analytized coupling can be 
estimated by expanding the two compensating terms in (\ref{SH2}) 
into inverse powers of $z=Q^2/\Lambda^2$ for large $z$ 
\cite{Alekseev:2000}
\bea
&&\Delta^{(2)}_{\rm p}(z)+\Delta^{(2)}_{\rm c}(z)=\frac{1}{\beta_0}
\sum_{n=1}^{\infty}\frac{c_n}{z^n}\lb{alek2}\\
&&\nn\\
&& c_n=-\frac{1}{2}-\int_0^\infty
d\xi\,\frac{\exp{\left[-nB_1(1+\xi)\right]}}
{\left(1+\xi-\ln\xi\right)^2 +\pi^2}\nn\,. 
\eea
Thus for $z>1$ non-perturbative contributions can be recasted analogously 
to (\ref{npt1}) as a convergent power series with all negative coefficients, 
approaching $-1/2\,$ 
(e.g. $c_1=0.535\,$, see refs. \cite{Alekseev:2000} 
and \cite{Alekseev:2002jb}).\\
As yet pointed out, to get the most accurate result at two-loop 
in the IR domain, one needs to start with the exact RG solution 
(\ref{RG2loop-ex}).  
To this aim first of all one has to extrapolate eq.(\ref{RG2loop-ex}) in
such a way to define a real analytic function regular for $-e^{-1}\le\zeta\le0\,$;
the two branches $W_1(\zeta)$ and $W_{-1}(\zeta)$ merge continuously on
$(-e^{-1},0)\,$, so that the correct recipe is to use eq.(\ref{RG2loop-ex}) 
for $\rm{Im}(\zeta)\ge0$ and the same equation with $W_{-1}(\zeta)$ replaced by
$W_1(\zeta)$ for $\rm{Im}(\zeta)<0\,$. 
The discontinuity of the two loop exact solution across the time-like cut defines 
the spectral density \cite{Magradze:2000hz} (see also 
\cite{Magradze:1999um}) 
\bea
&&\rho_{\rm{ex}}^{(2)}(\sigma)=-\frac{1}{\beta_0B_1}\mathrm{Im}\left[\frac{1}
{1+W_1(\zeta(t))}\right]\qquad t=\ln(\sigma/\Lambda^2)\lb{exspc2}\\
&&\nn\\
&&\quad\zeta(t)=\frac{1}{eB_1}\exp\left[-\frac{t}{B_1}+
i\pi\left(\frac{1}{B_1}-1\right)\right]\,.\nn
\eea 
Dispersion integral (\ref{alphaSH}) with 
$t=\ln(\sigma/\Lambda^2)$ now leads to the 
exact two-loop analytic coupling
\be
\alpha^{(2)}_{\rm{an}}(z)=\frac{1}{\pi}\int_{-\infty}^{\infty}dt\,
\frac{e^t}{e^t+z}\,\bar{\rho}_{\rm{ex}}^{(2)}(t)\,
\lb{exan2}
\ee
where $\bar{\rho}_{\rm{ex}}^{(2)}(t)=\rho_{\rm{ex}}^{(2)}(\sigma)\,$.
Numerical estimates of (\ref{exan2}) as well as for the analytic
iterative coupling (\ref{SH2}) have been performed at low scales 
with $n_f=3$ in \cite{Magradze:1999um} (see Tab.1), both normalized  
at the $\tau$ mass $M_{\tau}=1.777\,$GeV, 
$\alpha_{\mathrm{s}}(M_{\tau}^2)=0.36$; comparison reveals the 
relative error for the analytized solution (\ref{SH2}) to be around 
$1.8\%$ in the IR region.\\ 
In spite of its accuracy eq.(\ref{exan2}) cannot be easily
handled, even though it is a source of numerical information to
which simpler expressions have to be compared. Actually, for
practical aims, many useful two-loop approximate formulas have
been suggested; among them we recall here the ``one-loop-like''
model \cite{Solovtsov:1999in} 
\be
\bar{\alpha}_{\rm{an}}^{(2)}(l)=\frac{1}{\beta_0}\left[\frac{1}{l}-
\frac{1}{\exp{l}-1}\right]\,,
\quad l=\ln\left(\frac{Q^2}{\Lambda^2}\right)+B_1\ln\sqrt{\ln^2\left(
\frac{Q^2}{\Lambda^2}\right)+4\pi^2} 
\lb{SH2ap}
\ee 
suitable for analysis of rather low energy phenomena, 
since it approximates the exact two-loop analytic coupling with $1\%$ 
precision for $Q\geq 1\,$GeV, and correctly reproduces both the universal 
freezing value and the UV two-loop asymptotic behavior. However, its 
accuracy breaks down when taking into account flavor thresholds. To this 
end it has been suggested \cite{Bakulev:2004cu} to use 
eq.(\ref{SH2ap}) provided that the scaling constant and the 
coefficient $B_1$ are replaced by adjustable parameters (respectively 
$\Lambda_{21}$ and $c_{21}^{fit}$ listed in tab. III of ref.\cite{Bakulev:2004cu} 
for different initial $\Lambda^{(n_f=3)}$), as a result of an interpolation 
procedure; this yields an accuracy within $1\%$ in the whole space-like 
region. A ``one-loop-like'' model employing the scaling constant as a 
fitting parameter, suitable from $2$ to $100\,$ GeV, has been recently 
developed in \cite{Shirkov:2005sg}.\\
Increasing difficulties arise when dealing with even higher-loop 
level and it becomes prohibitive to achieve useful explicit formulas. 
Starting e.g. with the standard three or four-loop asymptotic solution 
(\ref{4loopEC}), one has to face with the leading singularity in $z=1$ 
of the form (\ref{IR3loop}), beside the IR log-of-log generated cut; 
terms accounting for these divergences acquire the 
form of cumbersome finite limits integral as in (\ref{SH2}). 
Nonetheless, the effects of non-perturbative contributions have been 
widely investigated up to four-loop \cite{Alekseev:2002jb,Alekseev:2002zn}, 
both in IR (where they play the most prominent role) and UV region, by 
using asymptotic solution (\ref{4loopEC}) as a perturbative input. 
While confirming at once IR stability due to the universal freezing value 
of the analytized coupling, its UV tail has been reduced \cite{Alekseev:2002zn} 
in the form of power type corrections analogous to (\ref{alek2}). Within this 
approximation the $n_f$-dependent coefficients $c_n$ are all 
negative up to four-loop, and their absolute values monotonously increase 
with $n$.  
In the large $Q^2$ limit, however,   
there is no need to sum a high number of terms,
and truncation of e.g. the three-loop non-perturbative series to first 
term yields the approximate expression 
\cite{Alekseev:2002zn}
\vspace{-0.1truecm} 
\bea
&&\bar{\alpha}^{(3)}_{\rm{an}}(z)=\alpha_{\rm{s}}^{(3)}(z)+\frac{1}{\beta_0} c_1
\frac{\Lambda^2}{Q^2}\lb{alek3}\\
&&\qquad c_1=-1+B_1(1-\gamma_{\rm E})-\frac{B_1^2}{2}\left[B_2-\frac{\pi^2}{6}+
(1-\gamma_{\rm E})^2\right]\nn
\eea
\vspace{-0.02truecm}
for the analytic coupling, with $1\%$ accuracy yet at $Q^2\simeq5\Lambda^2\,$. Here 
$\gamma_{\rm E}$ is the Euler constant, 
$B_2=\beta_0\beta_2/\beta_1^2$ 
and $\alpha_{\rm{s}}^{(3)}(z)$ is the perturbative counterpart as given 
by (\ref{4loopEC}); if $n_f=6$ we roughly have $c_1\simeq-0.52\,$.\\
By normalizing the three-loop analytized coupling at 
$\alpha_{\mathrm{s}}(M_Z^2)\simeq0.118$ (with perturbative input from 
eq.(\ref{4loopEC}) and numerical evaluation of dispersion integral 
(\ref{alphaSH})), one can extract the three-loop scaling constant 
$\Lambda^{(n_f=5)}\simeq210\,$MeV \cite{Alekseev:2002zn}, which
lies within the errors of the pure perturbative estimate (see
sec. 2.3), being the non-perturbative tail negligible around the
normalization point. Obviously the main discrepancies emerge in the 
low energy region, where non-perturbative contributions slow down the
rise of the curve. By using continuous matching up to three-loop, at
the $\overline{\rm{MS}}$ quark masses $m_b=4.3\,$GeV and $m_c=1.3\,$GeV,
for the three-loop analytized case one finds roughly \cite{Alekseev:2002zn}
\be
\Lambda^{(n_f=5)}\simeq210\,{\rm MeV}\,\,\to\,\,\Lambda^{(n_f=4)}\simeq299\,{\rm MeV}
\,\,\to\,\,\Lambda^{(n_f=3)}\simeq382\,{\rm MeV}\,,
\lb{L_an}
\ee
to be compared with the relative perturbative estimates\footnote{
The shifts of $\Lambda$s in (\ref{L_pt}) w.r.t. sec. 2.3 are mainly
due to the different normalization value, and partly to continuous 
matching used here; indeed there is little sensitivity to the 
implementation to trivial matching given by eq.(\ref{L3match}).}
\be
\Lambda^{(n_f=5)}\simeq210\,{\rm MeV}\,\,\to\,\,\Lambda^{(n_f=4)}\simeq290\,{\rm MeV}
\,\,\to\,\,\Lambda^{(n_f=3)}\simeq329\,{\rm MeV}\,.
\lb{L_pt}
\ee
These discrepancies at low scales can be translated into the values
of analytic and perturbative coupling at the $\tau$ mass 
$M_{\tau}=1.777\,$GeV, given respectively by 
$\alpha_{\rm an}^{(3)}(M_{\tau}^2)\simeq0.294$ and
$\alpha_{\rm{s}}^{(3)}(M_{\tau}^2)\simeq0.318\,$. 
Note that normalization at $\alpha_{\rm s}(M_{\tau})=0.35$ is 
sometimes adopted, and this leads to considerably higher values for 
the scaling constant. 
Last, it should be stressed the nearly indistinguishability of the
two to four-loop analytic curves (fig.\ref{alphas2}(a)), confirming 
higher order 
stability on the whole spacelike axis \cite{Shirkov:1997wi, Alekseev:2002zn}. 
Thus we argue the two-loop correction to be a well 
satisfactory improvement of the one-loop result, so that it is 
reasonable to resort to approximate formulas as given e.g. by 
(\ref{SH2ap}). However, in recent works \cite{Kurashev:2003pt, Magradze:2005ab} 
the multi-loop approximation (\ref{alpha^k}) as a power series in the two-loop 
exact coupling (\ref{RG2loop-ex}) has been exploited as an input in the dispersion
integral (\ref{alphaSH}), to yield high-accuracy three and four-loop 
analytic coupling of the form \cite{Kurashev:2003pt}
\be
\alpha_{\rm{an}}^{(k)}(Q^2)=\sum_{n\ge1} p_n^{(k)}
\left[\alpha_{\rm{an}}^{(2)}(Q^2)\right]^n
\lb{alpha^k_an}
\ee
with $\alpha_{\rm{an}}^{(2)}$ given by (\ref{exan2}) and coefficients 
the same as in (\ref{alpha^k}). Numerical values at low scales can be 
found in \cite{Kurashev:2003pt}, to which we refer for more details.

\vspace{-0.2truecm}
\subsection{The time-like coupling in the analytic approach}
\vspace{-0.3truecm}
Coming back to the issue of defining a reasonable expansion parameter
in the time-like domain ($q^2=s>0$), we start by noting that such a 
definition naturally arises in a self-consistent way within the 
framework of the analytic approach \cite{Milton:1997us, Solovtsov:1997at}. 
Further, it can be regarded as the final step in the procedure RKP 
for $\pi^2$-resummation outlined in sec. 2.5.\\
As yet noted, eq.(\ref{dispAdl}) and its formal inverse (\ref{invsAdl}) 
can be generalized to the proper tool for relating s- and t-channel
observables, that is for crossing the two distinct regions. On this 
ground, it has then been suggested \cite{Jones:1995rd, Milton:1997us} 
to use the same integral transformations in order to connect the 
time-like and space-like couplings, employing as a suitable regularization  
the analytically improved coupling (\ref{alphaSH}), free of 
unphysical singularities at any loop level. Then \cite{Milton:1997us}
\bea
&&\tilde{\alpha}(s)=\frac{i}{2\pi}\int_{s-i\varepsilon}^{s+i\varepsilon}
\frac{dq^2}{q^2}\,\alpha_{\rm an}(-q^2)\lb{int-r}\\
&&\alpha_{\rm an}(-q^2)=-q^2\int_0^{\infty}ds\,\frac{\tilde{\alpha}(s)}{(s-q^2)^2}
\lb{int-d}
\eea
with $q^2$ as usual in $\mathbb{C}-\{q^2=s>0\}\,$. Specifically, 
eq.(\ref{int-r}) can be assumed as a definition of the time-like 
coupling. Once the space-like singularities have been washed out 
the ambiguity about the path for (\ref{int-r}) disappears, and we 
find a one-to-one relation between the t- and s-channel couplings.
Note also that $\tilde \alpha(s)$ can be equally defined by the 
differential equation \cite{dokshitzer2}
\be
s {d \over ds} \tilde \alpha(s)\,
=\,-{1 \over \pi}\rho(s) \qquad {\rm with}\qquad  \tilde 
\alpha(\infty) = 0 \,,
\lb{dks6}
\ee
as can be immediately checked, e.g. by differentiating eq.(\ref{int-d}). 
Then, eq.(\ref{dks6}) immediately yields  
\be
\tilde{\alpha}(s)=\frac{1}{\pi}\,\int_s^{\infty}\frac{d\sigma}{\sigma}\,
\rho(\sigma)\lb{a_s}
\ee
and $\rho(\sigma)$ as given by (\ref{dens}). Moreover, eq.(\ref{dks6}) 
emphasizes the straightforward relation between the ``time-like 
$\beta$-function'' and the spectral density, thus reviving an old 
hypothesis due to Schwinger \cite{Schwinger:1975th}.\\ 
With the use of (\ref{dks6}), eq.(\ref{int-d}) can be also formally 
inverted as \cite{dokshitzer2}
\bea
&&\tilde \alpha(s)= { \sin (\pi {\cal P}) \over {\pi\cal P}}\,\alpha_{\rm an}(s) = 
\, \alpha_{\rm an}(s) - {1\over 3!} \left(\pi {\cal P}\right)^2 \alpha_{\rm an}(s) + \dots 
\lb{dks9}\\
&&\qquad=\alpha_{\rm{an}}(s)\left[1-\frac{\pi^2\beta_0^2}{3}\alpha^2_{\rm{an}}(s)-
\frac{5\pi^2\beta_0\beta_1}{6}\alpha_{\rm an}^3(s)+\dots\,\right]\nn
\eea
where ${\cal P}=s(d/ d s)$. Of course at one-loop 
(\ref{dks9}) coincides with (\ref{UValpha_RKP2}), and 
\begin{figure}[t]
\begin{tabular}{c c c}
\includegraphics[width=7.8cm,height=7.7cm]{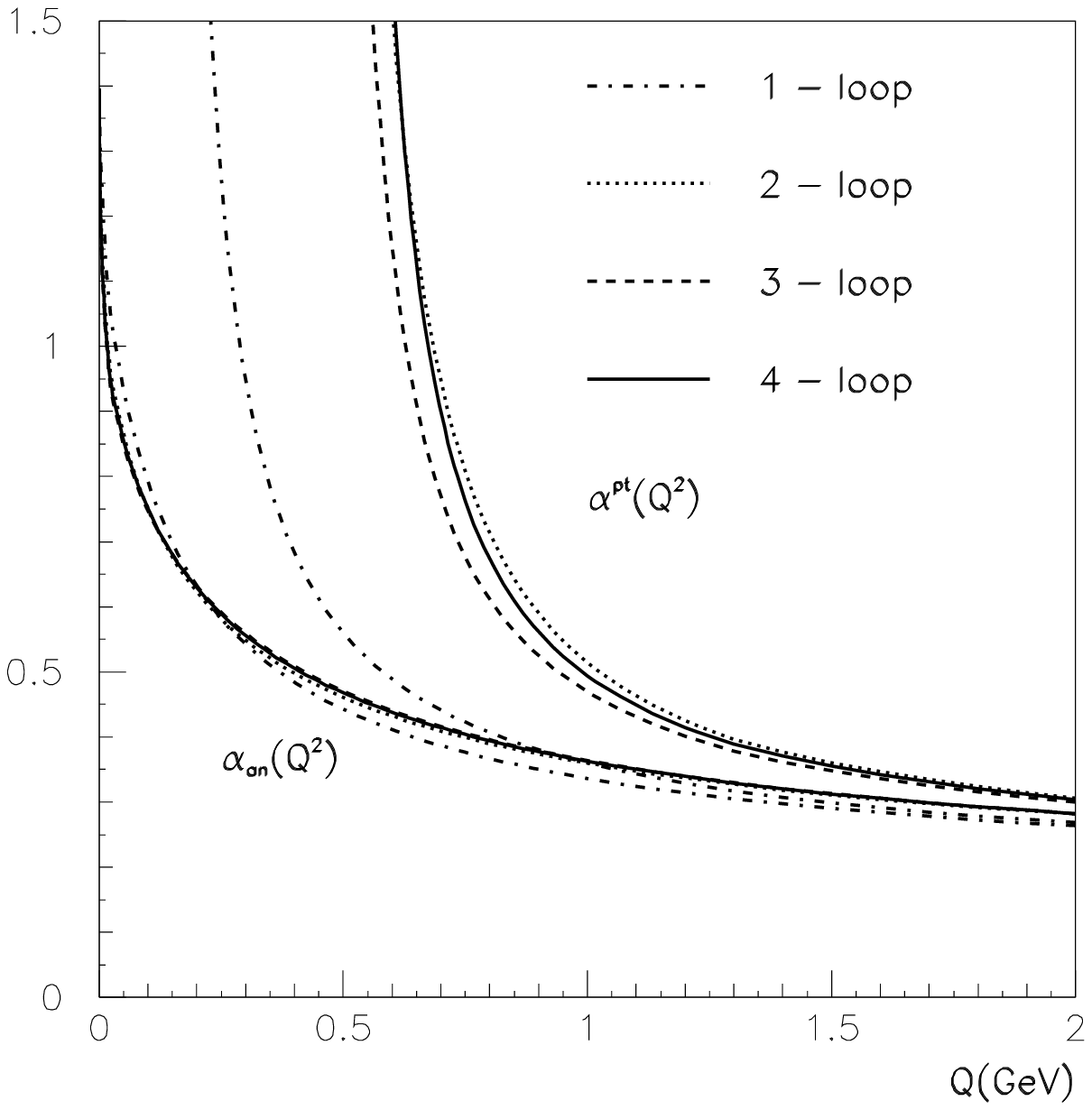}  & 
\hskip -0.5 cm\includegraphics[width=6.0cm,height=5.9cm]{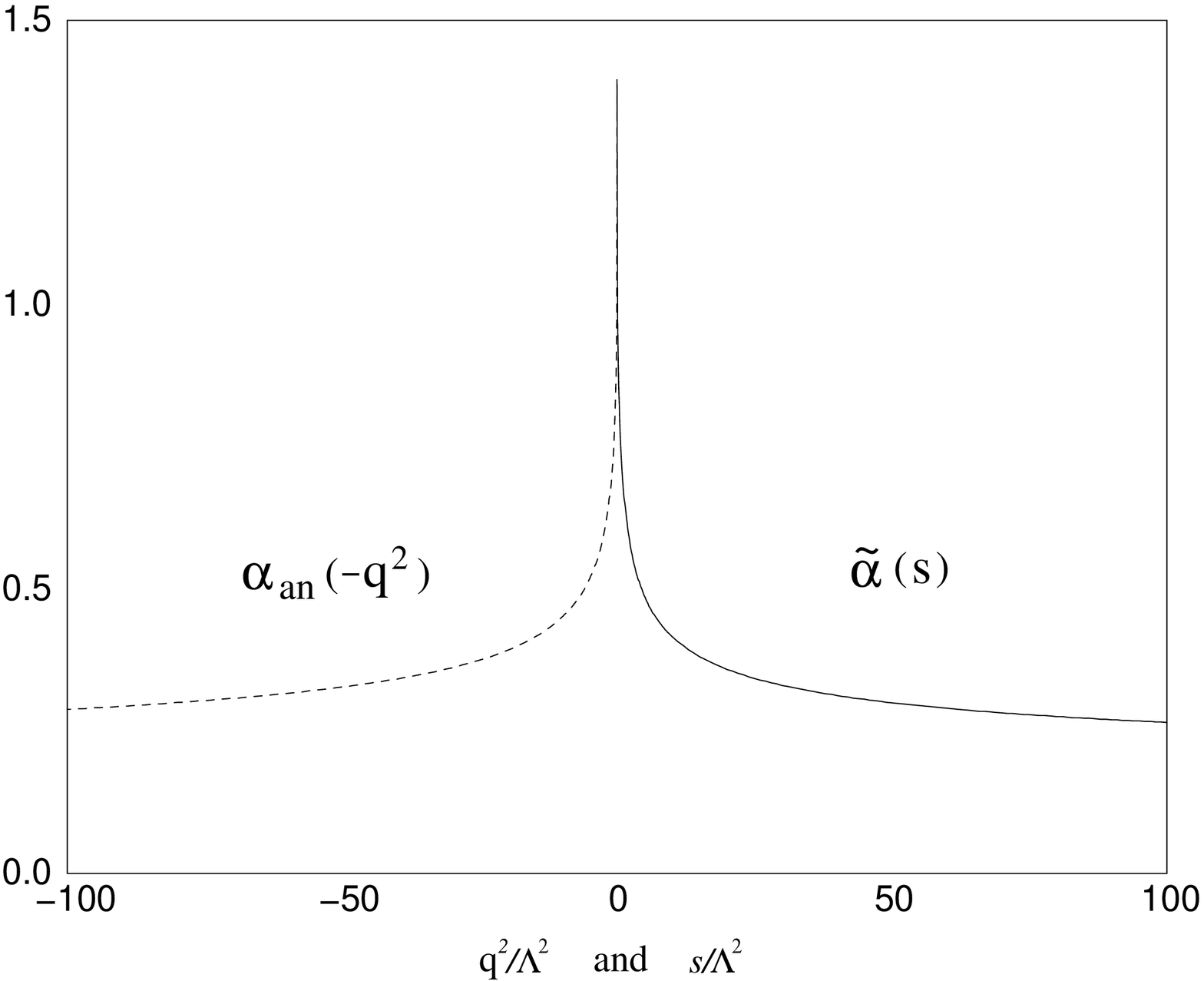} \\
  \footnotesize{(a)} & \footnotesize{(b)} \\
\end{tabular}
\caption{\footnotesize (a) Analytic coupling from \cite{Alekseev:2002zn} 
up to four-loop with the above normalizing conditions and threshold matching, 
compared to pure perturbative input (\ref{4loopEC}).
(b) From\cite{Milton:1997us}: comparison of one-loop timelike and spacelike
couplings (\ref{alpha_RKP}) and (\ref{alphaSH1}).} 
\label{alphas2}
\vspace{-0.5truecm}
\end{figure}
from eqs.(\ref{a_s}) and (\ref{dens1}) one finds   
again eq.(\ref{alpha_RKP}); however this now leads through  
(\ref{int-d}) to the starting space-like coupling (\ref{alphaSH1}), 
being the required analytic properties preserved within this framework
(they are both plotted in fig.\ref{alphas2}(b)).
Furthermore, at two-loop level inserting into eq.(\ref{a_s}) the spectral 
density (\ref{exspc2}) computed on the two-loop exact RG solution 
(\ref{RG2loop-ex}), unlike the space-like case, integral can be taken 
analytically \cite{Kurashev:2003pt}
\bea
&&\tilde{\alpha}^{(2)}(s)=-\frac{\beta_0}{\beta_1}+\frac{1}{\pi\beta_0}
\rm{Im}\left[1+W_1\left(\zeta(s)\right)\right]\lb{a_s2}\\
&&\nn\\
&&\zeta(s)=\frac{1}{eB_1}\exp\left[-\frac{1}{B_1}\ln\left(\frac{s}{\Lambda^2}\right)
+i\pi\left(\frac{1}{B_1}-1\right)\right]\nn\,.
\eea
The main feature \cite{Milton:1997us} of (\ref{a_s}) and 
(\ref{alphaSH}) is the common freezing value at the origin 
$\tilde{\alpha}(0)=\alpha_{\rm{an}}(0)=1/\beta_0$, independent of the 
loop level and of any adjustable parameter. Moreover, they exhibit 
similar leading UV behavior, constrained by asymptotic freedom, as can 
be seen by (\ref{dks9}) (and recalling that asymptotically (\ref{alphaSH}) 
reduces to the pure perturbative coupling $\alpha_{\rm s}$).\\ 
Nevertheless this approximate ``mirror symmetry'' is broken in the 
intermediate region, the discrepancy being about $9\%$ at one-loop, 
and slightly less at two and three-loop 
(see ref.\cite{Milton:1997us} for numerical comparisons).\\
We finally just note that ref.\cite{Milton:1998wi} exploits an argument 
against a possible exact symmetry ruling the ``t-s dual'' couplings, 
(\ref{alphaSH}) and (\ref{a_s}), on the ground of causality principle.

\vspace{-0.4truecm}
\subsection{Related models}
\vspace{-0.3truecm}
Since the last ten years there have been a number of  
different attempts to avoid Landau singularities invoking as 
well analyticity of the coupling in the space-like momentum region. 
Within the dispersive approach, it is remarkable the existence of 
models suggesting IR enhancement of the QCD coupling, whose most 
attractive feature is supposed to be a straightforward relation
with quark confining potential within the framework of one-gluon 
exchange (see sec. 3.1).\\ 
To this class belongs for instance the 
``synthetic coupling'' model recently developed in 
\cite{Alekseev:2004vx,Alekseev:2005vh}, which amounts to modify by 
hand the analytically improved coupling (\ref{alphaSH}) by additional 
non-perturbative pole-type terms; at one-loop it reads \cite{Alekseev:2005vh}
\be
\alpha_{\rm{syn}}(Q^2)=\frac{1}{\beta_0}\left[\frac{1}{\ln(Q^2/\Lambda^2)}
+\frac{\Lambda^2}{\Lambda^2-Q^2}+\frac{c\Lambda^2}{Q^2}+
\frac{(1-c)\Lambda^2}{Q^2+m_g^2}\right]\,.
\lb{a_syn}
\ee
where $m_g=\Lambda/\sqrt{c-1}\,$. 
Infrared enhancement, due to the pole term at $Q^2=0\,$ (cf.
eq.(\ref{richardson3})), is governed by one dimensionless 
parameter $c\in(1,\infty)$, which is meant to relate the 
scaling constant $\Lambda$ to the string tension $\sigma$ 
of potential models, through the same eq.(\ref{richardson4}) 
but with $\Lambda$ replaced by $\sqrt c\Lambda$. The pole term at 
$Q^2=-m_g^2<0$ corresponds to a non vanishing dynamical 
gluon mass, while leaving the 
analytical structure of eq.(\ref{alphaSH1}) along the 
space-like axis unchanged. The value of the $m_g$ parameter 
has been estimated \cite{Alekseev:2005vh} as $400-600\rm{MeV}\,$
(see sec. 3.1). Eq.(\ref{a_syn}) can be derived, analogously to 
(\ref{alphaSH1}), from a dispersion relation with a 
spectral density of the form (\ref{dens1}) plus 
two $\delta$-terms properly accounting for the poles. Along 
with the singular behavior as $1/Q^2$ at $Q^2\to0\,$, 
reproducing the linear confining part of the potential 
(\ref{richardson1}), construction of (\ref{a_syn}) is mainly 
motivated by the UV asymptotic \cite{Alekseev:2005vh} of its 
non-perturbative contribution of the form $1/(Q^2)^3\,$,   
decreasing faster than (\ref{npt1}) as $Q^2\to\infty\,$.\\
Quite analogous result has been achieved in 
\cite{Nesterenko:2001xa, Nesterenko:2003xb}, 
merging analyticity and IR slavery at zero momentum in the 
one-loop formula 
\be
\alpha_{\rm{N}}(Q^2)=\frac{1}{\beta_0}\frac{z-1}{z\ln z}
\lb{a_N}
\ee
where $z=Q^2/\Lambda^2\,$. The trick undertaken here 
is to impose analyticity ab initio on the whole perturbative 
expansion of the QCD $\beta$-function, and then to solve 
the ensuing `` analytized RG equation'' for the running 
coupling. This could be done in principle at any loop-level, 
formally \cite{Nesterenko:2001xa} 
\be
\mu^2\frac{d\ln\alpha_{\rm{N}}(\mu^2)}{d\mu^2}=-\left\{\sum_{j=0}^{l-1}\beta_j 
\alpha^{j+1}(\mu^2)\right\}_{\rm{an}}\,,
\lb{RG_N}
\ee
where  
the r.h.s. of 
(\ref{RG_N}) is achieved through the usual dispersion integral 
(\ref{alphaSH}), starting from its spectral density now given by 
the discontinuity of the expression of the l-loop $\beta$-function
as a whole across the 
time-like cut. Note that in the one-loop case the r.h.s. of 
(\ref{RG_N}) is merely proportional to the one-loop analytic 
coupling (\ref{alphaSH1}), so that further integration leads to the 
singularity at zero momentum. 
This ``new analytic invariant charge'' possesses the universal 
(i.e. loop-independent) asymptotic $\simeq\Lambda^2/Q^2$ as 
$Q^2\to0\,$, which again results in confining quark-antiquark 
potential \cite{Nesterenko:2003xb}. Besides, this variant of 
the dispersive approach, while disclosing in some measure the 
ambiguities suffered by the method, shares appealing features 
with the IR finite analytic coupling (\ref{alphaSH}); namely, 
it displays no unphysical singularities and no adjustable 
parameters, and exhibits good higher-loop and RS stability (see 
e.g. \cite{Nesterenko:2003xb} for technical details).
Analogous IR behavior of the QCD coupling has been 
found in \cite{Schrempp:2001ir} on the ground of different 
reasoning (see also \cite{Nesterenko:2001xa, Nesterenko:2003xb} 
and refs. therein for similar proposal and results). It 
is worth noting that in the most recent developments 
\cite{Nesterenko:2004ry} of the model in hand, eqs.(\ref{RG_N})
and (\ref{a_N}), inclusion of the lightest hadron 
masses ($\pi$ meson) is accomplished consistently with the 
dispersive approach, and relations with chiral symmetry breaking 
phenomena are also investigated \cite{Aguilar:2005sb}. It has 
been shown there how nonzero pion mass substantially affects low 
energy behavior of the invariant charge (\ref{a_N}), by slowing 
down the IR enhancement of the massless case to an IR finite value, 
depending in turn on the pion mass. As a result, this massive
running coupling displays a plateau-like freezing 
on the IR time-like axis, specifically for $\sqrt{s}$ in the interval 
between $0$ and the two-pion threshold (see also 
\cite{Nesterenko:2005nj}), in qualitative agreement with results 
of OPT \cite{stevenson} (sec. 3.3).\\
Finally we remind another attempt \cite{Webber:1998um} to modify 
eq.(\ref{alphaSH1}) in order to estimate non-perturbative power 
corrections (see sec 4.5). This is dictated by the need to cancel 
somehow the unwanted behavior (\ref{npt1}),  
and further to slow down the too large value of 
(\ref{alphaSH1}) at the origin (see sec. 5). This has been 
done by requiring no power corrections faster than $1/Q^{2p}\,$,  
and a number of adjustable parameters to remodel 
the IR behavior of the one-loop analytic coupling (\ref{alphaSH1}).  
Thus an useful generalization reads
\be
\alpha_{\rm W}(z)=\frac{1}{\beta_0}\left[\frac{1}{\ln z}+\frac{z+b}{(1-z)(1+b)}
\left(\frac{1+c}{z+c}\right)^p\right]\,.
\lb{a_W}
\ee
Setting $b=1/4$ and $c=p=4\,$ and $n_f=3\,$, $\Lambda=250\,$MeV 
eq.(\ref{a_W}) has a maximum at $0.4\,$GeV and then freezes to a 
considerably lower value than (\ref{alphaSH1}), fitting data on 
power corrections (see \cite{Webber:1998um} and refs. therein). 

\vspace{-0.2truecm}
\subsection{Power suppressed non-perturbative corrections}
\vspace{-0.3truecm}
As mentioned in the introduction, intrinsically non-perturbative
effects would manifest themselves in power-type corrections 
$A_\nu/ Q^\nu$ or 
$(A_\nu/Q^\nu)\ln(Q^2/Q^2_{\rm I})$, to the 
expression of various observables.
We want to discuss the subject in the context of
ref. \cite{dokshitzer2}, in which it was originally proposed, by
making use of the coupling $\alpha_{\rm SGD}(Q^2)$
we have introduced by eq. (\ref{dks3}).\\
Note that, as consequence of its definition,  $\alpha_{\rm SGD}(Q^2)$
must have only physical singularities in the $Q^2$ complex plane  and the
entire dispersive formalism discussed in this section can be applied to
it. Actually such formalism sprang from a common source
\cite{Dokshitzer:1993pf} and developed along parallel lines from
references \cite{Shirkov:1997wi} and \cite{dokshitzer2}. 
According to eq. (\ref{alphaSH}) we can write
\bea
&&\alpha_{\rm SGD}(Q^2) = {1 \over \pi} \int_0^\infty dm^2 {\rho (m^2)
\over m^2+Q^2}\nn\\ 
&&\,\qquad\qquad=\alpha_{\rm SGD}(0) - {Q^2 \over \pi} \int_0^\infty dm^2 
{\rho (m^2)\over m^2(m^2+Q^2)}\,.
\lb{dks20}
\eea
Then, the factor corresponding to a dressed gluon line in a Feynman
integral can be written
\be
{-i\,  \alpha_{\rm SGD}(-k^2) \over k^2+i0} = 
{-i \, \alpha_{\rm SGD}(0) \over k^2+i0}
- {1 \over \pi} \int_0^\infty \frac{dm^2}{m^2} \rho (m^2)
 {-i \over k^2-m^2+i0}\,.
\lb{dks21}
\ee
Let us now consider some hard process initiated by a quark and an 
inclusive infrared and collinear safe observable $V(Q^2,x)$ related 
to it. Let us assume, to be 
specific, the observable has a zero order expression $V_0 (Q^2,x)$ 
in terms of parton model and consider the first order QCD correction 
$V_1 (Q^2,x)$ in which dressed gluon lines are inserted in the original
skeleton graph. Let us denote by ${\cal F}_1(\epsilon,\,x)$ the Feynman
integral (or the sum of Feynman integrals) that gives such correction
but in which formally a mass $m$ is given to the gluon and it is set
$\epsilon={m^2 \over Q^2}$. Due to (\ref{dks21}) the insertion gives
\bea
&& V_1 (Q^2,x)= \alpha_{\rm SGD}(0){\cal F}_1(0,\,x)
- {1 \over \pi} \int_0^\infty \frac{dm^2}{m^2}\rho (m^2){\cal F}_1(\epsilon,\,x)=\lb{dks10} \\ 
&& \quad =\int_0^\infty {dm^2 \over m^2}  
\tilde \alpha_{\rm SGD}(m^2) \dot {\cal F}_1(\epsilon,\,x)\,,
\quad {\rm with}\quad \dot {\cal F}_1(\epsilon,\,x)  = -\epsilon 
{\partial \over \partial \epsilon} {\cal F}_1(\epsilon,\,x)
\,,
\nn
\eea
where we have introduced the time-like coupling
\footnote{Note that in the original papers the quantity 
$\tilde\alpha_{\rm SGD}(m^2)$ was denoted as $\alpha_{\rm eff}(m^2)\,$.} 
$\tilde\alpha_{\rm
 SGD}(m^2)\,$, related to $\alpha_{\rm SGD}(m^2)$ by the equation
$m^2(d\tilde\alpha_{\rm SGD}(m^2)/dm^2)=-\rho(m^2)/\pi$ (cf. 
eq.(\ref{dks6})).\\
Eq. (\ref{dks10}) gives a kind of weighted average of ${\cal F}_1(\epsilon,\,x)$ and
shows that the effect of the running coupling can be simulated giving an
appropriate effective mass $m_g\,$, and justifies the phenomenological
application based on such idea that we mentioned in sec. 3.1.
Let us now write $\alpha_{\rm SGD}(Q^2)$  and
correspondingly $\tilde\alpha_{\rm SGD}(Q^2)$ as the sum of a
perturbative and a non perturbative part
\be
\alpha_{\rm SGD}(Q^2) = \alpha_{\rm  SGD}^{\rm PT}(Q^2) +
\alpha_{\rm SGD}^{\rm NP}(Q^2) \,.
\lb{dks11a}
\ee
Since, however, perturbative theory appears to work very well down
to $Q\,\sim$ 1 or 2 GeV and in some particular case even better,
$\alpha_{\rm SGD}^{\rm NP}(Q^2)$ must vanish sufficiently
fast as $Q$ increases, let us say at least as $1/Q^6\,$.
Consequently, looking at eq.(\ref{int-d}) we must assume that 
at least the two first integer moments of 
$\tilde \alpha_{\rm SGD}(m^2)$ vanish,
\be
\int_0^\infty {dm^2 \over m^2} m^2 \tilde \alpha_{\rm SGD}^{\rm NP}(m^2)=0\,,
\, \,  \int_0^{\infty} {dm^2 \over m^2} m^4 \tilde \alpha_{\rm SGD}^{\rm NP}(m^2)=0\,.
\lb{dks10a}
\ee
Let us now restrict  to collinear and infrared safe observables. Then 
$\dot {\cal F}_1(\epsilon,\,x)$ must vanish conveniently for $\epsilon
\to 0$ or $\infty$. Let us assume for $\epsilon \to 0$ 
\bea
&&\dot {\cal F}_1(\epsilon,\,x)\to \qquad \qquad \\ 
&&\,\to \,{C_F \over 2 \pi} \epsilon^p
\left[ (f_0 + f_1 \ln \epsilon + f_2 \ln^2 \epsilon) + \epsilon (g_0
  + g_1 \ln \epsilon + g_2 \ln^2 \epsilon) + \dots \right]\,,\nn
\lb{dks11}
\eea
where $p$ can be integer or half integer and $ f_q,\, g_q 
, \dots$ depend on the particular process. For the various jet shape
variables in $e^+e^-$ annihilation (thrust, jet mass, $C$ parameter)   
e.g., $\dot {\cal F}_1(\epsilon,\,x)\to {C_F \over 2 \pi}f_V \sqrt
\epsilon $ with $f_V= 4,~2,~6\pi$ respectively.\\
As a consequence we have for $Q\to \infty$
\be
\int_0^\infty {dm^2 \over m^2}  
\tilde \alpha_{\rm SGD}^{\rm NP}(m^2) \dot {\cal F}_1(\epsilon,\,x)\, \to 
f_0 {A_{2p} \over Q^{2p}} + f_1 \left({A_{2p}^\prime \over Q^{2p}} 
- {A_{2p} \over Q^{2p}}\ln Q^2\right) + \dots \nn 
\lb{dks12}
\ee
or
\be
V(Q^2,x) = V_{\rm PT}(Q^2,x) + {1 \over Q^{2p}}[C_1 A_{2p} +
C_2 A_{2p}^\prime+ C_3 A_{2p}^{\prime\prime}]
\lb{dks13}
\ee
where we have introduced the moments
\bea
 && A_{2p}= {C_F \over 2 \pi}\int_0^\infty 
  {dm^2 \over m^2} m^{2p}\,  \tilde \alpha_{\rm SGD}^{\rm NP}(m^2)\,, \nn \\
&& A_{2p}^\prime = {C_F \over 2 \pi}\int_0^\infty {dm^2 \over m^2} 
   m^{2p} \ln m^2 \, \tilde \alpha_{\rm SGD}^{\rm NP}(m^2)\,,\\
&&  A_{2p}^{\prime\prime} ={C_F \over 2 \pi}\int_0^\infty 
{dm^2 \over m^2} m^{2p}\ln^2 m^2 \, \tilde \alpha_{\rm SGD}^{\rm NP}(m^2)\,.
\nn
\lb{dks14}
\eea
The coefficients $C_k$ are dimensionless, are in practice at most
linear in $\ln Q^2$ and are calculable but process dependent, $A_{2p},~ 
A_{2p}^\prime, ~\dots$ are theoretical unknown but should be
universal. They  could be determined in 
principle by studying the dependence on the hard scale $Q$ of  
appropriate observables, $V(Q^2,x)$.\\
Terms in $1/Q$ should emerge for what we have seen in various
$e^+e^-$ jet shape variables; correction of the type
$1/Q^2$ in the DIS structure function (see \cite{Dasgupta:2003iq}). The results can
be equivalently expressed in terms of the moments of 
$\alpha_{\rm SGD}^{\rm NP}(Q^2)$ rather then of $\tilde \alpha_{\rm
  SDG}^{\rm NP}(m^2)$. The experimental situation is not completely clear but
the data seem to be consistent with
\be
{\cal A}_1 = {C_F \over 2 \pi}\int_0^\infty {dQ^2 \over Q^2}
Q\,\alpha_{\rm SDG}^{\rm NP}(Q^2)={2 \over \pi} A_1 \approx 0.2-0.25 \,{\rm
  GeV}
\ee
and
\be
{\cal A}_2 = {C_F \over 2 \pi}\int_0^\infty {dQ^2 \over Q^2}
Q^2 \alpha_{\rm SDG}^{\rm NP}(Q^2)=- A_2^\prime \approx 0.2\, {\rm
  GeV}^2\,.
\ee
Note that in principle the quantity $V_{\rm PT}(Q^2,x)$ is well
defined, since  $ \alpha_{\rm SDG}$ has no unphysical
singularities, due to its definition (\ref{dks3}). To be consistent up
two-loop one should use equations for $ \alpha_{\rm SDG}^{\rm NP}$ of 
the type considered in sec. 4.1 or 4.2 with the appropriate value of
$\Lambda_{\rm SGD}$. However, often an infrared cutoff $Q_{\rm I}$ has been
introduced in the application and ${\cal A}_1$ related to the quantity
$\langle {\alpha_{\rm s} \over \pi} \rangle$ considered
in sec. 3.1.\\
Relations involving OPE 
and the non trivial IR structure of the theory have been considered in the 2D 
Gross-Neveu model in \cite{Langfeld:1995si}.\\
To conclude this section we should mention that the same physical
problem we have briefly discussed from the point of view of the running
coupling is the object of another very alive line of research that 
operates in the framework of the Borel summability and the renormalon 
singularities. Since there are obviously connections between the two
perspectives, we must invite the interested reader to consult some of 
the existing excellent general reviews, e.g. \cite{Beneke:2000kc},  
and among the most recent works we remind ref.\cite{Cvetic:2002qf}.  

\vspace{-0.3truecm}
\subsection{Analytic Pertubation Theory}
\vspace{-0.3truecm}
As explained at lenght, analytization displays 
a suitable method to get rid of unphysical singularities which
affect the standard expansion parameter in t- and s-channel. 
Thus the issue of how perturbation theory should be
accordingly modified naturally rises, and it has been 
investigated from phenomenological and theoretical point of
view (e.g. \cite{Shirkov:1998sb}). We briefly recall here a
consistent way to incorporate the ghost-free model for the IR 
finite couplings eqs.(\ref{alphaSH}) and (\ref{a_s}) 
within perturbation theory, known as Analytic Perturbation Theory 
(APT), and developed in 
\cite{Milton:1998cq}-\cite{Shirkov:2001sm}. 
The main requirement here is the subtraction of unphysical singularities 
in the RG improved series for physical observables as a whole, by 
computing their discontinuity across the time-like cut, as it has 
been done for the space-like coupling itself eqs.(\ref{dens}) and 
(\ref{alphaSH}). Specifically \cite{Shirkov:2000qv}, given a 
space-like observable perturbatively known 
\be 
D_{\mathrm{PT}}(Q^2)=1+\sum_{k\ge1}d_k \alpha_{\rm s}^k(Q^2)\,,
\lb{D_PT} 
\ee 
one can define the $k$-th spectral density 
\be
\rho_k(\sigma)=\mathrm{Im}\,\left[\alpha_{\rm s}^k(-\sigma)\right]\,. 
\lb{ro_k}
\ee 
Then eq.(\ref{D_PT}) is to be substituted by the ghost-free
expansion
\be
D_{\mathrm{APT}}(Q^2)=1+\sum_{k\ge1}d_k\mathcal{A}_k(Q^2)
\lb{D_APT} 
\ee 
where, leaving the loop level understood, 
\be 
\mathcal A_{k}(Q^2)=\frac{1}{\pi}\int_0^{\infty}\frac{d\sigma}
{\sigma+Q^2}\,\rho_k(\sigma)\,. 
\lb{A_k} 
\ee 
Then a standard power expansion is converted into a 
non power one. Clearly, there is now no unique expansion parameter, 
but an entire set of ghost-free expansion functions (\ref{A_k}) 
at any loop level, each defined by the analytization of subsequent 
powers of the perturbative coupling. Obviously from (\ref{A_k}) 
with $k=1$ the analytic coupling (\ref{alphaSH}) is recovered at 
each loop level. This recipe, working for space-like observables, 
is manifestly quite analogous to the RKP non-power expansion 
(\ref{R_rkp}) of a time-like observable; this in turn, within this 
framework, can be rexpressed via the $k$-th spectral density 
(\ref{ro_k}) \cite{Shirkov:2000qv} 
\be
R_{\mathrm{APT}}(s)=1+\sum_{k\ge1}d_k\mathfrak{A}_{k}(s) 
\lb{R_APT}
\ee 
and  
\be 
\mathfrak{A}_{k}(s)=\frac{1}{\pi}\int_s^{\infty}\frac{d\sigma}
{\sigma}\,\rho_k(\sigma) 
\lb{U_k} 
\ee 
which yields for $k=1$ the analytic time-like coupling itself 
(\ref{a_s}). The key point here \cite{Shirkov:2000qv} is 
that, due to the forced analyticity of the coupling and its 
analytized powers (\ref{A_k}), the two sets (\ref{A_k}) and 
(\ref{U_k}) are put into one-to-one relation 
by the linear integral transformations (\ref{dispAdl}) and 
(\ref{invsAdl}), in this context usually renamed $\mathbf{R}$ and 
$\mathbf{D}$ respectively, that is 
\be 
\mathcal A_k(Q^2)=\mathbf{D}[\mathfrak{A}_k(s)]\,,\qquad\mathfrak{A}_k(s)=
\mathbf{R}[\mathcal A_k(Q^2)]\,.
\lb{A-U}
\ee 
This yields a closed theoretical scheme for representing 
observables of any real argument, both space-like and time-like 
(for a quite recent review see \cite{Shirkov:2001sm} to which 
we also refer for technical details).
\begin{figure}
\begin{center}
\includegraphics[width=28pc]{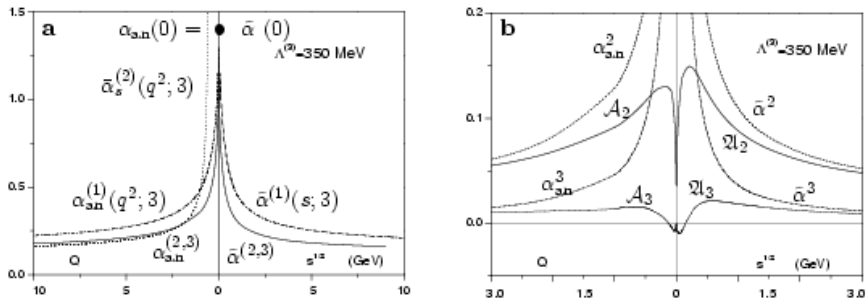} 
\end{center}
\caption{\footnotesize (a) Space-like and time-like global
analytic couplings in a few GeV domain with $n_f=3$ and $\Lambda=350\,$MeV; 
(b) ``Distorted mirror symmetry" for global expansion
functions, corresponding to exact two-loop solutions.} \label{fAA} 
\end{figure}
The main features of these functional sets are illustrated in
fig.\ref{fAA} taken from \cite{Shirkov:2001sm}. As yet noted the first 
function in both sets coincides
with the relative analytic coupling, respectively (\ref{alphaSH})
and (\ref{a_s}), while $\mathcal A_{k\ge2}$ and 
$\mathfrak{A}_{k\ge2}$ start with an IR zero and oscillate in the 
IR domain around $k-1$ zeros; furthermore they all obey the UV 
asymptotic $1/\ln^k z$ resembling the corresponding powers
$\alpha_{\rm s}^k(z)\,$ \cite{Shirkov:2001sm}. The differential 
recursion relations \cite{Shirkov:2005sg}
\be
\frac{1}{k}\frac{d\mathcal{A}_k^{(n)}(Q^2)}{d\ln Q^2}=
-\sum_{j=1}^{n}\beta_{j-1}\mathcal{A}_{k+j}^{(n)}(Q^2)\,,\qquad
\frac{1}{k}\frac{d\mathfrak{A}_k^{(n)}(s)}{d\ln s}=
-\sum_{j=1}^{n}\beta_{j-1}\mathfrak{A}_{k+j}^{(n)}(s)
\lb{rec_AU}
\ee
that hold in all orders \cite{Kurashev:2003pt}, allow to relate 
different analytized powers within each set, tough 
explicit expressions are reliable in a simple form only in 
the one-loop case (e.g.\cite{Shirkov:2005sg}). 
Meanwhile, $\mathcal A_k$ and $\mathfrak{A}_k$ for $k=1,2,3$ 
have been tabulated in \cite{Magradze:2000hz} up to three-loop, 
starting from the exact two-loop solution (\ref{RG2loop-ex}), 
and the Pade' approximant technique in the three-loop case; 
further improved approximations as eq.(\ref{alpha^k}) 
have been exploited at low scales in \cite{Kurashev:2003pt} 
(see also \cite{Magradze:2005ab}). These numerical values are 
reproduced quite well in the range $(2,100)\,$GeV by ``one-loop'' 
inspired approximate formulas, derived for practical aims in 
\cite{Shirkov:2005sg}, and depending on an effective scaling 
constant $\Lambda_{\mathrm{eff}}^{n_f}$ as a fitting parameter 
(see Tab.1 in ref.\cite{Shirkov:2005sg}). Computation of the 
functional sets has been recently extended to include analytic 
images of any real power of the coupling (Fractional Analytic 
Perturbation Theory) on the grounds of properties of the 
transcendental Lerch function\cite{Bakulev:2005fp}.\\
The APT algorithm can also include thresholds effects
\cite{Shirkov:2000qv} in a real description, by modifying 
the $k$-th spectral density (\ref{ro_k}) discontinuously at the 
heavy quark thresholds $m_f$
\be
\rho_k(\sigma)=\rho_k(\sigma,3)+\sum_{n_f\ge4}\theta(\sigma-m^2_f)
\left[\rho_k(\sigma,n_f)-\rho_k(\sigma,n_f-1)\right]\,, 
\lb{ro_kf}
\ee 
descending from the trivial matching condition (cf. sec. 4.2).  
The global functions resulting from densities (\ref{ro_kf}), 
$\mathcal A_{k}$ and $\mathfrak{A}_{k}\,$, can be obtained from the 
local ones with $n_f$ fixed, by adding specific shift constants 
$c_k(n_f)\,$, not negligible in the $n_f=3,4$ region (see 
\cite{Shirkov:2000qv} for numerical evaluations; for instance, in 
both t- and s-channel it has been estimated $c_1(3)\simeq0.02\,$).\\  
The main tests of APT being obviously at low
and intermediate scales, a number of applications to 
specific processes, both in the space-like and time-like (low and 
high energy) domain, have been quite recently performed. 
For instance, Bjorken and Gross-Llywellin-Smith sum rules 
\cite{Milton:1998cq}, $\tau$-lepton \cite{Milton:2000fi} (see also 
\cite{Geshkenbein:2001mn}) and Ypsilon decay \cite{Shirkov:2005sg}, 
$e^+e^-$-annihilation into hadrons 
\cite{Solovtsov:1997at, Solovtsov:1999in, Shirkov:2001sm} and 
hadronic form factors \cite{Stefanis:2000vd, Bakulev:2004cu}. 
As a result (see tab.1 in ref.\cite{Shirkov:2005kj}), the main 
advantages of the APT approach are 
better convergence properties of the ghost-free non-power 
expansion than the usual power one, since the three-loop term 
is always strongly suppressed (even less than data errors). 
This feature yields a reduced scheme and loop-dependence. 
Finally, transition from Euclidean 
(space-like momentum) to distance picture has been also 
developed in \cite{Shirkov:2002td} to which we refer for 
details, involving a suitable modified sine-Fourier 
transformation, consistently with the APT logic.\\

\vspace{-0.7truecm}
\section{Lattice Theory}
\begin{figure}[t]
\begin{center}
\begin{tabular}{c c c}
\includegraphics[width=14.5pc]{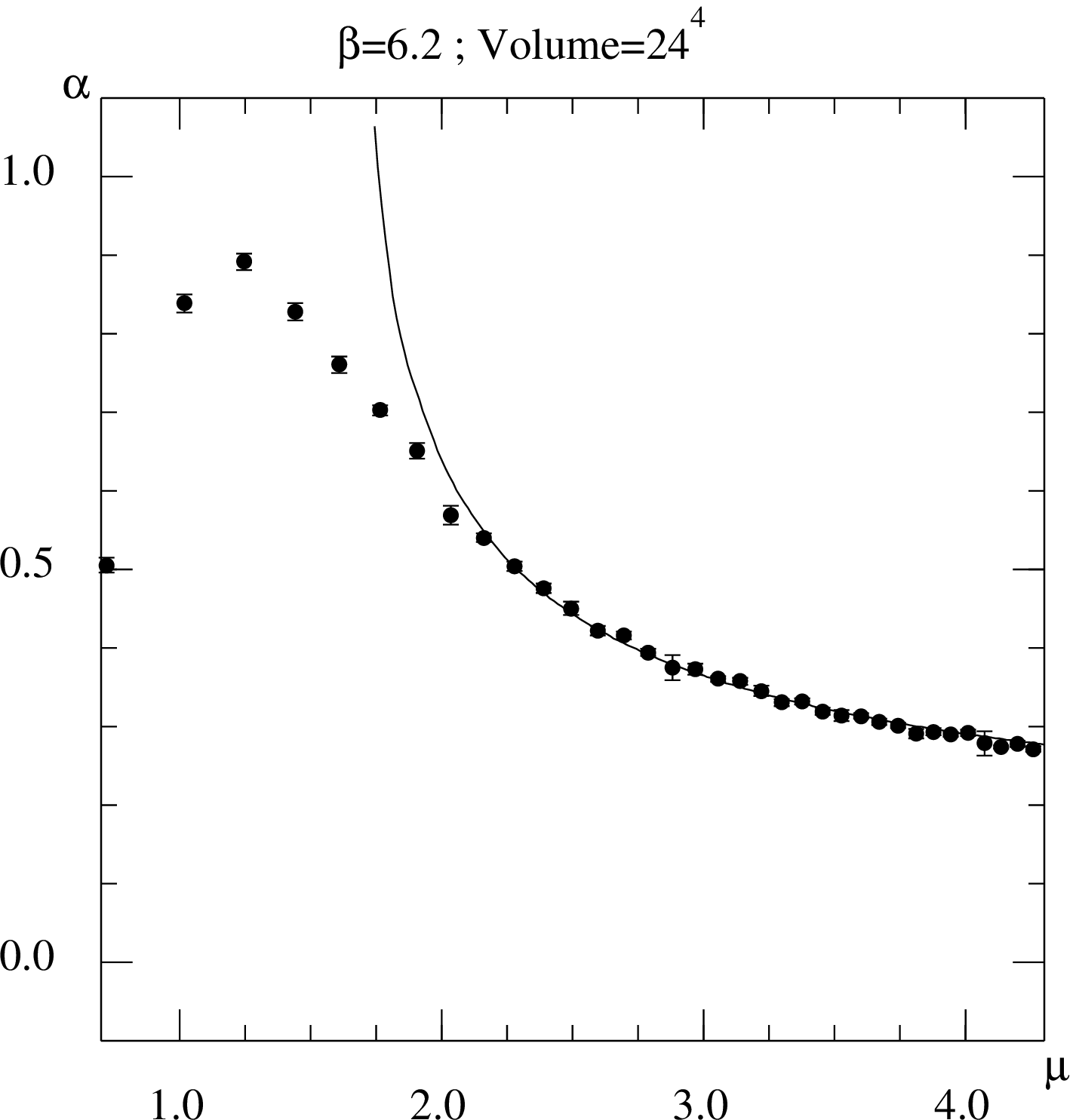}  & 
\hskip 0.5 cm\includegraphics[width=14.5pc]{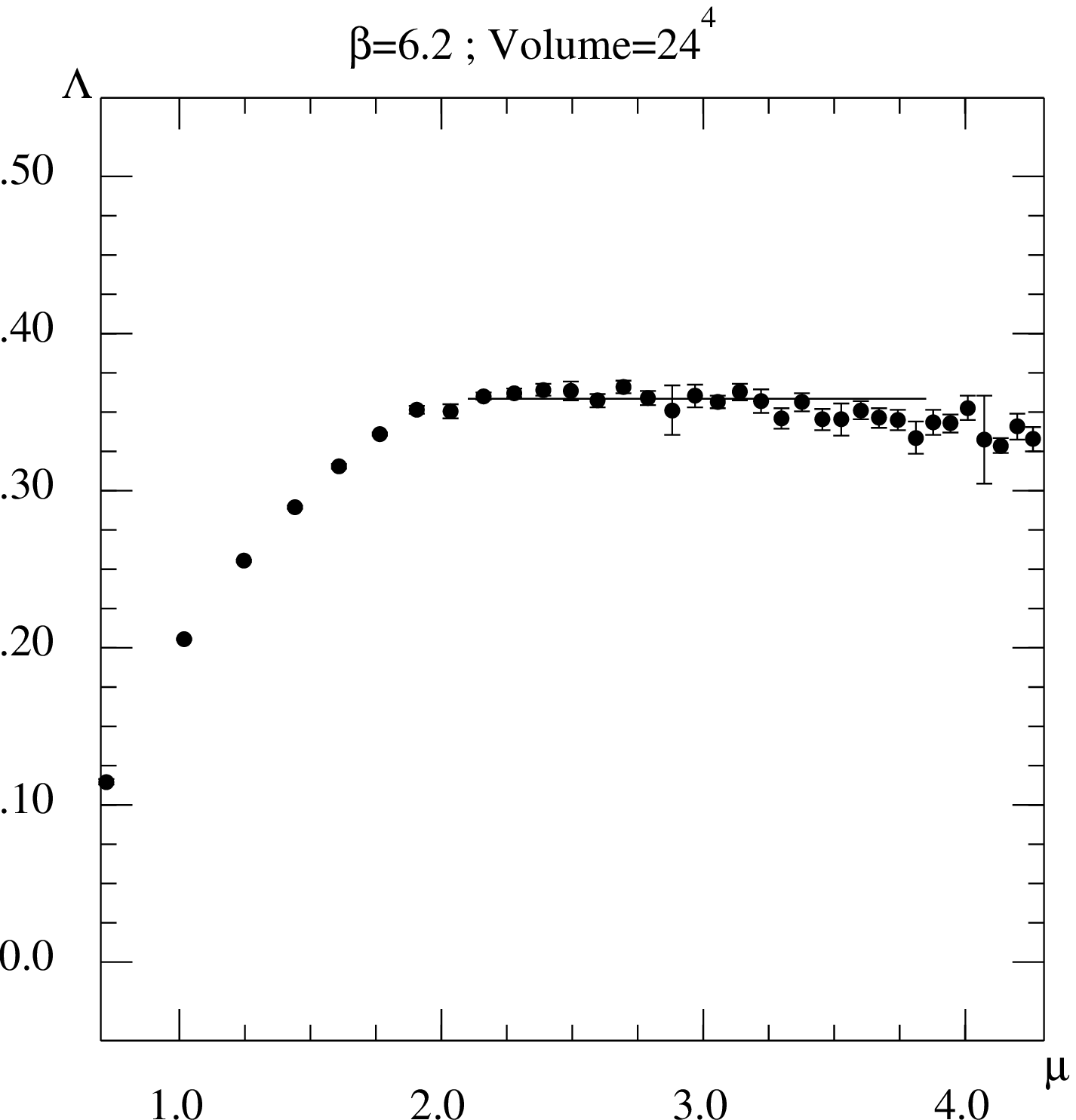} \\
  a &    b \\
\end{tabular}
\caption{\footnotesize (a)The QCD coupling constant $\alpha_{\rm M\widetilde{O}M}$ from
  \cite{pene98}, $\mu$ is in $\rm GeV$. The full line is the three-loop running. 
 (b)$\Lambda$ from \cite{pene98}.
\vspace{-0.5truecm}} 
\label{alpha_pene}
\end{center}
\end{figure}

\begin{figure}[t]
\begin{center}
\begin{tabular}{c c c}
\includegraphics[width=14.5pc]{alpha_s.eps}  & 
\hskip 0.5 cm\includegraphics[width=16.0pc,height=4.6cm]{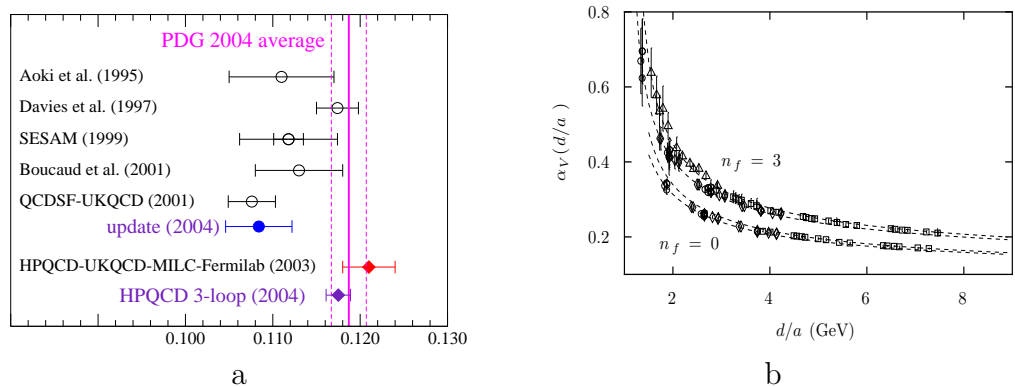} \\
  a &    b \\
\end{tabular}
\caption{\footnotesize (a) Strong coupling constant $\alpha_s(M_Z)$.
     Lattice data are from 
     Aoki et al.\protect\cite{aoki},
     Davies et al.\protect\cite{NRQCD},
     SESAM\protect\cite{NRQCD},
     Boucaud et al.\protect\cite{pene2001},
     QCDSF-UKQCD\protect\cite{QCDSF-UKQCD} and its
     update\protect\cite{Gockeler:2004},
     HPQCD-UKQCD-MILC-Fermilab\protect\cite{HPQCD2003/4} and its
     update\protect\cite{HPQCD2003/4}.
 (b)Values for $\alpha_V$ versus $d/a$ from each short-distance
  quantity from \cite{lastalpha}, with (top) and without (bottom) light-quark vacuum
  polarization. The dashed lines show the predictions of the
  perturbative running which best fit the distributions.
\vspace{-0.5truecm}} 
\label{alpha-plot}
\end{center}
\end{figure}

\begin{figure}[t] 
\begin{center} 
\includegraphics[width=32pc]{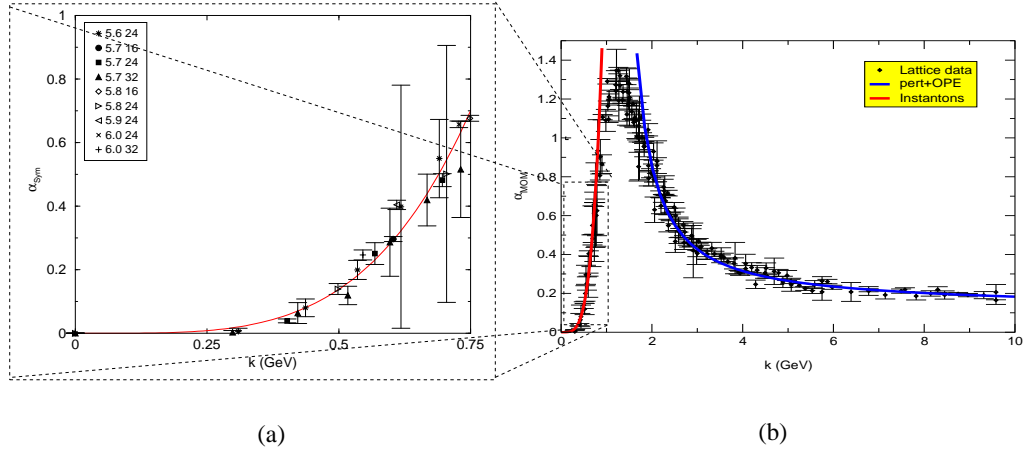} 
\caption{\footnotesize  
(b) Symmetric MOM coupling constant for different lattice settings 
and fits to perturbative expression plus power corrections in the high 
momenta region and to the expression of eq.(\ref{alphainst}) 
for small momenta. (a) Region of small momenta is zoomed.
\vspace{-0.7truecm}} 
\label{alpha_sym} 
\end{center} 
\end{figure}

\vspace{-0.5truecm}
In all development above one starts from manipulation on the
perturbative expansion and 
try to remedy to its lack of 
convergence, but one  can never attain to really non
perturbative effects related to the singularity of the theory in 
$\alpha_{\rm s}=0$.\\
Up to now, the only general technique to handle (even if with many
limitations)  non perturbative problems is Lattice Theory.\\
We want here to mention essential
ideas and summarize the present status of art for what concerns the
running coupling.\\
We refer to the   standard formulation   as due to Wilson \cite{Wilson74}, a
classical textbook is e.g. \cite{creutz}.\\
On the Euclidean lattice  the  spacing $a$  plays the role of the
UV cutoff of the continuum formulations. In momenta space the fields
are then defined on the Brillouen zone but, for practical purposes, the
lattice is finite, thus an IR cutoff appears too and the momenta
become finite. 
The quantum theory is obtained via path integral quantization and
the result is similar to a statistical mechanics formulation in $D=4$, at an
inverse temperature $\beta$ which turns out to be $2N/g_0^2$, for the
$\rm{SU(N)}$
gauge theory and where $g_0$ is the bare coupling constant.
 The lattice is a gauge invariant regulator but for finite lattice the functional integral is perfectly convergent, and
no gauge-fixing procedure is in principle necessary to evaluate gauge
invariant quantities. Very used are
improved actions, namely actions with extra terms of physical
dimensions higher than four, with coefficients suitable chosen in
order to have a better approach to the continuum limit (see
\cite{nieder} for a review). 
A complication of the lattice formulation is the well
known doubling problem of the fermions, in order to have in the
continuum limit  the correct number of fermions one has to use an
action which breaks the chiral symmetry for finite $a$.
The chiral symmetry can be restored by considering a critical
value for the hopping parameter for any lattice spacing. 
Recipes for the  fermionic action which differ from the Wilsonian 
one are widely used, each one with its own problems.
To get answers from the lattice usually one does not compute the
functional integral but perform Monte-Carlo simulations.
The computation of the running coupling constant on the lattice starts
from its non perturbative definition in the low energy region; there
are a lot of definitions which have been actually used, and in this
region they can have a very different behavior with the energy scale, but
extrapolated in some way  at high energy they must behave similarly, and it
has to be possible to correlate one to another using perturbation
theory. That this happens is a consistency check of the theory on all
the energy scales. In particular at high energy it is possible to
commute to the  $\overline{\mathrm{MS}}$ scheme in dimensional
regularization, defined only in perturbation theory.
An exhaustive review on the various possible approaches is in
\cite{Weisz-rev}.\\ 
From a physical point of view the  framework of the computation is fixed    
by setting the scale. In the pure gauge theory one needs, from the
experiments, the knowledge of a physical quantity of mass dimension $M_H$,
and on the lattice, starting from a reasonable $g_0$, or equivalently
$\beta$, one has to be able to compute the number $aM_H$, and this determines
the spacing $a$ in physical units. The window of energies on the
lattice, in which $M_H$ has to fall is then $[1/L,1/a]$. $L/a$ is the
number of sites for side of the lattice and so is fixed by the
computer power, typical values are $16-64$.\\ 
The asymptotic freedom can
be formulated also in terms of bare coupling constant which vanishes for
$a\rightarrow 0$, for instance we have the two loop formula
\begin{equation}
\frac{1}{\alpha_0}=\beta_0 \rm ln(a^{-2} \Lambda _L^{-2}) + (\beta_1/\beta_0)
\rm ln(\rm ln(a^{-2} \Lambda _L^{-2})
\label{Go_run}
\end{equation}
(the ratios $\Lambda _L/\Lambda$, where $\Lambda $ refers to a suitable
continuum scheme, $e.g.$ $ {\overline{\mathrm{MS}}}$, have been
evaluated in  lattice perturbation theory long time ago
\cite{HH80}, \cite{DG81}).
Being interested in the continuum limit one would like to have $a$ and
$g_0$
as small as possible but one needs to take into account the limitations
of the window energy. The dependence on  the chosen observable should
be not important as far as we are ``close'' to the continuum limit
(scaling region), used quantities are the string tension, the mass
splitting in heavy quarkonium or the hadronic radius $r_0$ \cite{sommer94}. 
In the nineties there have been several determinations of the running
coupling constant in the pure gauge theory or in quenched QCD 
(\cite {bali-scill-93}-\cite{pene98}).\\
Among the papers, \cite{alles-97} and \cite{pene98} are closer in the
formalism to the other parts of this review, indeed the definition of the
renormalized coupling constant  uses a condition on the trigluon
vertex,  therefore  very similar to that employed in  usual
perturbative procedures. In principle such a definition of the
coupling can however be used in the full infrared region, the only limitations
being due to the lattice size effects.
The disadvantages of such an approach are that
$A_{\mu}$ on the lattice  is a
rather unnatural quantity, 
defined in conventional way through the fundamental link gauge variables;
moreover, to compute the Green functions, the gauge fixing is needed.
For the implementation of the gauge fixing on the lattice a good review
is  \cite{giusti2001}.
It is worth to mention that even the lattice formulation
of the Landau condition suffers for the problem of the Gribov
ambiguity. Actually one could pick out a unique element in each gauge orbit by
finding
the absolute minimum of a suitable functional with respect to gauge 
trasformations but, from a numerical
point of view, to distinguish the absolute minimum for instance from
the local ones is a very difficult task. Beyond \cite{giusti2001} a
clear analysis of the numerical situation is in
\cite{cucchieri-gribov}, relations with other approaches are in the
review \cite{alkofer-rev}, devoted to the infrared behavior of the QCD
Green functions.
On the lattice, by Monte Carlo
average, after performing the Fourier transform and in the Landau gauge,
 one computes the unrenormalized Green functions
\begin{equation}
G^{(n)}_{U\; \mu_1...\; \mu_n}(p_1,..\; ,p_n)= \langle A_{\mu_1}(p_1)..\;  ,A_{\mu_n}(p_n)\rangle\,.
\label{greenfunct} 
\end{equation} 
In \cite{alles-97} and \cite{pene98} the scale is set by a
previous determination of the string tension \cite{bali-scill-93},
then $\beta=6$ and correspondingly $a^{-1}=1.9\pm0.1\;\rm GeV$ in a
hypercubic lattice are taken in \cite{alles-97}, more lattices are used
in \cite{pene98}.\\
Assuming that finite-volume effects and discretization errors are 
under control, for suitable small momenta one can adopt the formalism of the
continuum QCD and then to write for the gluonic propagator
\begin{equation}
G_{U\;\mu\nu}(p)=G_{U}(p^2)(\delta_{\mu\nu}-\frac{p_{\mu} p_{\nu}}{p^2})
\label{propag}
\end{equation} 
and for the three point function (after factorization of the color
tensor $f^{a\;b\;c}$)
\begin{equation}
G_{U\;\mu\nu\rho}^{(3)}(p_1,p_2,p_3)=
\Gamma_{U\;\alpha\beta\gamma}^{(3)}(p_1,p_2,p_3)G_{U\;\alpha\mu}(p_1)G_{U\;\beta\nu}(p_2)G_{U\;\gamma\rho}(p_3)\,.
\label{G3}
\end{equation}
Where $p_1+p_2+p_3=0$.
On the lattice of course one measures the Green functions, so
differently from the usual perturbative approach the renormalization
conditions are expressed on the Green functions and not on the 1-P-I
functions.
Thus
\begin{equation}
G_R(p)|_{p^2=\mu^2}=Z^{-1}_g(a\mu)G_U(pa)|_{p^2=\mu^2}=\frac{1}{\mu^2}
\label{G2rencond}
\end{equation}
is the condition which fixes the wave-function renormalization
constant in non perturbative way.
For the vertex the most obvious approach would be to impose a
condition on the most symmetric point ($\rm MOM$ scheme), but from a
numerical point of view it turns out that  an asymmetrical
condition in which one of the momenta vanish works better ($\rm
M\widetilde{O}M$
 in continuum QCD \cite{HH80} ).
Defining the vertex renormalization constant according to
\begin{equation}
\frac{\sum_{\mu=1}^{4} G_{U\;\mu\nu\mu}^{(3)}(pa,0,pa)}
{G_U(pa)^2 G_U(0)}=6iZ_V^{-1}(pa)g_0 p_{\nu}\,,
\label{vertexfunct}
\end{equation}
the vertex renormalization condition reads
\begin{equation}
g(\mu)=Z_{g}^{3/2}(a\mu)Z_{V}^{-1}(a\mu)g_0\,.
\label{vertexRC}
\end{equation}
In a massless theory the Green functions at exceptional momenta are
potentially divergent, but in this case all is finite \cite{chet00},\cite{ball-chiu}. 
Note that the tensorial structure of the previous formulae is not
obvious because on the lattice $\rm{O(4)}$ symmetry no longer holds; a
test of  consistency of their numerical computation performed by the
authors is to check the previous structures for some values of $pa$ .
As a matter of fact the parametrization of the formulae (\ref{propag})
and (\ref{G3}) cannot be the most suitable in order to extract the
coupling,
lattice perturbation theory would suggest using  
$\hat p_{\mu}= \frac{2}{a}\sin(\frac{p_{\mu}a}{2})  $ instead of
$p_{\mu}$, indeed the authors from \cite{pene98} claim that only in
this case
they are able to
find evidence of the running of the coupling.
In order to prove the running of the coupling one  fits the data
of the simulation with the curve which express $\Lambda$ through
$g(\mu)$, see Fig.\ref{alpha_pene};  using the two loop formula (but in \cite{pene98} they  control the three loop effect)
\begin{equation}
\Lambda_{\rm {M\widetilde{O}M}}=\mu \exp(-\frac{1}{2\beta_0 \alpha(\mu)})(\beta_0
  \alpha(\mu))^{-\frac{\beta_1}{2\beta_0^2}}
\label{Lambda2loop}
\end{equation}
in both the papers one finds an evidence of plateau close to
$\rm 2\, GeV$, see Fig.\ref{alpha_pene}(b), beyond $\Lambda(g)$ decreases due to cutoff effect.
The effects of finite size are controlled by performing the computation
using more lattices with the same lattice spacing.\\
Having computed $\Lambda_{{\rm  M\widetilde{O}M}}$  one can get $\Lambda_
{\overline{\mathrm{MS}}}$. To this end one has to do a one-loop
\cite{Gonsalves79} calculation which expresses perturbatively the relation between
$g_{\overline{\mathrm{MS}}}(\mu)$ and $g_{\rm {M\widetilde{O}M}}(\mu)$.
Indeed a discrete renormalization group transformation which connects the
Green functions computed on the lattice and those computed in
${\overline{\mathrm{MS}}}$ scheme must exist. Being this scheme on the
lattice specified by conditions on the Green functions, the transformation is easily determined
by the one loop ${\overline{\mathrm{MS}}}$ continuum calculation of the trigluon vertex and of the
propagator  at the ${\rm {M\widetilde{O}M}}$ point and in
Landau gauge. 
More difficult computations in lattice perturbative theory are not
needed
in this approach.\\  
In \cite{alles-97} 
it is obtained $\Lambda_{\overline{\mathrm{MS}}}=0.34\pm0.05\;\rm GeV$;
the authors of \cite{pene98}, with a writing which distinguishes the possible errors
coming from the lattice spacing, instead get 
$\Lambda_{\overline{\mathrm{MS}}}=(0.303\pm0.05\,
GeV)\frac{a^{-1}}{1.97\;\rm GeV}$.\\
The same renormalization scheme has been applied to measure the
coupling using the gluon-quark vertex in quenched simulations in  \cite{skull}.\\ 
In order to avoid the major source of
systematic errors,  in spite of the computer cost, simulations with
dynamical quarks have been addressed.
The task has been undertaken using NRQCD \cite {NRQCD} in the past; also
the authors of \cite{pene98} published a  computation with
$\rm{n_f=2}$ dynamical quarks \cite{pene2001} in the MOM scheme
previously described. 
Interestingly the matching with the perturbative behavior of the
coupling, and then the extraction of $\Lambda$, was made taking into
account an effect of $\rm{OPE}$ condensate arising from the gluonic propagator;
therefore the data computed on the lattice for the coupling where
compared not with $\alpha_{s,pert}(\frac{\mu}{\Lambda})$ but with
$\alpha_{s,pert}(\frac{\mu}{\Lambda})(1+\frac{c}{\mu^2})$, where $c$ has
to be adjusted in a combined fit. The form of this correction, similar
to those introduced in eq.(\ref{a_syn}), is based
on previous lattice computations and already utilized in order to improve
the determination of $\alpha_s$ in quenched simulation \cite{peneOPE}.\\
They also perform the extrapolation of the coupling to $\rm{n_f}=3$
and then the renormalization group evolution to the value of $M_Z$ of
the energy through the thresholds, Fig.\ref{alpha-plot}(a) quotes their result.\\
A recent review \cite{Hashimoto} analyzes the results from lattice
calculations up to 2004 with emphasis to the theoretical issues in
dynamical quark simulations. In particular a comparison among 
determinations of the various coupling constant is presented, besides
the already cited papers, Fig.\ref{alpha-plot}(a) includes some
newer 
\cite{HPQCD2003/4},
\cite{QCDSF-UKQCD} and
some older result \cite{aoki}. 
The most recent result in Fig.\ref{alpha-plot}(a), obtained by the
$\rm{HPQCD}$ collaboration (2004), is 
$\alpha_s(M_Z)=0.1175(15)$ in good agreement with the $\rm{PDG}$ 2004
average $0.1182(20)$ \cite{pdg}. It has be noted that the results 
obtained using Wilson fermions
\cite{QCDSF-UKQCD}, \cite{Gockeler:2004} are
significantly lower than those obtained using staggered fermions \cite{HPQCD2003/4}.\\
An original approach to the lattice computations has been developed by
the $\rm{ALPHA}$ collaboration which  recently performed a determination
of the strong coupling constant with two dynamical flavors
\cite{alpha2005}. Exposition of  the method is in \cite{luscher97-98}, 
a detailed discussion of the mass renormalization
in \cite{alphamass}.\\
One  of the computational problems  in the usual
approaches is that the window energy  must be wide to contain both the matching
to the hadronic scale and the asymptotic behavior of the coupling.
In the approach of the  $\rm{ALPHA}$ collaboration one can determine a recurrence
relation between quantities referring to double energy each step, and
thus one constructs a ladder from the hadronic to the perturbative regime.\\
The technical tool is an intermediate renormalization scheme in which
the renormalized coupling constant is defined by derivative 
 with respect to parameters characterizing the boundary
conditions of
the Schroedinger functional, namely the amplitude for a transition from a
given field configuration at a given time,  $t=0$, to another one at $t=T$. Moreover the
boundary conditions yield a natural infrared cut-off so it is possible
to perform simulations at vanishing quark mass \cite{alphamass} 
(using Wilson fermions the mass
renormalization is additive, and imposed using the $\rm{PCAC}$
relation). 
The renormalization of the coupling is  defined on the
massless theory thus  a non perturbative mass independent
scheme is obtained. 
In other computations the light quark masses are actually rather large,
and chiral extrapolations are needed in order to compute the coupling to be
compared to the  $\overline{\mathrm{MS}}$ one.\\
Taking $T=L$, being $a$ and $L$ the
only dimensional quantities, the following relation between bare and
renormalized couplings holds 
$\overline{\mathrm{g}}_L=\overline{\mathrm{g}}(a/L,g_0)$ .  
Here $1/L$
plays the role of the finite scale of energy. 
The same procedure can be
applied by doubling the size, $L\rightarrow 2L$, and after
elimination of $g_0$ one finds a relation between
the two renormalized coupling constants. At the end, by
extrapolating  the limit $a/L \rightarrow0$, they define the ''step
scaling function'' which connects the two couplings:
\begin{equation}
\sigma(u_{L/2})=u_L\,.
\label{sigma}
\end{equation}
In (\ref{sigma}) $u={\overline{\mathrm{g}}}^2$ and the two couplings
refer to lattices of size $L/2$ and $L$  at fixed Physics, namely
bare parameters. The previous equation is actually a discrete
renormalization group transformation, indeed a relation between
the $\beta$  and the $\sigma$ functions is determined. 
Of course the step scaling function is constructed by a 
numerical procedure; 
consistency with  perturbation theory requires $\sigma(u)\sim u$
as $u\rightarrow0$, for larger $u$,  $\sigma(u)$  is computed
by a suitable interpolation starting from the set of points determined
on the lattice (a discussion of the systematic error arising from this
interpolation, which it is claimed to be negligible in a suitable
range of coupling, is in \cite{capitani}).\\
Starting with a given
coupling and a given size ($u_{max},L_{max})$ using the step
scaling function one computes 
$u_i={\overline{\mathrm{g}}}^2(2^{-i}L_{max})$ 
and for sufficient high energy, with the identification $\mu=
2^{i}/L_{max}$, one matches the pertubative regime
evaluating $\Lambda$ (the ratio between this $\Lambda$ and
$\Lambda_{\overline{\mathrm{MS}}}$ is known). \\
The starting point of the iteration is a priori unknown, it is
determined
by matching some hadronic quantity, as a practical approach it was 
computed $r_0$ in the chiral limit; 
assuming $r_0=0.5\,\rm fm$ it turns out
$\Lambda^{\rm{n_f}=2}_{\overline{\mathrm{MS}}}=245(16)(16)\;\rm MeV$.
For the low energy behavior of this coupling see \cite{alphainfrared}.\\
The most recent determination of the running coupling constant has been obtained
by a $\rm {HPQCD-UKQCD}$ collaboration \cite{lastalpha}. Using a new
discretization of the light quark action the authors perform the first
computation with  three dynamical quarks. After having fixed the
lattice parameters ($m_u=m_d$, $ m_s$, $m_c$, $m_b$ and $ a$) matching
five experimental quantities the determination of the coupling is
obtained by an optimization of lattice perturbation theory, in which
the coupling must fit 28 short-distance quantities calculated on the
lattice. The final estimate is $\alpha_s(M_Z)=0.1170(12)$, which
slightly corrects the already reported result of \cite{HPQCD2003/4}.\\
The figure \ref{alpha-plot}(b), from the cited paper, shows the remarkable 
effect of the vacuum polarization of
the three light quarks on the coupling $\alpha_V$ introduced in
eq.(\ref{richardson1})(the momentum is represented in units of the
inverse lattice spacing).\\ 
Among the many lattice results, we would like now to pay a little
attention to those concerning  the behavior of the coupling in the
very low energy region.\\ 
The $\rm MOM$ definition of the coupling is likely the best suited to
this aim, indeed the already cited authors using this approach
addressed this task in \cite{pene-instantons}. In order to reach low
momenta also  lattices with a rather large lattice spacing, and then 
low $\beta$, have  been used. The results for the pure gauge theory are
shown in Fig.\ref{alpha_sym}.   
The number of data in fig.\ref{alpha_sym}(b) is sufficient to check the theoretical model of the vacuum as
instanton liquid. According to this model the coupling must have a
$p^4$ dependence for small energy, the data in Fig.\ref{alpha_sym}(a)
are indeed fitted by this power low, the proportionality coefficient
yields the instanton density $n$:
\begin{equation}
\alpha_{\rm MOM}(p)=\frac{1}{18\pi}n^{-1}p^4\,.
\label{alphainst}
\end{equation}
Such a behavior  is  not a universal feature, from a glance at
the figures 8 and 9 in the second paper of \cite{skull}, in which the
coupling is defined in $\rm MOM$ scheme but using the quark-gluon vertex,
 one realizes that there the coupling is still vanishing for small
 momenta
but, in spite of the rather little number of data, it does not seem
with the $p^4$ low. 
On the contrary the coupling $\alpha_{\rm V}(Q)$
introduced in eq.(\ref{richardson1}) should instead diverge for
$Q\rightarrow0$.\\
In the last years infrared features of QCD has
been addressed using the Dyson-Schwinger equations. This approach,
which starts from suitable truncations of the DSE, seems in particular
effective in the study of the low energy behavior of the propagators
(see \cite{alkofer-rev} for a review). In this way infrared exponents for the
gluon-gluon and ghost-ghost two point functions have been determined.
In the Landau gauge using the ghost and the gluon propagators one can
give a definition of the invariant coupling (which then resembles the familiar
definition of QED) and  it turns out 
that 
it  has a finite infrared limit \cite{DSE}. 
It is found 
evidence for this fixed point at least in the context of
pure $\rm{SU(2)}$ lattice gauge theory \cite{Bloch},
 however  this conclusion does not seem 
 supported in other  recent lattice $\rm{SU(3)}$
 simulations \cite{furui}-\cite{sternbeck}.  The two results are compared and
 discussed 
 in \cite{fischer}, very recent results also in \cite{fischer2}. 
As yet noted, such a lack of universality of the running coupling constant in
the low energy regime is somewhat expected, however on such different
behaviors,
which are a little disturbing, some comment is in order.
In \cite{Shirkov-IR2002} a  comparison of various approaches is  made 
with reference to
singular RG trasformations.
The RG
transformations are associated to a change of variable $g\rightarrow
g'=f(g)$, which in general depends on the dimensional
parameters too; in the  perturbative framework  $f(g)$
is usually supposed to be an analytic function in the origin $g=0$. 
Now  considering the $\rm MOM$ schemes, it is well known that in
perturbation theory  
the definitions of the coupling are vertex
dependent \cite{Gonsalves79} but, at least in the massless case, this
dependence turns out to be weak, namely the function $f$ is close to
the identity. It is conceivable that in the infrared region,
all more reason taking into account the different dependence of the
coupling  on the quark light masses, the
function $f$ can have non trivial consequences.\\ 
Some proposals that we
have examined which tries to extend the perturbative series in the
low energy region can be seen as a not trivial RG transformation, at
the beginning defined in full perturbative region, but such that the perturbative
series in the new variable does not suffer from some pathology of the
first one. Thus formula (\ref{alphaSH1}), which avoids the Landau pole,
can be viewed, as already noticed, as the result of the non-analytic
 transformation $\alpha_s\rightarrow \alpha'_s =\alpha_s +
\frac{1}{\beta_0}(1-\exp(\frac{1}{\beta_0 \alpha_s}))^{-1}$.\\
Every devisable transformation should share with the previous one the
two nice properties to enjoy the asymptotic freedom, $\alpha_s\approx0^+
\rightarrow\alpha'_s\approx0^+$, and to be free of ghost singularity, 
$\alpha_s=\infty\rightarrow\alpha'_s= finite$. A priori there are not
other compelling requirements, and one can imagine a situation in which,
with respect to a given problem, a definition, and then a consequent
perturbative expansion, has better properties than another.
The considerations in \cite{Shirkov-IR2002} show that with other
transformations one can induce various infrared behaviors in the new
coupling. 
The
conclusion is that 
``there is not direct physical sense in attempts to
establish some ``correct IR behavior'' of the perturbative QCD invariant 
coupling''.
\vspace{-0.5truecm}
\section{Comparison with the data and conclusive remarks}
\begin{figure}[t]
\begin{center}
\begin{tabular}{c c}
\hskip -0.4cm\includegraphics[width=6.3cm,height=7.2cm]{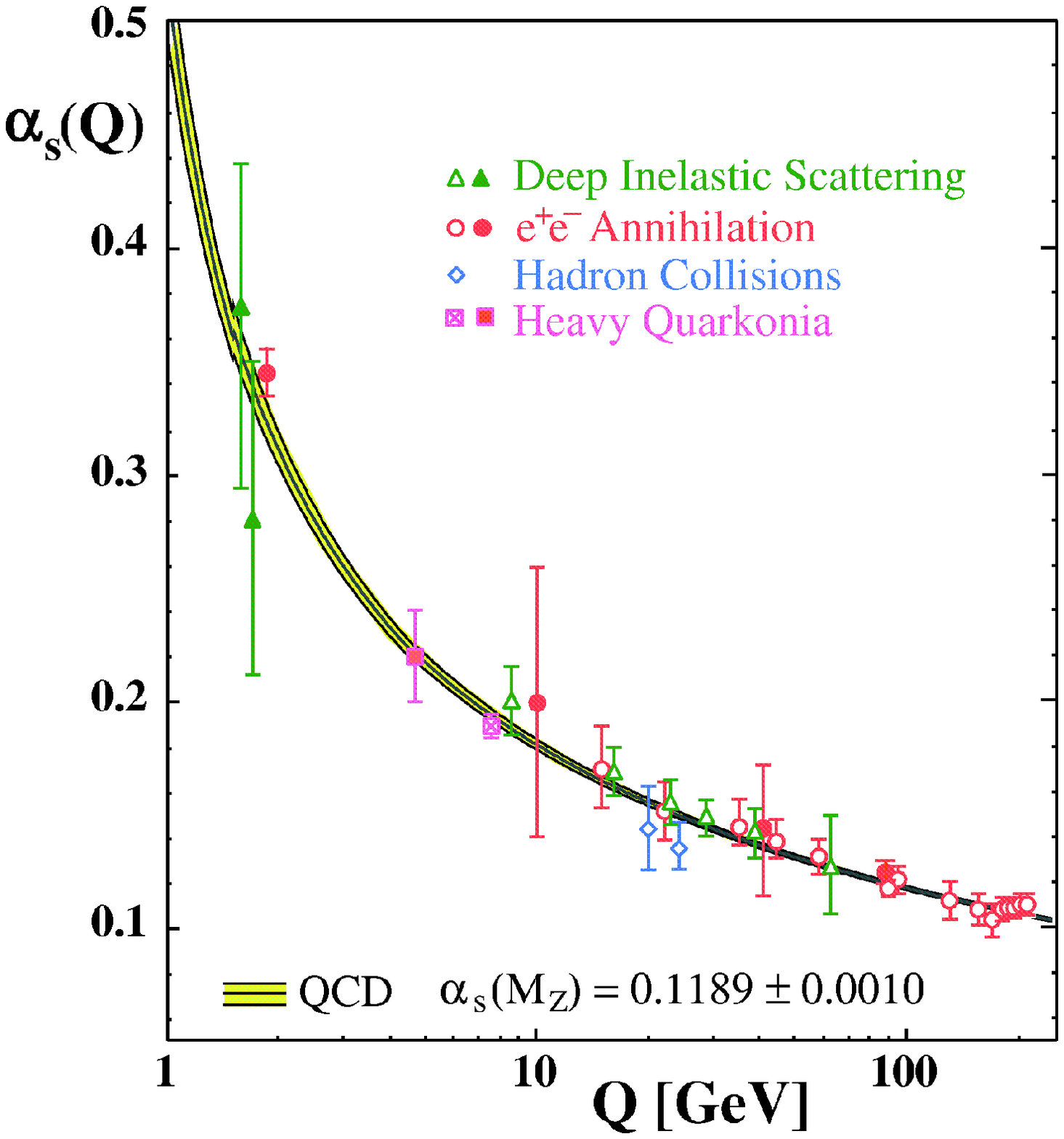}  & 
\hskip -0.2cm\includegraphics[width=7.2cm,height=7.2cm]{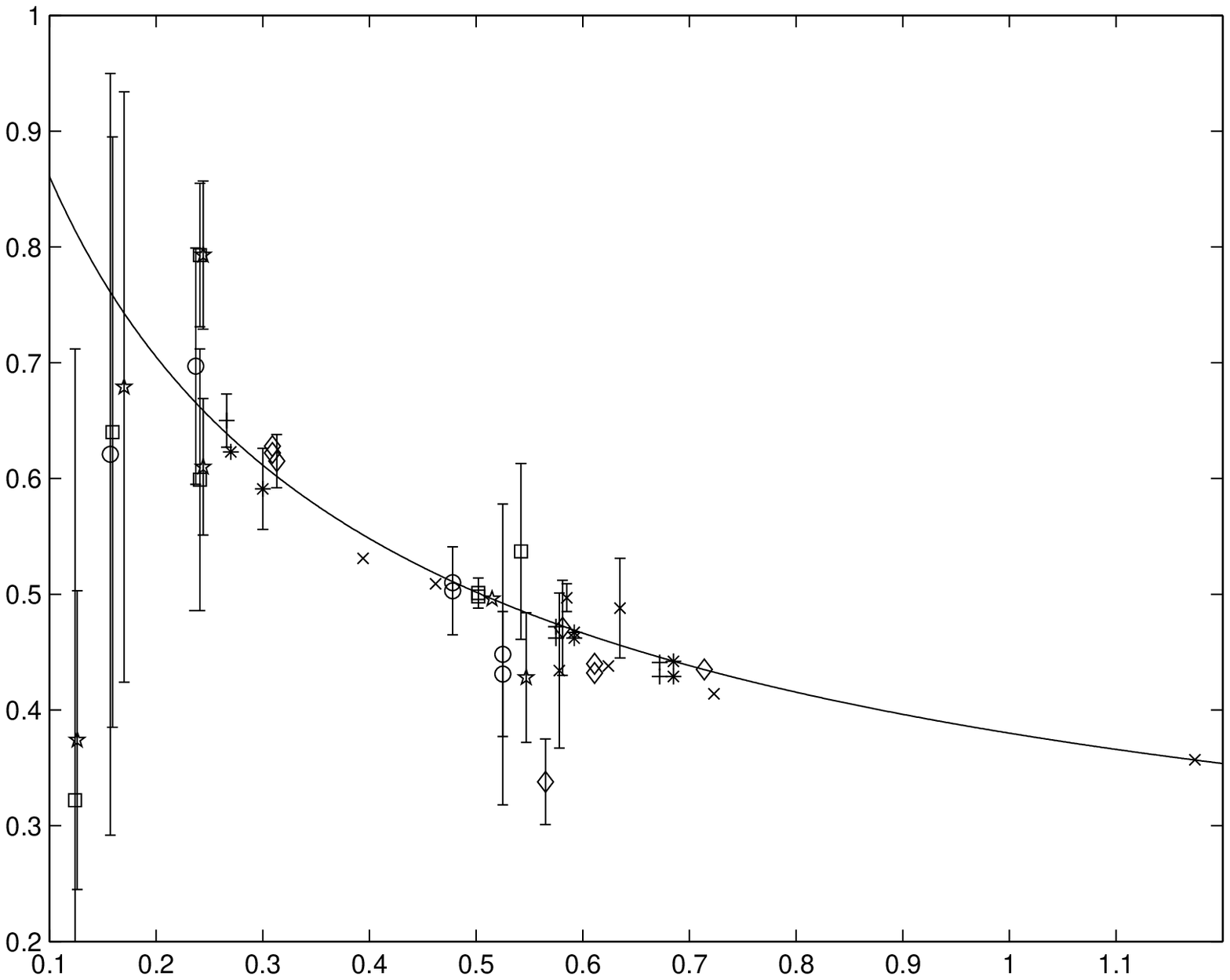} 
\put(-175,180){ {\footnotesize $\alpha_{\rm s}(Q^2)$} }
\put(-55,20){ {\footnotesize $Q\,\,$[GeV]} } \\
 \footnotesize (a) &  \footnotesize (b) \\
\end{tabular}
\caption{\footnotesize (a) From \cite{bethke}: summary of measurements 
of $\alpha_s(Q^2)\,$.
(b) From \cite{ip}: comparison between the 1-loop analytic coupling 
(\ref{alphaSH1}) with $\Lambda^{(n_f=3)}=206\,$MeV, and the values of $\alpha_{\rm s}$ 
fitting experimental data on quarkonium spectrum within Salpeter formalism (circles,
pentagrams and squares refer to light-light states, diamonds and crosses to heavy-heavy,
plus signes and asterisks to light-heavy states).
\vspace{-0.7truecm}} 
\label{asq}
\end{center}
\end{figure}

\vspace{-0.3truecm}
For completeness in this concluding section we want to  
comment briefly on the experimental situation as continuously kept updated by
\cite{pdg} and other specialized reviews (see e.g. \cite{bethke}).
Tab.1 in the contribution of S.Bethke in this issue summarizes the 
determination of $\alpha_{\rm s}(Q^2)$ at various energies between 
1.58 and 206 GeV extracted by various types of 
experiments: deep inelastic scattering 
(DIS), $e^+e^-$ annihilation (hadron cross section, structure function,
jets and jet shapes), hadron-hadron collision, Z, $\Upsilon$ and
$\tau$ decay, heavy quark bound states.\\
In fig. \ref{asq}(a) from \cite{bethke} such results are compared with 
the 4-loop
perturbative expression of $\alpha_{\rm \overline{MS}}(Q^2)$ as given
by (\ref{4loopEC}) together with threshold matching. 
As it can be seen the agreement is very good inside the
errors and this constitutes a very significant test for QCD. Note,
however, that assuming all expressions equally normalized at $M_Z=91.2\,$GeV
the difference between the 3-loop and the 4-loop expressions is of
order 1/10000 in the interval from 10 to 200 GeV, it becomes of the 
order 1/1000 between 2-loop and 4-loop and of few per cent between 
1-loop and 4-loop.
We have in this way the proof that at three loops the theory is
practically at convergence in comparison with the precision of the 
experimental data and that two loops (naturally after the appropriate
rescaling of $\Lambda$) is already a very good approximation in the
considered range 
\footnote{Obviously the difference explodes near the singularity.}.
The choice of a particular RS or another is essentially 
immaterial. However, under a scale of few $\Lambda ^{(n_f=3)}$ (let us say
2 GeV), the $\overline{\rm MS}$ becomes useless, due to the Landau 
singularities, and we have to refer to any alternative scheme free of
unphysical singularities and to follow it very consistently. A
comparison with the data becomes RS dependent and would have no
meaning out of the well defined framework.\\
Note that the analytic scheme discussed in sec. 4 seems particularly
convenient at this aim. It shares the simplicity and the universality
of the  $\overline{\rm {MS}}$ scheme at a large extent. It provides a
coupling which is regular in the entire interval $0<Q^2<\infty$ and
has a finite limit ${1 \over \beta_0}$ for $Q\to 0$ independent of the
number of loops used. As a consequence of this fact its expression
converges much faster as the number of loops increases. The difference
between the 2-loop and the 3-loop expression is again of order 1/1000
but this time more or less in the entire range from 0 to
$+\infty$. Even the difference between the 1-loop and the 3-loop is of
order of 1/100 and may be a fraction of this in limited ranges. \\
For construction $\alpha_{\rm an}(Q^2)$ coincides asymptotically with 
$\alpha_{\rm \overline{MS}}(Q^2)$ for large $Q$ and $\alpha_{\rm an}
^{(2)}(Q^2)$ fits the data of Tab.1 from \cite{bethke} nearly as well 
as $\alpha_{\rm \overline{MS}}^{(3)}(Q^2)$. Note however that, as seen 
in sec. 4.2, we have more precisely 
\be
\alpha_{\rm an}^{(p)}(Q^2) \to \alpha_{\rm \overline{MS}}^{(p)}(Q^2)
  + {c_1^{(p)} \over Q^2}\,.
\lb{concl1}
\ee
The coefficients $c_1^{(p)}$ seem to decrease as the number $p$ of the
loops increases. However for the value of $p$ we actually use the
power term in (\ref{concl1}) can be non negligible at some
intermediate scale and may be disturbing in particular with reference
to the problematic discussed in sec 4.5. As we have seen they may be
eliminated for every given $p$ by choosing an appropriate non
perturbative term $\alpha_{\rm an}^{\rm NP}(Q^2)$ to be added to  
the ``perturbative'' $\alpha_{\rm an}^{\rm PT}(Q^2)$. As we have seen
phenomenologically modified expressions for $\alpha_{\rm an}(Q^2)$
have also been proposed to eliminate such term.\\
Information on the QCD interaction under few GeV can be obtained from
power corrections to various observable at some intermediate energy,
as we have seen. Other information can come from relativistic
calculation of the spectrum of light-light, light-heavy and from the
highly excited heavy-heavy quarkonium states.\\
In \cite{Baldicchi:2004wj}, e.g., a reasonable reproduction of the entire calculable
spectrum and in particular of $\pi-\rho$, the $K-K^*$ and
$\eta_c - J/\psi$ was obtained. Such analysis was performed in the
framework of a Salpeter formalism constructed using as input only a 
generalization of the ansatz (\ref{ansatz}) and a 1-loop analytic coupling
with $n_f=3$  and $\Lambda^{(n_f=3)}=180$ MeV. Note that such  
$\alpha_{\rm an}^{(1)}(Q^2)$ differs from $\alpha_{\rm an}^{(2)}(Q^2)$ 
with $\Lambda^{(n_f=3)}=375\,$MeV by less than 0.5 \%.\\
Actually the developments reported in sec 4.5 were made in the
framework of the SGD scheme and this scheme may be the most natural
even for bound states problems. In fact the inclusion of certain second
order corrections would simply amount to a rescaling of
$\Lambda$. However the SGD and the analytic scheme are strictly
related and have even a common root in a sense \cite{Dokshitzer:1993pf}. They are
both based on dispersive techniques and can be made to coincide up to
two loops.\\
In any way according to a more recent preliminary analysis, made in the
context of the mentioned relativistic determination of the quarkonium
spectrum \cite{ip}, it seems that the curve $\alpha_{\rm an}^{(1)}(Q^2)$ 
(with $\Lambda^{(n_f=3)}=206$MeV) is in very good agreement with the data 
up to $200\,$MeV corresponding e.g. to the $c\bar{c}$(1D) state (see 
fig.{\ref{asq}}(b)). On the contrary for higher excited states, 
corresponding to 
$100<Q<200$MeV, the values of $\alpha_{\rm s}$ that reproduce the  
data seem to be somewhat lower than $\alpha_{\rm an}^{(1)}(Q^2)$ and
this could confirm the need of a non perturbative contribution to be added
to $\alpha_{\rm an}(Q^2)$ as given in sec. 4.5 \cite{ip}. 
However, the experimental situation is so uncomplete and confused in this range
and the applicability of the theoretical method so questionable that no real
significant conclusion can be drawn.\\ 
In conclusion, the use in QCD of a RS, in which the
running coupling constant is free from Landau singularities, seems to
offer a framework in which all existing phenomenology can be discussed 
and non perturbative corrections usefully parametrized. It is even a
framework in which a comparison with Lattice Theory results is easier
and more natural.
\vspace{-0.5truecm}
\section{Acknowledgments and apologies}
\vspace{-0.4truecm}
We grateful acknowledge very fruitful discussions we have had on the subject
with our colleagues D.V. Shirkov, G. Marchesini, S.Forte, M. Baldicchi, 
N. Brambilla, A. Vairo.\\
On the contrary we regret for all very interesting papers to
which we were not able to give sufficient attention for lack of space 
or that we have possibly completely overlooked. They are certainly
many and again we apologize for that.    



\vspace{-0.5truecm}
\begin{thebibliography}{100}


\bibitem{monograph} 
  N.~N.~Bogoliubov and D.~V.~Shirkov, 
  \emph{Introduction to the Theory of Quantized Field}, 
  Wiley-Intersc., N.Y. (1980);
  S.~Weinberg,
  \emph{The Quantum Theory of Fields, Vol.II, Modern Applications},
  Cambridge University Press, Cambridge (1996).
\bibitem{Shirkov:1999hj}
  D.~V.~Shirkov and V.~F.~Kovalev,
  Phys.\ Rept.\  {\bf 352} (2001) 219;
  S.~Weinberg,
  Phys.\ Rev.\ D {\bf 8} (1973) 3497.
\bibitem{Brodsky:1982gc}
  S.~J.~Brodsky, G.~P.~Lepage and P.~B.~Mackenzie,
  Phys.\ Rev.\ D {\bf 28} (1983) 228;
  S.~J.~Brodsky, C.~R.~Ji, A.~Pang and D.~G.~Robertson,
  Phys.\ Rev.\ D {\bf 57} (1998) 245;
  G.~Grunberg,
  Phys.\ Rev.\ D {\bf 46} (1992) 2228.
\bibitem{Grunberg:1980ja}
  G.~Grunberg,
  Phys.\ Rev.\ D {\bf 29} (1984) 2315;  
  Phys.\ Lett.\ B {\bf 95} (1980) 70
  [Erratum-ibid.\ B {\bf 110} (1982) 501];
  Phys.\ Rev.\ D {\bf 40} (1989) 680.
\bibitem{Gribov:1977wm}
  V.~N.~Gribov,
  Nucl.\ Phys.\ B {\bf 139} (1978) 1; 
  L.~Baulieu and M.~Schaden,
  Int.\ J.\ Mod.\ Phys.\ A {\bf 13} (1998) 985 and references therein.
\bibitem{bethke}
  S.~Bethke,
  (in this issue) arXiv:hep-ex/0606035; 
  Nucl.\ Phys.\ Proc.\ Suppl.\  {\bf 135} (2004) 345; 
  {\bf 121} (2003) 74;
  J.\ Phys.\ G {\bf 26} (2000) R27.


\bibitem{Caswell:1974cj}
  W.~E.~Caswell and F.~Wilczek,
  Phys.\ Lett.\ B {\bf 49} (1974) 291.
\bibitem{Gross:1973id}
  D.~J.~Gross and F.~Wilczek,
  Phys.\ Rev.\ Lett.\  {\bf 30} (1973) 1343.
\bibitem{Caswell:1974gg}
  W.~E.~Caswell,
  Phys.\ Rev.\ Lett.\  {\bf 33} (1974) 244;
  D.~R.~T.~Jones,
  Nucl.\ Phys.\ B {\bf 75} (1974) 531.
\bibitem{Tarasov:1980au}
  O.~V.~Tarasov, A.~A.~Vladimirov and A.~Y.~Zharkov,
  Phys.\ Lett.\ B {\bf 93} (1980) 429.
\bibitem{Larin:1993tp}
  S.~A.~Larin and J.~A.~M.~Vermaseren,
  Phys.\ Lett.\ B {\bf 303} (1993) 334.
\bibitem{vanRitbergen:1997va}
  T.~van Ritbergen, J.~A.~M.~Vermaseren and S.~A.~Larin,
  Phys.\ Lett.\ B {\bf 400} (1997) 379.
\bibitem{Czakon:2004bu}
  M.~Czakon,
  Nucl.\ Phys.\ B {\bf 710} (2005) 485; 
  T.~R.~Morris and O.~J.~Rosten,
  arXiv:hep-th/0606189.


\bibitem{Bardeen:1978yd}
  W.~A.~Bardeen, A.~J.~Buras, D.~W.~Duke and T.~Muta,
  Phys.\ Rev.\ D {\bf 18} (1978) 3998.
\bibitem{Collins} 
  J.~C.~Collins, 
  \emph{Renormalization}, Cambridge University Press (1984).
\bibitem{Corless:1996}
  R.~M.~Corless, G.~H.~Gonnet, D.~E.~G.~Hare, D.~J.~Jeffrey, and D.~E.~Knuth,
  Advances in Computation Mathematics,\ V {\bf 5} (1996) 329.
\bibitem{Grunberg98}
  E.~Gardi, G.~Grunberg and M.~Karliner,
  JHEP {\bf 9807} (1998) 007.
\bibitem{pdg}
  S.~Eidelman {\it et al.}  [Particle Data Group],
  Phys.\ Lett.\ B {\bf 592} (2004) 1.
\bibitem{Kourashev:1999ye}
  D.~S.~Kourashev,
  arXiv:hep-ph/9912410.
\bibitem{Gonsalves79}
  W.~Celmaster and R.~J.~Gonsalves,
  Phys.\ Rev.\ D {\bf 20} (1979) 1420.


\bibitem{Bernreuther:1997jn}
  W.~Bernreuther, A.~Brandenburg and P.~Uwer,
  Phys.\ Rev.\ Lett.\  {\bf 79} (1997) 189; 
  K.~G.~Chetyrkin and J.~H.~Kuhn,
  Phys.\ Lett.\ B {\bf 406} (1997) 102.
\bibitem{Vermaseren:1997fq}
  J.~A.~M.~Vermaseren, S.~A.~Larin and T.~van Ritbergen,
  Phys.\ Lett.\ B {\bf 405} (1997) 327; 
  K.~G.~Chetyrkin and A.~Retey,
  Nucl.\ Phys.\ B {\bf 583} (2000) 3.
\bibitem{Melnikov:2000qh}
  K.~Melnikov and T.~v.~Ritbergen,
  Phys.\ Lett.\ B {\bf 482} (2000) 99.
\bibitem{Appelquist:1974tg}
  T.~Appelquist and J.~Carazzone,
  Phys.\ Rev.\ D {\bf 11} (1975) 2856.
\bibitem{Bernreuther:1983zp}
  W.~Bernreuther,
  Annals Phys.\  {\bf 151} (1983) 127;
  W.~Bernreuther and W.~Wetzel,
  Nucl.\ Phys.\ B {\bf 197} (1982) 228
  [Erratum-ibid.\ B {\bf 513} (1998) 758]; 
  W.~J.~Marciano,
  Phys.\ Rev.\ D {\bf 29} (1984) 580.
\bibitem{Larin:1994va}
  S.~A.~Larin, T.~van Ritbergen and J.~A.~M.~Vermaseren,
  Nucl.\ Phys.\ B {\bf 438} (1995) 278.
\bibitem{Chetyrkin:1997sg}
  K.~G.~Chetyrkin, B.~A.~Kniehl and M.~Steinhauser,
  Phys.\ Rev.\ Lett.\  {\bf 79} (1997) 2184.
\bibitem{Chetyrkin:1997un}
  K.~G.~Chetyrkin, B.~A.~Kniehl and M.~Steinhauser,
  Nucl.\ Phys.\ B {\bf 510} (1998) 61;
  G.~Rodrigo, A.~Pich and A.~Santamaria,
  Phys.\ Lett.\ B {\bf 424} (1998) 367;
  G.~Rodrigo and A.~Santamaria,
  Phys.\ Lett.\ B {\bf 313} (1993) 441.
\bibitem{Shirkov:1990vw}
  D.~V.~Shirkov,
  Nucl.\ Phys.\ B {\bf 371} (1992) 467;
  D.~V.~Shirkov and S.~V.~Mikhailov,
  Z.\ Phys.\ C {\bf 63} (1994) 463.


\bibitem{Moorhouse:1976qq}
  R.~G.~Moorhouse, M.~R.~Pennington and G.~G.~Ross,
  Nucl.\ Phys.\ B {\bf 124} (1977) 285.
\bibitem{Yndurain} 
  F.~J.~Yndurain, 
  \emph{The Theory of Quark and Gluon Interactions} 
  (Springer-Verlag, 1999). 
\bibitem{Adler:1974gd}
  S.~L.~Adler,
  Phys.\ Rev.\ D {\bf 10} (1974) 3714.
\bibitem{Chetyrkin:1979bj}
  K.~G.~Chetyrkin, A.~L.~Kataev and F.~V.~Tkachov,
  Phys.\ Lett.\ B {\bf 85} (1979) 277; 
  W.~Celmaster, R.~J.~Gonsalves,
  Phys.\ Rev.\ Lett.\  {\bf 44} (1980) 560; 
  M.~Dine, J.~R.~Sapirstein,
  Phys.\ Rev.\ Lett.\  {\bf 43} (1979) 668.
\bibitem{Gorishnii:1990vf}
  S.~G.~Gorishnii, A.~L.~Kataev and S.~A.~Larin,
  Phys.\ Lett.\ B {\bf 259} (1991) 144; 
  L.~R.~Surguladze and M.~A.~Samuel,
  Phys.\ Rev.\ Lett.\  {\bf 66} (1991) 560
  [Erratum-ibid.\  {\bf 66} (1991) 2416].
\bibitem{Kataev:1995vh}
  A.~L.~Kataev and V.~V.~Starshenko,
  Mod.\ Phys.\ Lett.\ A {\bf 10} (1995) 235.
\bibitem{Bjorken:1989xw}
  J.~D.~Bjorken,
  SLAC-PUB-5103
  {\it {Invited lectures given at 1989 Cargese Summer Inst. in Particle Physics, Cargese, 
  France, Jul 18 - Aug 4, 1989}}.
\bibitem{Pennington:1981cw}
  M.~R.~Pennington and G.~G.~Ross,
  Phys.\ Lett.\ B {\bf 102} (1981) 167.
\bibitem{Pennington:1983rz}
  M.~R.~Pennington, R.~G.~Roberts and G.~G.~Ross,
  Nucl.\ Phys.\ B {\bf 242} (1984) 69. 
\bibitem{Parisi:1980jy}
  G.~Parisi and R.~Petronzio,
  Phys.\ Lett.\ B {\bf 94} (1980) 51. 
\bibitem{Radyushkin:1982kg}
  A.~V.~Radyushkin,
  JINR Rapid Commun.\  {\bf 78} (1996) 96
  [arXiv:hep-ph/9907228].
\bibitem{Krasnikov:1982fx}
  N.~V.~Krasnikov and A.~A.~Pivovarov,
  Phys.\ Lett.\ B {\bf 116} (1982) 168.
\bibitem{Jones:1995rd}
  H.~F.~Jones and I.~L.~Solovtsov,
  Phys.\ Lett.\ B {\bf 349} (1995) 519.


\bibitem{richardson}
  J. L, Richarson,
  Phys.\ Lett.\ B {\bf 82} (1979) 272.
\bibitem{buchmuller}
  W.~Buchm\"uller, G.~Grunberg and S.-H.~H.~Tye,
  Phys.\ Rev.\ Lett.\ {\bf 45} (1980) 103; 
  Phys.\ Rev.\ Lett.\ {\bf 45} (1980) 587.
\bibitem{fischler}
  W. Fischler, Nucl.\ Phys. \ B {\bf 129} (1977) 157; 
  A. Billoire, Phys.\ Lett. \ B {\bf 92} (1980) 343;  
  M. Peter, Phys.\ Rev.\ Lett. \ {\bf 78} (1997) 602; 
  Y.~Schroder, Phys.\ Lett.\ B {\bf 447} (1999) 321
\bibitem{appelquist}
  T. Appelquist, M. Dine  and I. J. Muzinich, Phys.\ Rev.\ 
  D {\bf 17} (1978), 2074; Phys. Lett. B {\bf 69} (1977) 231;
  N.~Brambilla, A.~Pineda, J.~Soto and A.~Vairo,
  Phys.\ Rev.\ D {\bf 60} (1999) 091502;
  Nucl.\ Phys.\ B {\bf 566} (2000) 275.
\bibitem{eichten}
  E.~Eichten and F.~Feinberg,
  Phys.\ Rev.\  D {\bf 23} (1989) 2724.
\bibitem{bcp}
  N.~Brambilla, P.~Consoli and G.~M.~Prosperi,
  Phys.\ Rev.\ D {\bf 50} (1994) 5878; 
  A. Barchielli, N. Brambilla, G. M. Prosperi, 
  Nuovo\ Cimento\ A {\bf 103} (1990) 59; 
  A. Barchielli, E. Montaldi, G. M. Prosperi, 
  Nucl.\ Phys.\ B {\bf 296} (1988) 625. 
\bibitem{simonov}
  Yu.~A.~Simonov,
  Nucl.\ Phys.\ B {\bf 324} (1989) 67.
\bibitem{dosh}
  H.~G.~Dosh and Yu.~A.~Simonov,
  Phys.\ Lett.\ B {\bf 205} (1988) 339.
\bibitem{baker}
  M. Baker, J. S. Ball, N, Brambilla, G. M. Prosperi, F. Zachariasen,
  Phys.\ Rev.\ D {\bf 54} (1996) 2829; 
  M. Baker, J. S. Ball, F. Zachariasen,
  Phys.\ Rev.\ D {\bf 51} (1995) 1968.  
\bibitem{nikolaev92}
  N. N. Nikolaev and B. G. Zakharov, Z.\ Phys.
  C {\bf 49} (1991) 607;
  Z.\ Phys.\ C {\bf 53} (1992) 331;
  V.~Barone, M.~Genovese, N.~N.~Nikolaev, E.~Predazzi and B.~G.~Zakharov,
  Z.\ Phys.\ C {\bf 58} (1993) 541;
  G.~M.~Prosperi and M.~Baldicchi,
  Fizika B {\bf 8} (1999) 251.
\bibitem{Higashijima:1983gx}
  K.~Higashijima,
  Phys.\ Rev.\ D {\bf 29} (1984) 1228.
\bibitem{isgur} 
  S. Godfrey and N. Isgur,
  Phys.\ Rev.\ D {\bf 32} (1985) 189;
  T. Barnes, Z.\ Phys.\ C {\bf 11} (1981) 135;
  T. Barnes, F. E. Close and S. Monaghan, 
  Nucl.\ Phys.\ B {\bf 198} (1983) 380.
\bibitem{Zhang}
  T. Zhang and R. Koniuk,
  Phys.\ Lett.\ B  {\bf 261} (1991) 311;
  T. Zhang, 
  Phys.\ Rev.\ D {\bf 42} (1990) 3764.
\bibitem{Halzen:1992vd}
  F.~Halzen, G.~I.~Krein and A.~A.~Natale,
  Phys.\ Rev.\ D {\bf 47} (1993) 295.
\bibitem{Ball:1995ni}
  P.~Ball, M.~Beneke and V.~M.~Braun,
  Nucl.\ Phys.\ B {\bf 452} (1995) 563;
  M.~Beneke and V.~M.~Braun, V. I. Zakharov,
  Phys.\ Rev\ Lett.\ {\bf 73} (1994) 3058;
  M.~Beneke and V.~M.~Braun,
  Nucl.\ Phys.\ B {\bf 454} (1995) 253;
  B. R. Webber,
  Phys.\ Lett.\ B {\bf 339} (1994) 148.
\bibitem{Cornwall:1981zr}
  J.~M.~Cornwall,
  Phys.\ Rev.\ D {\bf 26} (1982) 1453;
  J.~M.~Cornwall and J.~Papavassiliou,
  Phys.\ Rev.\ D {\bf 40} (1989) 3474;
  J.~Papavassiliou and J.~M.~Cornwall,
  Phys.\ Rev.\ D {\bf 44} (1991) 1285.
\bibitem{Field:2001iu}
  J.~H.~Field,
  Phys.\ Rev.\ D {\bf 66} (2002) 013013;
  D.~V.~Shirkov,
  Phys.\ Atom.\ Nucl.\  {\bf 62} (1999) 1928
  [Yad.\ Fiz.\  {\bf 62} (1999) 2082];
  A.~M.~Badalian and D.~S.~Kuzmenko,
  Phys.\ Rev.\ D {\bf 65} (2002) 016004.


\bibitem{Beneke:1994qe}
  M.~Beneke and V.~M.~Braun,
  Phys.\ Lett.\ B {\bf 348} (1995) 513;
  P.~Ball, M.~Beneke and V.~M.~Braun,
  Nucl.\ Phys.\ B {\bf 452} (1995) 563.
\bibitem{Brodsky:1994eh}
  S.~J.~Brodsky and H.~J.~Lu,
  Phys.\ Rev.\ D {\bf 51} (1995) 3652.
\bibitem{cmw}
  S.~Catani, G.~Marchesini and B.~R.~Webber,
  Nucl.\ Phys.\ B {\bf 349} (1991) 635.
\bibitem{dokshitzer1}
  Yu.~L.~Dokshitzer, V.~A.~Khoze and S.~L.~Troyan,
  Phys.\ Rev.\ D {\bf 53} (1996) 89.
\bibitem{dokshitzer2}
  Yu.~L.~Dokshitzer, G.~Marchesini and B.~R.~Webber,
  Nucl.\ Phys.\ B {\bf 469} (1996) 93.
\bibitem{brodsky03}
  S.~J.~Brodsky, S.~Menke, C.~Merino and J.~Ratsham,
  Phys.\ Rev.\ D {\bf 67} (2003) 055008 and references therein.
\bibitem{ALEPH}
  (ALEPH Collaboration), R. Barate et al.
  Eur.\ Phys.\ J. {\bf C 4} (1998) 409;
  (OPAL Collaboration), K. Ackerstaff et al.
  Eur.\ Phys.\ J. {\bf C 7} (1999) 571.
  E.~Braaten, S.~Narison and A.~Pich,
  Nucl.\ Phys.\ B {\bf 373} (1992) 581.
\bibitem{Marciano:1988vm}
  W.~J.~Marciano and A.~Sirlin,
  Phys.\ Rev.\ Lett.\  {\bf 61} (1988) 1815.
\bibitem{pich}
  F.~Le Diberder and A.~Pich,
  Nucl.\ Phys.\ B {\bf 289} (1992) 165.
\bibitem{Grunberg:2006jx}
  G.~Grunberg,
  Phys. Rev. D {\bf 73} (2006) 091901; arXiv:hep-ph/0601140; see also
  E.~Gardi and R.~G.~Roberts,
  Nucl.\ Phys.\ B {\bf 653} (2003) 227.
\bibitem{sudakov}
  V.~V.~Sudakov, 
  Sov.\ Phys.\ JETP {\bf 3} (1956) 65;
  S.~Catani, L.~Trentadue,
  Nucl.\ Phys.\ B {\bf 327 } (1989) 323;
  G.~Sterman,
  Nucl.\ Phys.\ B {\bf 281} (1987) 310;
  S.~Forte and G.~Ridolfi,
  Nucl.\ Phys.\ B {\bf 650} (2003) 229.
\bibitem{Watson:1996fg}
  N.~J.~Watson,
  Nucl.\ Phys.\ B {\bf 494} (1997) 388;
  Nucl.\ Phys.\ B {\bf 552} (1999) 461;
  J.~Papavassiliou,
  Phys.\ Rev.\ D {\bf 62} (2000) 045006;
  D.~Binosi and J.~Papavassiliou,
  Nucl.\ Phys.\ Proc.\ Suppl.\  {\bf 121} (2003) 281.
 

\bibitem{stevenson}
 A. C. Mattingly and P. M. Stevenson,
 Phys.\ Rev.\ D {\bf 49} (1994) 437.
\bibitem{OPT}
 P. M. Stevenson, 
 Phys. Rev. D {\bf 23} (1981) 2916.
\bibitem{optult}
  P. M. Stevenson, 
  Nucl. Phys. B {\bf 231} (1984) 65.
\bibitem{KSS}
  J. Kubo, S. Sakakibara, and P. M. Stevenson, 
  Phys. Rev. D {\bf 29} (1984) 1682.
\bibitem{chyla2}
  J. Ch\'yla, A. Kataev, and S. Larin,  
  Phys. Lett. B {\bf 267} (1991) 269.
\bibitem{us}
  A. C. Mattingly and P. M. Stevenson, 
  Phys. Rev. Lett. {\bf 69} (1992) 1320.
\bibitem{higgs}
  A. L. Kataev, S. A. Larin, and  L. R. Surguladze, 
  Phys. Rev. D {\bf 43} (1991) 1633.


\bibitem{Shirkov:1997wi}
  D.~V.~Shirkov and I.~L.~Solovtsov,
  Phys.\ Rev.\ Lett.\  {\bf 79} (1997) 1209;
  Nucl.\ Phys.\ Proc.\ Suppl.\  {\bf 64} (1998) 106.
\bibitem{Ginzburg:1966}
  I.~F.~Ginzburg and D.~V.~Shirkov,
  Sov.\ Phys.\ JEPT\  {\bf 22} (1966) 234.
\bibitem{Redmond:1958pe}
  P.~J.~Redmond,
  Phys.\ Rev.\  {\bf 112} (1958) 1404;
  P.~J.~Redmond and J.~L.~Uretsky,
  Phys.\ Rev.\ Lett.\  {\bf 1} (1958) 147;
  N.~N.~Bogoliubov, A.~A.~Logunov and D.~V.~Shirkov,
  Sov.\ Phys.\ JETP {\bf 37} (1960) 574.
 \bibitem{Solovtsov:1999in}
  I.~L.~Solovtsov and D.~V.~Shirkov,
  Theor.\ Math.\ Phys.\  {\bf 120} (1999) 1220
  [Teor.\ Mat.\ Fiz.\  {\bf 120} (1999) 482].
\bibitem{Milton:1997mi}
  K.~A.~Milton, I.~L.~Solovtsov and O.~P.~Solovtsova,
  Phys.\ Lett.\ B {\bf 415} (1997) 104.
\bibitem{Krasnikov:1995is}
  N.~V.~Krasnikov and A.~A.~Pivovarov,
  Phys.\ Atom.\ Nucl.\  {\bf 64} (2001) 1500
  [Yad.\ Fiz.\  {\bf 64} (2001) 1576].
\bibitem{Alekseev:2000}
  A.~I.~Alekseev,
  Phys.\ Rev.\ D {\bf 61} (2000) 114005.
\bibitem{Alekseev:2002jb}
  A.~I.~Alekseev,
  Phys.\ Atom.\ Nucl.\  {\bf 65} (2002) 1678
  [Yad.\ Fiz.\  {\bf 65} (2002) 1722]; 
  J.\ Phys.\ G {\bf 27} (2001) L117.
\bibitem{Alekseev:2002zn}
  A.~I.~Alekseev,
  Few Body Syst.\  {\bf 32} (2003) 193.
\bibitem{Magradze:1999um}
  B.~A.~Magradze,
  Int.\ J.\ Mod.\ Phys.\ A {\bf 15} (2000) 2715;
  arXiv:hep-ph/9808247.
\bibitem{Shirkov:2005sg}
  D.~V.~Shirkov and A.~V.~Zayakin,
  arXiv:hep-ph/0512325.
\bibitem{Magradze:2000hz}
  B.~A.~Magradze,
  arXiv:hep-ph/0010070.
\bibitem{Kurashev:2003pt}
  D.~S.~Kurashev and B.~A.~Magradze,
  Theor.\ Math.\ Phys.\  {\bf 135} (2003) 531
  [Teor.\ Mat.\ Fiz.\  {\bf 135} (2003) 95]; 
  B.~A.~Magradze,
  arXiv:hep-ph/0305020.
\bibitem{Milton:1997us}
  K.~A.~Milton and O.~P.~Solovtsova,
  Phys.\ Rev.\ D {\bf 57} (1998) 5402; 
  K.~A.~Milton and I.~L.~Solovtsov,
  Phys.\ Rev.\ D {\bf 55} (1997) 5295. 
\bibitem{Solovtsov:1997at}
  I.~L.~Solovtsov and D.~V.~Shirkov,
  Phys.\ Lett.\ B {\bf 442} (1998) 344.
\bibitem{Schwinger:1975th}
  J.~S.~Schwinger,
  Proc.\ Nat.\ Acad.\ Sci.\  {\bf 71}, 5047 (1974);
  K.~A.~Milton,
  Phys.\ Rev.\ D {\bf 10} (1974) 4247.
\bibitem{Milton:1998wi}
  K.~A.~Milton and I.~L.~Solovtsov,
  Phys.\ Rev.\ D {\bf 59} (1999) 107701.

 
\bibitem{Alekseev:2004vx}
  A.~I.~Alekseev and B.~A.~Arbuzov,
  Mod.\ Phys.\ Lett.\ A {\bf 20} (2005) 103; 
  Mod.\ Phys.\ Lett.\ A {\bf 13} (1998) 1747.
\bibitem{Alekseev:2005vh}
  A.~I.~Alekseev,
  Theor.\ Math.\ Phys.\  {\bf 145} (2005) 1559
  [Teor.\ Mat.\ Fiz.\  {\bf 145} (2005) 221]; 
  arXiv:hep-ph/0503242.
\bibitem{Nesterenko:2001xa}
  A.~V.~Nesterenko and I.~L.~Solovtsov,
  Mod.\ Phys.\ Lett.\ A {\bf 16} (2001) 2517; 
  A.~V.~Nesterenko,
  Phys.\ Rev.\ D {\bf 62} (2000) 094028.
\bibitem{Nesterenko:2003xb}
  A.~V.~Nesterenko,
  Int.\ J.\ Mod.\ Phys.\ A {\bf 18} (2003) 5475; 
  Int.\ J.\ Mod.\ Phys.\ A {\bf 19} (2004) 3471.
\bibitem{Schrempp:2001ir}
  F.~Schrempp,
  J.\ Phys.\ G {\bf 28} (2002) 915.
\bibitem{Nesterenko:2004ry}
  A.~V.~Nesterenko and J.~Papavassiliou,
  Nucl.\ Phys.\ Proc.\ Suppl.\  {\bf 152} (2005) 47;
  Phys.\ Rev.\ D {\bf 71} (2005) 016009.
\bibitem{Aguilar:2005sb}
  A.~C.~Aguilar, A.~V.~Nesterenko and J.~Papavassiliou,
  J.\ Phys.\ G {\bf 31} (2005) 997.
\bibitem{Nesterenko:2005nj}
  A.~V.~Nesterenko and J.~Papavassiliou,
  arXiv:hep-ph/0507320;
  J. Phys. G {\bf 32} (2006) 1025.
\bibitem{Webber:1998um}
  B.~R.~Webber,
  JHEP {\bf 9810} (1998) 012.



\bibitem{Dokshitzer:1993pf}
  Y.~L.~Dokshitzer and D.~V.~Shirkov,
  Z.\ Phys.\ C {\bf 67} (1995) 449.
\bibitem{Dasgupta:2003iq}
  Yu.~L.~Dokshitzer, B.~R.~Webber,
  Phys.\ Lett.\ B {\bf 404} (1997) 321;
  S.~Catani, B.~R.~Webber,
  Phys.\ Lett.\ B {\bf 427} (1998) 377;
  Yu.~L.~Dokshitzer, A.~Lucenti, G.~Marchesini, G.~P.~Salam
  Nucl.\ Phys.\ B {\bf 511} (1998) 396;
  JHEP {\bf 05} (1998) 003;
  M.~Dasgupta and G.~P.~Salam,
  J.\ Phys.\ G {\bf 30} (2004) R143;
  JHEP {\bf 0208} (2002) 032;
  M.~Dasgupta,
  J.\ Phys.\ G {\bf 28} (2002) 907;  
  S.~S.~Agaev,
  Phys.\ Rev.\ D {\bf 69} (2004) 094010; 
  S.~Kluth,
  arXiv:hep-ex/0606046;
  T.~Kluge,
  arXiv:hep-ex/0606053.
\bibitem{Langfeld:1995si}
  K.~Langfeld, L.~von Smekal and H.~Reinhardt,
   Phys.\ Lett.\ B {\bf 362} (1995) 128.
\bibitem{Beneke:2000kc}
  M.~Beneke and V.~M.~Braun,
  arXiv:hep-ph/0010208;
  M.~Beneke,
  Phys.\ Rept.\  {\bf 317} (1999) 1
\bibitem{Cvetic:2002qf}
  G.~Cvetic,
  Phys.\ Rev.\ D {\bf 67} (2003) 074022;
  A.~L.~Kataev,
  JETP Lett.\  {\bf 81} (2005) 608
  [Pisma Zh.\ Eksp.\ Teor.\ Fiz.\  {\bf 81} (2005) 744].


\bibitem{Shirkov:1998sb}
  D.~V.~Shirkov,
  Theor.\ Math.\ Phys.\  {\bf 119} (1999) 438
  [Teor.\ Mat.\ Fiz.\  {\bf 119} (1999) 55]
\bibitem{Milton:1998cq}
  K.~A.~Milton, I.~L.~Solovtsov and O.~P.~Solovtsova,
  Phys.\ Lett.\ B {\bf 439} (1998) 421;
  Phys.\ Rev.\ D {\bf 60} (1999) 016001.
\bibitem{Milton:2000fi}
  K.~A.~Milton, I.~L.~Solovtsov, O.~P.~Solovtsova and V.~I.~Yasnov,
  Eur.\ Phys.\ J.\ C {\bf 14} (2000) 495;
  K.~A.~Milton, I.~L.~Solovtsov and O.~P.~Solovtsova,
  Phys.\ Rev.\ D {\bf 64} (2001) 016005;
  O.~P.~Solovtsova,
  Theor.\ Math.\ Phys.\  {\bf 134} (2003) 365
  [Teor.\ Mat.\ Fiz.\  {\bf 134} (2003) 416].
\bibitem{Shirkov:2000qv}
  D.~V.~Shirkov,
  Theor.\ Math.\ Phys.\  {\bf 127} (2001) 409; 
  arXiv:hep-ph/0003242.
\bibitem{Shirkov:2001sm}
  D.~V.~Shirkov,
  Eur.\ Phys.\ J.\ C {\bf 22} (2001) 331.
\bibitem{Geshkenbein:2001mn}
  B.~V.~Geshkenbein, B.~L.~Ioffe and K.~N.~Zyablyuk,
  Phys.\ Rev.\ D {\bf 64} (2001) 093009.
\bibitem{Stefanis:2000vd}
  N.~G.~Stefanis, W.~Schroers and H.~C.~Kim,
  Eur.\ Phys.\ J.\ C {\bf 18} (2000) 137;
  A.~I.~Karanikas and N.~G.~Stefanis,
  Phys.\ Lett.\ B {\bf 504} (2001) 225;
  N.~G.~Stefanis,
  Nucl. Phys. Proc. Suppl. {\bf 152} (2006) 245.
\bibitem{Bakulev:2004cu}
  A.~P.~Bakulev, K.~Passek-Kumericki, W.~Schroers and N.~G.~Stefanis,
  Phys.\ Rev.\ D {\bf 70} (2004) 033014
  [Erratum-ibid.\ D {\bf 70} (2004) 079906].
\bibitem{Shirkov:2005kj}
  D.~V.~Shirkov,
  AIP Conf.\ Proc.\  {\bf 806} (2006) 97
  [arXiv:hep-ph/0510247].
\bibitem{Magradze:2005ab}
  B.~A.~Magradze,
  arXiv:hep-ph/0512374.
\bibitem{Bakulev:2005fp}
  A.~P.~Bakulev, A.~I.~Karanikas and N.~G.~Stefanis,
  Phys.\ Rev.\ D {\bf 72} (2005) 074015; 
  A.~P.~Bakulev, S.~V.~Mikhailov and N.~G.~Stefanis,
  Phys.\ Rev.\ D {\bf 72} (2005) 074014
  [Erratum-ibid.\ D {\bf 72} (2005) 119908];
  N.~G.~Stefanis, A.~P.~Bakulev, K.~I.~Karanikas, S.~V.~Mikhailov, 
  arXiv:hep-ph/0601270.
\bibitem{Shirkov:2002td}
  D.~V.~Shirkov,
  Theor.\ Math.\ Phys.\  {\bf 136} (2003) 893
  [Teor.\ Mat.\ Fiz.\  {\bf 136} (2003) 3];
  D.~V.~Shirkov,
  arXiv:hep-ph/0408272.


\bibitem{Wilson74} 
  K.Wilson, Phys.\ Rev.\ D {\bf 10} (1974) 2445 and in \emph {New
  Phenomena in Subnuclear Physics}, edited by A.Zichichi (Plenum, New
  York,1977).
\bibitem{creutz} 
  M.Creutz, \emph {Quarks, Gluons and Lattice}, Cambridge
  University Press, 1983.
\bibitem{nieder}
  F.Niedermayer, Nucl.\ Phys.\ B (Proc.Suppl.) {\bf 53} (1997) 56. 
\bibitem{Weisz-rev}
  P.Weisz, Nucl.\ Phys.\  B (Proc.Suppl.) {\bf 47} (1996) 71.  
\bibitem{HH80}
  A.Hasenfratz and P.Hasenfratz, \ Phys.\ Lett.\  B {\bf 93}  (1980) 165. 
\bibitem{DG81}
  R.Dashen and D.J.Gross, Phys.\  Rev.\  D {\bf23} (1981) 2340.
\bibitem{sommer94}
  R.Sommer, Nucl.\ Phys.\  B {\bf411} (1994) 839. 
\bibitem{bali-scill-93}
  G.S.Bali and K.Schilling, Phys.\  Rev.\  D {\bf47} (1993) 661. 
\bibitem{Kadra-92}
  A.X.El-Khadra  et al., Phys.\ Rev.\ Lett.\  {\bf 69} (1992) 729.
\bibitem{Luscher94}
  M.L\"uscher et al., Nucl.\  Phys.\  B {\bf 413} (1994) 481. 
\bibitem{Booth}
  S.P.Booth et al., Phys.\ Lett.\  B {\bf 294} (1992) 385. 
\bibitem{lepagemackenzie}
  G.P.Lepage and P.Mackenzie, Phys.\ Rev.\  D {\bf 48} (1993) 2250.
\bibitem{Divitiis}
  G. de Divitiis et al., Nucl.\ Phys.\   B {\bf 433} (1995) 390;
  Nucl.\ Phys.\  B {\bf 437} (1995) 447. 
\bibitem{alles-97}
  B.Alles et al., Nucl.\ Phys.\ B {\bf 502} (1997) 325. 
\bibitem{pene98}
  Ph. Boucaud et al.,JHEP\  {\bf 10} (1998) 017;  
  JHEP {\bf 12} (1998) 004.  
\bibitem{giusti2001}
  L.Giusti et al., Int.\ J.\ Mod.\ Phys.\  A {\bf 16} (2001) 3487. 
\bibitem{cucchieri-gribov}
  A.Cucchieri, Nucl.Phys. B {\bf 508} (1997) 353.
\bibitem{alkofer-rev}
  R.Alkofer and L.von Smekal, Phys. Rep.{\bf 353} (2001) 281.
\bibitem{chet00} 
  K.G.Chetyrkin and A.R\'etey, [arXiv:hep-ph/0007088].
\bibitem{ball-chiu} 
  J.S.Ball and T.W.Chiu, Phys.\  Rev.\  D {\bf 22} (1980) 165.
\bibitem{skull}
  J.I.Skullerud, Nucl.\ Phys.\ B (Proc.Suppl.) {\bf 63} (1998) 242; 
  J.I.Skullerud and A.Kizilers\" u, 
  JHEP \ {\bf 09} (2002) 013. 
\bibitem{NRQCD}
  A.X El-Khadra, proceedings of the XXXIst
  Rencontre de Moriond on Electroweak Interactions and Unified
  Theories, Les Arcs 1800, France, March 16-23, 1996 [arXiv:hep-ph/9608220];
  C.T.H Davies et al., Phys.\  Rev.\  D {\bf 56} (1997) 2755; 
  A.Spitz et al. (SESAM coll.), Phys.\  Rev.\  D {\bf 60} (1999) 074502.  
\bibitem{pene2001}
  Ph. Boucaud et al.,  JHEP \ {\bf 01} (2002) 046. 
\bibitem{peneOPE}
  Ph. Boucaud et al., Phys.\ Lett.\  B {\bf 49}3 (2000) 315;
  Ph. Boucaud et al.,  JHEP \ {\bf 04} (2000) 006. 
\bibitem{Hashimoto} 
  S.Hashimoto, Int.\ J.\ Mod.\ Phys.\  A {\bf 20} (2005) 5133. 
\bibitem{HPQCD2003/4}
  C.T.H.Davies et al., Phys.\ Rev.\ Lett. {\bf 92} (2004) 022001; 
  Q.Mason et al., Nucl.\ Phys.\  B (Proc.Suppl.) {\bf 140} (2005) 713.
\bibitem{QCDSF-UKQCD} 
  S.Booth et al.\ [QCDSF-UKQCD collab.],
  Phys.\ Lett.\  B {\bf 519} (2001) 229. 
\bibitem {Gockeler:2004}
  M.Gockeler et al. [QCDSF collab.],
  Nucl.\ Phys.\ Proc.\ Suppl.\  {\bf 140} (2005) 228. 
\bibitem{aoki}
  S.Aoki et al., Phys.\ Rev.\ Lett.\  {\bf 74} (1995) 22. 
\bibitem{alpha2005}
  M.Della Morte et al., Nucl.\ Phys.\ B {\bf 713} (2005) 378. 
\bibitem{luscher97-98}
  M.L\"uscher, Talk given at the International
  Symposium on Lepton-Photon Interactions, Hamburg, July
  1997; 
  M.L\"uscher, Advanced Lattice QCD, Lectures given at
  Les Huches Summer School in Theoretical Physics, Les Huches, France, 1997.
\bibitem{alphamass}
  M.Della Morte et al., Nucl.\ Phys.\ B {\bf 729} (2005) 117. 
\bibitem{capitani}
  S.Capitani et al., Nucl.\ Phys.\ B {\bf 544} (1999) 669. 
\bibitem{lastalpha}
  Q.Mason et al., Phys.\ Rev.\ Lett.\ {\bf 95} (2005) 052002. 
\bibitem{pene-instantons}
  Ph.Boucaud et al., JHEP\ {\bf 04} (2003) 005. 
\bibitem{alphainfrared}
  J.Heitger et al., Nucl.\ Phys.\ B (Proc.Suppl.) {\bf 106} (2002) 859. 
\bibitem{DSE}
  C.S.Fischer and L.Alkofer, Phys.\ Lett.\ B {\bf 536} (2002) 177; 
  Phys.\ Rev.\ D {\bf 67} (2003) 094020; 
  C.S.Fischer, L.Alkofer and H.Reinhardt, Phys.\ Rev.\ D {\bf 65} (2002) 094008;
  C.~S.~Fischer,
  J.\ Phys.\ G {\bf 32} (2006) R253.
\bibitem{Bloch}
  J.C.R.Bloch, A.Cucchieri, K.Langfeld and T.Mendes,
  Nucl.Phys.B {\bf 687} (2004) 76. 
\bibitem{furui}
  S.Furui and H.Nakajima, Phys.\ Rev.\ D {\bf69} (2004) 074505. 
\bibitem{sternbeck}
  A.Sternbeck et al., PoS (LAT2005) 333 [arXiv:hep-lat/0509090];
  Phys.\ Rev.\ D {\bf72} (2005) 014507.
\bibitem{fischer}
  C.S.Fischer, PoS (LAT2005) 330 [arXiv:hep-lat/0509031].
\bibitem{fischer2}
  C.S.Fischer, B.Gr\"uter and R.Alkofer, Ann. of Phys. {\bf 321} (2006) 1918;
  O.~Oliveira and P.~J.~Silva,
  arXiv:hep-lat/0609027.
\bibitem{Shirkov-IR2002}
  D.V.Shirkov,
  Theor.\ Math.\ Phys.\ {\bf132} (2002) 1309 [Teor.\ Mat.\ Fis.\ {\bf 132} (2002) 484]. 


\bibitem{Baldicchi:2004wj}
  M.~Baldicchi and G.~M.~Prosperi,
  AIP Conf.\ Proc.\  {\bf 756} (2005) 152;
  Phys.\ Rev.\ D {\bf 66} (2002), 074008.
\bibitem{ip}
  M.~Baldicchi, G.~M.~Prosperi, D.~V.~Shirkov and C.~Simolo, in preparation.

\end{thebibliography}
\end{document}